\documentclass[11pt,a4paper]{article}

\usepackage[ruled,vlined,linesnumbered]{algorithm2e}
\SetAlgorithmName{Pseudocode}{Pseudocode}{List of Pseudocodes}
\newcommand{\pragma}[1]{%
  {\ttfamily\#pragma omp #1}%
}

\usepackage[numbers,sort&compress]{natbib}
\usepackage{tikz}
\usepackage{adjustbox}
\usetikzlibrary{
    arrows.meta,
    positioning,
    calc,
    fit,
    backgrounds
}
\usepackage{listings}
\usepackage{microtype}
\usepackage{silence}
\usepackage{array}
\usepackage{tabularx}
\usepackage{booktabs}

\usepackage[
    left=20mm,
    right=20mm,
    top=22mm,
    bottom=22mm
]{geometry}

\usepackage[T1]{fontenc}
\usepackage[utf8]{inputenc}
\usepackage{lmodern}

\usepackage{amsmath,amssymb,amsfonts,amsthm}

\usepackage{graphicx}
\usepackage{float}
\usepackage{caption}
\usepackage{subcaption}
\usepackage{booktabs}
\usepackage{multirow}
\usepackage{array}
\usepackage{tabularx}

\usepackage{enumitem}
\usepackage{xcolor}
\usepackage{microtype}

\usepackage[numbers,sort&compress]{natbib}
\usepackage{url}
\usepackage[
    colorlinks=true,
    linkcolor=black,
    citecolor=black,
    urlcolor=blue
]{hyperref}
\usepackage{orcidlink}

\newcommand{\InteriorRef}[1]{%
  \hyperref[alg:interior-kernel]{%
    \textsc{Interior}\ensuremath{\left(#1\right)}%
  }%
}

\newcommand{\BoundaryRef}[1]{%
  \hyperref[alg:boundary-kernel]{%
    \textsc{Boundary}\ensuremath{\left(#1\right)}%
  }%
}

\newcolumntype{C}{>{\centering\arraybackslash}X}

\title{Evaluating OpenMP Offloading for Intra-node Multi-GPU Programming across NVIDIA, AMD, and Intel Architectures: A 3D Heat Transfer Case Study}

\author{%
Ezhilmathi Krishnasamy\,\orcidlink{0000-0002-1971-4973}$^{1,2,3,*}$\\[6pt]
\small $^{1}$ Rudolfovo-Science and Technology Centre Novo mesto \\
\small $^{2}$ Faculty of Mechanical Engineering, University of Ljubljana \\
\small $^{3}$ Department of Computer Science, FSTM, University of Luxembourg\\
\small $^{*}$ Corresponding author:
\href{mailto:ezhilmathi.krishnasamy@rudolfovo.eu}{ezhilmathi.krishnasamy@rudolfovo.eu}%
}

\date{}

\begin{document}

\maketitle

\begin{abstract}
Currently, most supercomputers are equipped with GPUs from manufacturers such as NVIDIA, AMD, or Intel, which provide substantial parallelism and high throughput. It is common for a single compute node (intranode) to host multiple GPUs, typically four or more. Therefore, effectively leveraging all these GPUs within a single compute node is essential for applications in scientific and engineering domains. However, several factors must be considered before utilizing these GPUs for scientific computing, including the implementation of data communication, the programming models available for use across these GPUs, and the level of performance that can be achieved with a single codebase across different GPU architectures and configurations within a single compute node. OpenMP Offloading is a prominent directive-based programming model that can be executed on all three GPU types: NVIDIA, AMD, and Intel. In this research, we present an analysis of the benefits and performance challenges of using OpenMP Offloading to address the 3D heat equations, which involve both primary computation, as well as halo computation and communication. For additional comparison and scalability study, we also consider the Conjugate Gradient method. We investigate how performance varies in relation to native GPU programming models—CUDA for NVIDIA, HIP for AMD, and SYCL for Intel. Furthermore, we demonstrate that OpenMP Offloading can achieve performance improvements of approximately 2x for 2 GPUs and around 4x for 4 GPUs when compared to single-GPU OpenMP Offloading implementations across all three GPU types. This analysis is conducted systematically through various OpenMP Offloading implementations that utilize different low-level APIs for memory allocation, memory transfer options (synchronous, asynchronous, and peer-to-peer), and other native GPU programming models such as CUDA (NVIDIA), HIP (AMD), and SYCL (Intel).
\end{abstract}

\vspace{4pt}
\noindent\textbf{Keywords:}
Multi-GPU; OpenMP Offloading; Performance Portability; Thread Mapping; Host-Device Data Transfer; Asynchronous Offloading; Compute-Communication Overlap; 3D Heat Equation; Stencil Computation; Sparse Matrix; Conjugate Gradient; CUDA; HIP; SYCL; NVIDIA; AMD; Intel GPU


\section{Introduction}
In recent decades, it has become increasingly common for supercomputers and small clusters equipped with Graphics Processing Units (GPUs) to accelerate computational applications. Typically, GPUs are employed to enhance the speed of these applications, a process often referred to as Offloading computation to GPUs. However, the diversity of hardware and the range of programming options available can present developers with numerous challenges. While this variety offers benefits, it can also necessitate significant adjustments or even complete rewrites of the original code to accommodate different GPU architectures using languages such as CUDA~\cite{nvidia_cuda_programming_guide}, SYCL~\cite{khronos_sycl_2020_spec}, and OpenCL~\cite{khronos_opencl_api_spec}. For instance, code designed for execution on an NVIDIA GPU using CUDA cannot be directly run on an AMD GPU; however, it can be relatively easily converted into a HIP~\cite{amd_hip_programming_guide} version for execution. Nevertheless, questions continue to arise about whether similar performance levels can be achieved on AMD GPUs as on NVIDIA GPUs, primarily due to the architectural differences between the two. For example, in the CUDA framework, threads are organized into groups called warps, consisting of 32 threads, whereas in AMD's HIP, 64 threads are grouped into wavefronts. Consequently, when converting code from CUDA to HIP, one must manually adjust the thread blocks to align with their respective architectures to achieve optimal performance. Moreover, CUDA programming requires a deep understanding of thread organization and other CUDA-specific syntax. Should we consider Intel GPUs, utilizing their architecture necessitates either SYCL or OpenMP Offloading. Thus, it is essential to explore alternative programming models that can be easily executed across NVIDIA, AMD, and Intel GPUs. Fortunately, such programming models do indeed exist. Low-level programming models, like SYCL and OpenCL, as well as directive-based programming models such as OpenMP Offloading, can be employed on these GPUs. However, both SYCL and OpenCL require extensive syntax learning and a deeper understanding of thread management for GPU kernels. In contrast, OpenMP Offloading integrates seamlessly into existing code with minimal changes, facilitating execution across all these GPUs. Therefore, in this research, we investigate the OpenMP Offloading~\cite{openacc_specification,openmp_6_offload_overview} programming model on intra-node (single compute node) GPUs, using the 3D steady-state heat equation as a test case. The 3D heat equation demands intensive kernel computation and halo exchanges, which are typical in many scientific computing routines. In addition to this, we also examine the conjugate gradient method, which is quite different from stencil computation. This will allow us to determine whether OpenMP Offloading exhibits similar performance behavior to the 3D stencil heat problem.

\subsection{Research Objectives}
Directive-based programming models have existed even before GPUs became a significant component of scientific computing. When GPUs began to play a major role in this field, there was an initial surge in demand for learning low-level programming models such as CUDA and OpenCL. However, over time, other programming models have emerged to support GPU programming, including OpenACC~\cite{openacc_34_spec,openacc_specification}, OpenMP Offloading, and SYCL.

While SYCL, CUDA, and OpenCL require a comprehensive understanding of thread organization to execute kernels or primary computations on the GPU, OpenACC and OpenMP Offloading simplify this process through their directive-based programming models. These models extend OpenMP syntax, providing a more straightforward programming approach and reducing complexity. Generally, GPU programming involves mapping parallel threads across the available GPU cores, based on the Single Instruction Multiple Threads (SIMT)~\cite{lindholm2008nvidia} architecture. CUDA, HIP, and SYCL effectively utilize this concept by employing threads and thread blocks, which include warps and 1D, 2D, and 3D thread block configurations. In contrast, directive-based programming models streamline this process with simplified syntax, significantly diminishing the amount of code required compared to programming models such as CUDA, OpenCL, and SYCL.

Although OpenACC is a directive-based programming model, it unfortunately suffers from limited compiler support for AMD and Intel GPUs. It can accommodate FORTRAN programming models on AMD but has limited support for C/C++~\cite{hpe_openacc_2024, Krishnasamy2026MPIXIntegration}. Therefore, this research aims to examine the feasibility of OpenMP Offloading across various well-established GPUs available in the market, including NVIDIA, AMD, and Intel GPUs within a single compute node. We will evaluate whether there are any performance variations among different GPUs when using the same source code with OpenMP Offloading. To this end, we focus on a key application: 3D heat transfer, which is commonly encountered in compute-intensive applications. In multi-GPU scenarios, this would necessitate complete computations, along with main and halo computations, as well as halo data exchange. Furthermore, we investigate conjugate gradient methods that address the sparse matrix problems arising from unstructured grids, in contrast to the structured grids used in stencil computations. This exploration will help determine whether we can draw general conclusions about the performance behavior of OpenMP Offloading, rather than basing our conclusions solely on a single application of a 3D stencil problem.

\subsection{Contributions}  
This research makes several significant contributions to the field, which are outlined below:  
\begin{itemize}  
    \item A comprehensive analysis of single-node multi-GPU OpenMP Offloading implementations, specifically focusing on three-dimensional heat equation solvers. Additionally, the conjugate gradient method is investigated to support the general performance behavior of OpenMP Offloading.  
    \item A detailed examination of data transfer methodologies between GPUs within a single compute node, utilizing benchmark cases proposed for evaluation.  
    \item The development of a performance model designed to analyze single-node multi-GPU systems across various hardware platforms, including NVIDIA, AMD, and Intel architectures.  
    \item The provision of source code and research artifacts to enhance the reproducibility of the findings; detailed instructions for accessing and reproducing these artifacts are provided in Artifact Description Appendix~\ref{app:artifact}.  
\end{itemize}

\subsection{Organization of the Paper}
This paper is organized as follows. Section~\ref{related-work} reviews several significant research contributions related to OpenMP Offloading. In Section~\ref{equation}, we present the mathematical details of the 3D heat equation, which serves as the use case for this research. Section~\ref{platform} offers a comprehensive overview of the hardware utilized, including details about the compiler and its relevant flags, ensuring reproducibility of our results. Section~\ref{methodology} provides an in-depth discussion of key proposals that enhance the feasibility of OpenMP Offloading across various GPUs in multi-GPU contexts. This section includes illustrative code snippets demonstrating different OpenMP Offloading techniques, along with the derivation of performance models. Finally, Section~\ref{discussion} delivers a thorough analysis of multiple implementations of OpenMP Offloading, assessing their feasibility across various GPU architectures, including those from NVIDIA, AMD, and Intel. This section also features an extensive discussion of the performance models. The paper concludes with a summary of the principal findings derived from this research.

\section{Related Work}
\label{related-work}
Numerous studies have investigated OpenMP Offloading, focusing on single GPU configurations, multiple GPUs within a single compute node (intra-node), and multiple nodes (inter-node). This section reviews a selection of the most pertinent research in this area. E. Krishnasamy et al.~\cite{Krishnasamy2024TsunamiOpenMP} examined the single GPU OpenMP Offloading option for parallelizing the Tsunami scientific code, originally written in FORTRAN. Their findings indicated that implementing OpenMP Offloading results in a significant speedup compared to traditional CPU-based multicore OpenMP implementations when utilizing an NVIDIA GPU. Fridman et al.~\cite{fridman2023portability} explored the application of OpenMP Offloading in conjunction with OpenMP Validation and Verification (OMPVV) as a test case on both NVIDIA and Intel GPUs, specifically focusing on single GPU scenarios. Furthermore, E. Krishnasamy et al.~\cite{Krishnasamy2026MPIXIntegration} assessed OpenMP Offloading across various GPUs in the context of matrix-vector multiplication. This research utilized low-level Application Programming Interfaces (APIs) for memory allocation and synchronized memory transfers, considering both GPU-aware and non-GPU-aware methodologies. Additional research has validated the advantages of utilizing low-level APIs for memory transfer and computational kernel optimizations, with significant investigations conducted by E. Krishnasamy et al.~\cite{Krishnasamy2025OpenMPAMDandNVIDIA, Krishnasamy2026ComparativeCUDAOpenMP} on NVIDIA and AMD GPUs. Moreover, Ferat et al.~\cite{Ferat2022EnhancingMPI} evaluated hidden helper thread configurations within heterogeneous MPI+OpenMP applications, while Tian et al.~\cite{Tian2022HiddenHelperThreads} investigated the concurrent execution of deferred OpenMP target tasks. However, no research has systematically explored the utilization of multiple GPUs within a single compute node using OpenMP Offloading—especially utilizing low-level APIs—across three distinct platforms: NVIDIA, AMD, and Intel. This research aims to address this gap by investigating various approaches related to low-level API memory management, diverse parallelization strategies, performance models, data transfer between the CPU and GPU, as well as inter-GPU communication, all within the context of typical scientific applications such as 3D heat transfer.

\section{GPU Programming Models}
\label{programming-model}
GPU programming models facilitate significant computational speedups by leveraging the capabilities of GPUs. These models can be broadly categorized into two types: low-level programming models and high-level programming models. Notable examples of low-level programming models include CUDA, HIP, OpenCL, and SYCL. In contrast, OpenACC and OpenMP Offloading are classified as high-level or directive-based programming models. Both OpenACC and OpenMP Offloading support hybrid programming approaches, which enable effective targeting of both CPUs and GPUs.

Like other programming models, OpenACC and OpenMP Offloading continuously improve their APIs with each release, introducing new features related to compute kernels, memory allocation, and data transfer~\cite{openacc_34_spec, openacc_specification, openmp_6_offload_overview}. These models provide three types of memory transfer between CPUs and GPUs: Structured High-Level API, Unstructured High-Level API, and Low-Level API for Data Movement~\cite{Krishnasamy2025OpenMPAMDandNVIDIA, Krishnasamy2026ComparativeCUDAOpenMP, Krishnasamy2025OpenACCCUDA}.

\begin{itemize}
    \item Structured High-Level API: This API is user-friendly, though it may be insufficient for larger computations.
    \item Unstructured High-Level API: While this API is also easy to use, its performance has some limitations.
    \item Low-Level API for Data Movement: This API delivers optimal performance, as it is closely aligned with low-level GPU programming models.
\end{itemize}

Data movement remains one of the primary bottlenecks in scientific computing. Computational workloads are often classified as either compute-bound or memory-bound. Compute-bound problems are primarily limited by the available computational throughput of the hardware and the arithmetic requirements of the task. In contrast, memory-bound problems are restricted mainly by the speed at which data can be transferred through the memory hierarchy. Their performance depends on various factors, including the nature of the problem, memory access patterns, hardware bandwidth and latency, cache utilization, and the efficiency of the software interfaces and implementations used for data transfer.

Previous research by E. Krishnasamy et al. \cite{Krishnasamy2025OpenMPAMDandNVIDIA, Krishnasamy2026ComparativeCUDAOpenMP, Krishnasamy2025OpenACCCUDA} demonstrated that utilizing low-level APIs for memory allocation and testing various combinations of optimized compute kernel options revealed that the default parallel constructs and clauses offer the best computational efficiency for Basic Linear Algebra Subroutines (BLAS). Consequently, this work will primarily focus on low-level APIs for memory allocation alongside default parallel constructs and clauses for compute kernels.

\section{Mathematical Details}
\label{equation}
To test our intra-node multi-GPU parallelization using OpenMP Offloading, we aimed to select an application that demands an intensive computational kernel and involves communication between the CPU and GPU, as well as among GPUs, such as in halo exchanges. A prime example of this need is the requirement for halo exchanges during each iteration across a 3D domain, a common scenario in scientific computing simulations. Given these considerations, we chose the 3D heat equation. Below, we present the mathematical details of the 3D model that we considered in this work:

\begin{equation}
\frac{\partial u}{\partial t}
=
\nabla \cdot \left( \kappa(x,y,z)\nabla u \right) + f(x,y,z),
\end{equation}

where $u(x,y,z,t)$ is the scalar field and $\kappa(x,y,z)$ is the spatially varying
diffusion coefficient. In this simulation, the spatially varying coefficient reads

\begin{equation}
\kappa(x,y,z) = 0.5 + 0.45\sin(\pi x)\sin(\pi y)\sin(\pi z),
\end{equation}

and the source term is defined by

\begin{equation}
f(x,y,z) = 1 + \cos(\pi x)\cos(\pi y)\cos(\pi z).
\end{equation}

The equation is solved on a uniform grid with spacing $h$ and time step $\Delta t$.
Let $u^m_{i,j,k}$ denote the numerical solution at time level $t_m$.
	Using explicit time stepping, the update scheme for interior grid points is

	\begin{equation}
	u^{m+1}_{i,j,k}
	=
	u^{m}_{i,j,k}
	+ \Delta t\,f_{i,j,k}
	+ \frac{\Delta t}{2h^2}\left(D_x + D_y + D_z\right),
	\end{equation}

	Boundary condition values are fixed to zero (Dirichlet boundary conditions), and the
	initial condition is

	\begin{equation}
	u(x,y,z,0)=\sin(\pi x)\sin(\pi y)\sin(\pi z).
	\end{equation}

	The solution is advanced in time using an explicit time-stepping scheme
	until the final simulation time $T$. For numerical stability, the time step
	must satisfy the CFL-type condition

	\begin{equation}
	\frac{\Delta t}{h^2} < 0.5 .
	\end{equation}

	\begin{algorithm}[htbp]
\DontPrintSemicolon
\SetAlgoLined

\caption{OpenMP parallel explicit solver for the 3D
variable-coefficient heat equation.}
\label{alg:heat3d}

\KwIn{Grid size $n$ and final time $T$}
\KwOut{Numerical solution $u$}

$h \gets 1/(n-1),\quad
 \kappa_{\max} \gets 0.95,\quad
 \Delta t \gets h^2/(12\kappa_{\max}),\quad
 \beta \gets \Delta t/(2h^2)$\;

\BlankLine

\tcp{Initialize the solution and coefficient arrays}
\pragma{parallel for collapse(3)}\;

\ForEach{grid node $(i,j,k)$}{

    $f_{i,j,k}
    \gets
    \Delta t
    \left[
        1+
        \cos(\pi x_i)
        \cos(\pi y_j)
        \cos(\pi z_k)
    \right]$\;

    $\kappa_{i,j,k}
    \gets
    0.5+
    0.45
    \sin(\pi x_i)
    \sin(\pi y_j)
    \sin(\pi z_k)$\;

    \eIf{$(i,j,k)$ is a boundary node}{
        $u_{i,j,k} \gets 0$\;
    }{
        $u_{i,j,k}
        \gets
        \sin(\pi x_i)
        \sin(\pi y_j)
        \sin(\pi z_k)$\;
    }
}

\BlankLine

$t \gets 0$\;

\pragma{parallel}\;
\textbf{\{}\;

\Indp

\While{$t<T$}{

    \tcp{Distribute the stencil update among the existing threads}
    \pragma{for collapse(3)}\;

    \ForEach{interior node $(i,j,k)$}{
        $u^{\mathrm{new}}_{i,j,k}
        \gets
        u_{i,j,k}
        +
        f_{i,j,k}
        +
        \beta\mathcal{L}(u)_{i,j,k}$\;
    }

    \BlankLine

    \tcp{One thread advances the shared time level}
    \pragma{single}\;
    \textbf{\{}\;

    \Indp
    $\operatorname{swap}(u,u^{\mathrm{new}})$\;
    $t \gets t+\Delta t$\;
    \Indm

    \textbf{\}}\;
}

\Indm

\textbf{\}}\;

\Return{$u$}\;

\end{algorithm}

	\begin{table}[htbp]
\caption{Comparison of technical specifications of H100, MI250X, and Max 1550 accelerators.\label{tab:h100-mi250x-max1550}}
\centering
\small
\setlength{\tabcolsep}{4pt}

\begin{tabularx}{\textwidth}{%
    >{\raggedright\arraybackslash\hsize=1.30\hsize}X
    >{\centering\arraybackslash\hsize=0.90\hsize}X
    >{\centering\arraybackslash\hsize=0.90\hsize}X
    >{\centering\arraybackslash\hsize=0.90\hsize}X
}
\toprule

\textbf{Property}
& \textbf{H100~\cite{nvidia_h100,bsc_mn5}}
& \textbf{MI250X~\cite{amd_mi250x,lumi_g}}
& \textbf{Max 1550~\cite{intel_max1550,lrz_supermucng2}}\\

\midrule
\multicolumn{4}{@{}l}{\textit{Compute \& memory (per accelerator package)}}\\
Compute units (vendor terminology)	& 132 SMs		& 220 CUs			& 128 Xe-cores\\
Compute dies/tiles					& 1				& 2 GCDs			& 2 tiles\\
Peak FP64 vector (GFLOP/s)			& 34{,}000		& 47{,}900			& $\sim$52{,}000\\
Memory bandwidth (GB/s)				& 3{,}350		& 3{,}276.8			& 3{,}276.8\\
			\midrule
\multicolumn{4}{@{}l}{\textit{GPU--GPU interconnect}}\\
Interconnect technology					& NVLink 4	& Infinity Fabric	& Xe Link\\
Links/ports per accelerator pair		& NV6		& 2 IF links		& 4 Xe Link ports\\
Total P2P links/ports per accelerator	& 18		& 5 P2P links		& 12 active ports\\
			\midrule
\multicolumn{4}{@{}l}{\textit{Host interconnect (GPU--CPU)}}\\
Host link							& PCIe Gen5\,$\times$16	& Infinity Fabric	& PCIe Gen5\,$\times$16\\
Host-link bandwidth, 1 dir. (GB/s)	& $\sim$64				& $\sim$36			& $\sim$64\\
			\bottomrule
		\end{tabularx}
\end{table}

	\begin{figure}[htbp]
    \centering
    \includegraphics[width=\linewidth]{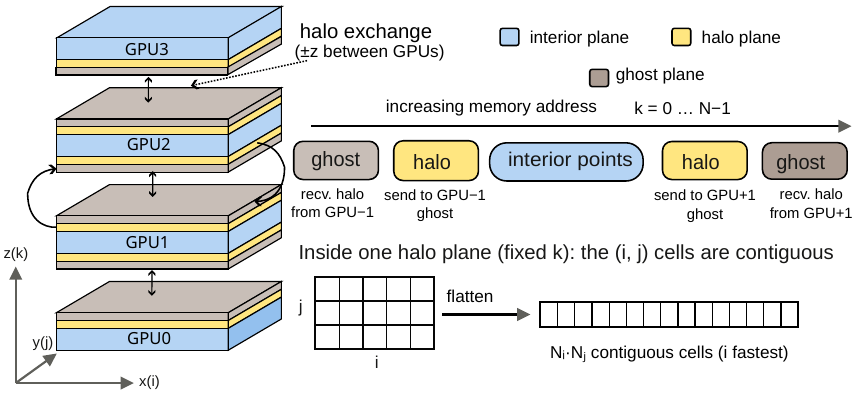}
    \caption{Schematic overview of the data partitioning in a 3D domain, illustrating halo data transfer during halo and interior-point computations.}
    \label{fig:3d-domain}
\end{figure}

	\begin{figure}[htbp]
    \centering

    \subfloat[H100 (BSC).\label{fig:h100_case}]{%
        \includegraphics[width=0.32\textwidth]{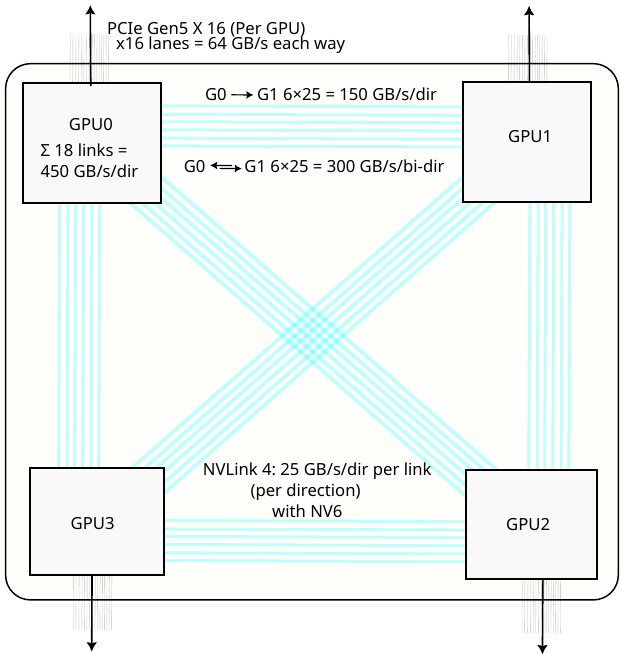}%
    }
    \hfill
    \subfloat[AMD MI250X (LUMI-G).\label{fig:amd_case}]{%
        \includegraphics[width=0.32\textwidth]{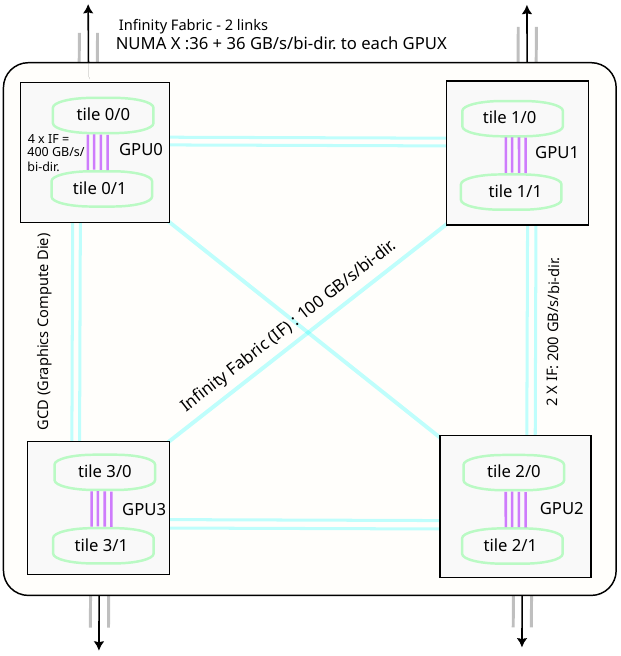}%
    }
    \hfill
    \subfloat[Intel Max 1550 (SuperMUC-NG).\label{fig:intel_case}]{%
        \includegraphics[width=0.32\textwidth]{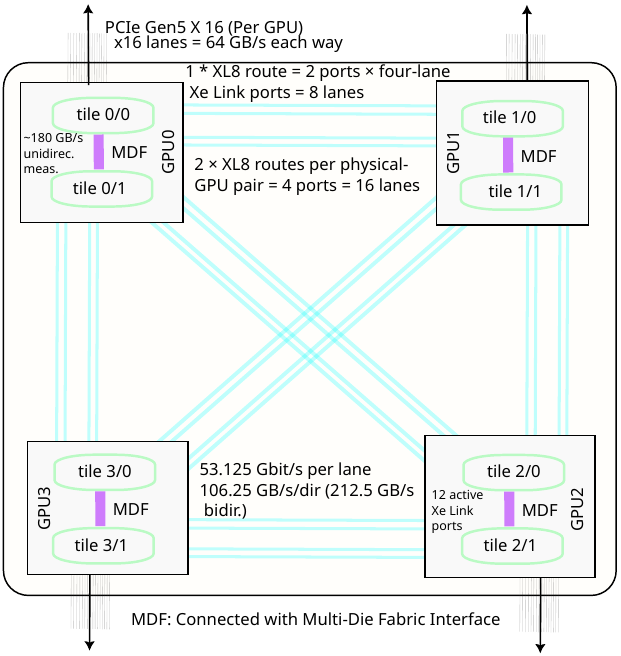}%
    }

\caption{Comprehensive comparison of the three GPUs, including theoretical memory bandwidth, CPU–GPU connectivity, and inter-GPU communication.}
\label{fig:gpu-architecture}
\end{figure}

\section{Testing Platforms and Experimental Setups}
\label{platform}
In this research, we examined three categories of GPUs: the NVIDIA H100 (BSC, Spain), the AMD MI250X (LUMI-G, Finland), and the Intel Data Center 1550 (SuperMUC-NG, Germany). Each GPU features a distinct architecture that provides varying levels of computational performance and memory bandwidth, as well as end-to-end network connectivity between the CPU and GPU, and among the GPUs on the NUMA node. Figure~\ref{fig:gpu-architecture} illustrates the data transfer network connectivity and theoretical bandwidths of these GPUs. Key specifications for each of the GPUs are presented in Table~\ref{tab:h100-mi250x-max1550}. For details regarding the compiler suite utilized for various tasks, please refer to Table~\ref{tab:compiler-support}. To evaluate the efficacy of our proposed methodologies, we implemented the 3D heat equation along with its multi-GPU parallelization strategies, which are described in Sections~\ref{equation} and~\ref{methodology}. This implementation was used to test both kernel and halo computations, as well as the necessary halo communication. We maintained a consistent time step of 500 while exploring various 3D domain sizes, ranging from \(512^3\) to \(1280^3\), with increments of \(128\). This parameterization enabled us to assess the runtime across the different GPU programming models and their respective implementations examined in this research.
\begin{table}[htbp]
\caption{Hardware and software environment per platform and programming models.\label{tab:compiler-support}}
\centering
\small
\setlength{\tabcolsep}{4pt}

\begin{tabularx}{\textwidth}{%
    >{\raggedright\arraybackslash\hsize=0.60\hsize}X
    >{\raggedright\arraybackslash\hsize=0.70\hsize}X
    >{\raggedright\arraybackslash\hsize=0.85\hsize}X
    >{\raggedright\arraybackslash\hsize=1.85\hsize}X
}
\toprule

\textbf{Platform}
& \textbf{GPU/arch}
& \textbf{Compiler}
& \textbf{Toolkit Stack \& Key Flags}\\

\midrule
\multirow[m]{5}{=}{NVIDIA H100}
	& \multirow[m]{5}{=}{Hopper, \texttt{sm\_90}}
	& NVHPC 25.3 (\texttt{nvcc})
	& CUDA: \texttt{NVHPC/25.3-CUDA-12.8.0}; \texttt{-O2 -arch=sm\_90 -{}-fmad=false -{}-prec-div=true -{}-prec-sqrt=true -{}-ftz=false}\\
			\cmidrule(lr){3-4}
	& & Clang 18.1.8
	& OpenMP: \texttt{clang/18.1.8-cuda12.8}, CUDA 12.8, GCC 13.2.0; \texttt{-O3 -fopenmp -{}-offload-arch=sm\_90 -ffp-contract=off}\\
			\midrule
AMD MI250X
	& CDNA2, \texttt{gfx90a} (2 GCDs/card)
	& AMDClang (ROCm 6.4.4)
	& HIP \& OpenMP: LUMI/25.09, PrgEnv-amd, \texttt{craype-accel-amd-gfx90a}, ROCm 6.4.4; OpenMP \texttt{-O3 -fopenmp -ffp-contract=off} (via \texttt{cc})\\
			\midrule
\multirow[m]{5}{=}{Intel Max 1550}
	& \multirow[m]{5}{=}{Xe-HPC, 2 stacks}
	& Intel oneAPI \texttt{icpx} (SYCL)
	& SYCL: \texttt{intel-toolkit} (+\texttt{intel-dpct}); \texttt{-fsycl -fiopenmp -ffp-contract=off}; SPIR-V JIT; \texttt{ZE\_FLAT\_DEVICE\_HIERARCHY=COMPOSITE}\\
			\cmidrule(lr){3-4}
	& & Intel oneAPI \texttt{icx} (OpenMP)
	& OpenMP: \texttt{intel-toolkit}; \texttt{-fiopenmp -fopenmp-targets=spir64 -ffp-contract=off}; \texttt{COMPOSITE}\\
			\bottomrule
		\end{tabularx}
\end{table}

\section{Methodology}
\label{methodology}
The primary objective of this research is to investigate the feasibility of utilizing OpenMP Offloading as a unified source code across the three GPUs examined in this research. Specifically, we aim to assess the performance enhancements achieved when employing one, two, and four GPUs to solve scientific computing problems, particularly in the context of three-dimensional (3D) heat transfer. Additionally, we seek to develop a performance model that can predict the runtime requirements for the current problem as well as for similar computational challenges.

To address these objectives, we propose the following key approaches. First, we will implement various versions of multi-GPU OpenMP Offloading to solve the heat equations on a single compute node (intra-node), which will allow us to identify the most effective optimization strategies. Second, we will explore additional optimization opportunities provided by OpenMP, which can enhance asynchronous operations related to multiple kernels and data transfer tasks when utilizing multi-GPU configurations within a single compute node.

Third, we will present detailed benchmark cases for data transfer that will enable us to evaluate performance under realistic conditions. Finally, we aim to develop a performance model that considers various data transfer scenarios and diverse OpenMP multi-GPU implementations. This model will assist in predicting runtimes for 3D heat transfer and similar computational problems.

\subsection{Multi-GPU OpenMP Offloading Versions}
We initialize the 3D computational domain in the order of k, j, i and subsequently partition the domain along the k direction. Algorithm~\ref{alg:heat3d} presents an OpenMP implementation of the 3D heat equations, where we delineate the primary computational segment designated for offloading to the GPUs. Figure~\ref{fig:3d-domain} illustrates the discretization of the 3D domain and clarifies how such a partitioning can enhance computational efficiency in a multi-GPU environment. It is crucial to note that omitting the split at the final index prevents the acquisition of flattened halo data. The core principle of this approach is to facilitate efficient halo computations while simply transferring contiguous data values for subsequent iterations.

To achieve efficient OpenMP Offloading, we draw from previous implementations of OpenMP Offloading \cite{Krishnasamy2025OpenMPAMDandNVIDIA, Krishnasamy2026ComparativeCUDAOpenMP}, which include low-level APIs specifically designed for memory allocation and data transfer between the Host and Device (HtoD) and Device and Host (DtoH). Furthermore, OpenMP Offloading provides options for synchronous, asynchronous, and Peer-to-Peer (P2P) data transfer, as detailed in Table~\ref{tab:high-low-spec}.

To evaluate the performance portability of OpenMP Offloading, we developed nine distinct variants: one version for single GPU configurations, four versions tailored for dual GPU setups, and four versions designed for quad GPU configurations:

\begin{itemize}
\item \texttt{OpenMP Offloading 1-GPU Baseline:} A baseline implementation utilizing a single GPU.
\item \texttt{OpenMP Offloading \{2,4\}-GPU Version 1:} A basic multi-GPU implementation.
\begin{itemize}
    \item A single host thread oversees all program calls, which encompass the execution of interior point computational kernels, halo computations, memory calls, and data synchronization associated with the copy APIs.
\end{itemize}
\item \texttt{OpenMP Offloading \{2,4\}-GPU Version 2:} An asynchronous operation that employs streams, directed by a single host thread.
\begin{itemize}
    \item This version uses a single host thread along with a streaming concept for asynchronous operations related to both interior and halo computations, as well as for data transfer.
\end{itemize}
\item \texttt{OpenMP Offloading \{2,4\}-GPU Version 3:} Utilization of multiple host threads to manage the GPUs in conjunction with streams, where each thread is responsible for an individual GPU.
\begin{itemize}
    \item Multiple host threads each govern a GPU, managing its respective kernel execution and data transfer while employing asynchronous data copying APIs.
\end{itemize}
\item \texttt{OpenMP Offloading \{2,4\}-GPU Version 4:} Facilitating P2P data transfer between GPUs using streams, managed by multiple host threads.
\begin{itemize}
    \item A scheme in which multiple host threads supervise the GPUs, with each thread responsible for kernel execution and data transfer of a single GPU, utilizing asynchronous data copying with P2P functionality APIs.
\end{itemize}
\end{itemize}

To facilitate a comparative analysis, we also developed equivalent implementations using CUDA, HIP, and SYCL, which will be benchmarked against the OpenMP Offloading versions. While one could theorize that the P2P implementation may yield superior performance, it is essential to evaluate it against the other three versions (i.e., Version 1: synchronous, Version 2: asynchronous, and Version 3: multiple host threads with asynchronous execution). To ensure a thorough comparison, we have also implemented three additional variants alongside Version 4 (P2P). Table~\ref{tab:implementation-comparison} provides a straightforward interpretation of these versions.
\begin{table}[htbp]
\caption{Summary of the comparison of the 2-GPU and 4-GPU parallel strategy implementations of \texttt{OpenMP Offloading \{2,4\}-GPU Versions}.\label{tab:implementation-comparison}}
\centering
\small
\setlength{\tabcolsep}{3pt}

\begin{tabularx}{\textwidth}{@{}
    >{\raggedright\arraybackslash}p{7.0cm}
    >{\centering\arraybackslash}p{2.2cm}
    >{\centering\arraybackslash}p{4.3cm}
    >{\raggedright\arraybackslash}X
@{}}
\toprule
\textbf{Version} &
\textbf{Host Threads} &
\textbf{Halo-Transfer Path} &
\textbf{Overlap with Computation} \\
\midrule
\texttt{OpenMP Offloading \{2,4\}-GPU Version 1} & 1    & GPU $\rightarrow$ host $\rightarrow$ GPU & None \\
\texttt{OpenMP Offloading \{2,4\}-GPU Version 2} & 1    & GPU $\rightarrow$ host $\rightarrow$ GPU & Limited \\
\texttt{OpenMP Offloading \{2,4\}-GPU Version 3} & 2, 4 & GPU $\rightarrow$ host $\rightarrow$ GPU & Yes, via host \\
\texttt{OpenMP Offloading \{2,4\}-GPU Version 4} & 2, 4 & GPU $\rightarrow$ GPU                    & Yes, via P2P \\
\bottomrule
\end{tabularx}
\end{table}

\subsubsection{OpenMP Offloading 1-GPU Baseline}
This implementation does not incorporate data partitioning, which eliminates the need for halo computations and halo data communication. OpenMP Offloading allows for comparable control over thread block configuration in targeted regions through the `num\_teams` and `thread\_limit` clauses, similar to CUDA's grid and block dimensions (refer to Table 4 in \cite{Krishnasamy2026ComparativeCUDAOpenMP}). Furthermore, OpenMP Offloading offers limited combinations of compute kernels with parallel constructs and clauses that, in theory, should enhance performance. This interplay has been examined in previous work by E. Krishnasamy \cite{Krishnasamy2026ComparativeCUDAOpenMP, Krishnasamy2025OpenMPAMDandNVIDIA}, as shown in Table 7 of \cite{Krishnasamy2025OpenMPAMDandNVIDIA}. Notably, our investigations into various configurations for compute kernels revealed that using the default compute constructs and clauses resulted in superior performance compared to manually adjusting the number of threads and block configurations. This finding aligns with earlier studies conducted by E. Krishnasamy \cite{Krishnasamy2026ComparativeCUDAOpenMP, Krishnasamy2025OpenMPAMDandNVIDIA}. Consequently, this research refrains from engaging in manual thread and grid block configuration. For compute kernel directives, we employed the `collapse(3)` directive to facilitate interior-point computation on the 3D dataset. For data transfer between the CPU and GPU, we utilized the low-level API, which has been shown to be more efficient than high-level APIs for data transfers \cite{Krishnasamy2025OpenMPAMDandNVIDIA, Krishnasamy2026ComparativeCUDAOpenMP}. Detailed specifications for both main computation and kernel computation are clearly outlined in Pseudocode \ref{alg:1gpu-base}.

\SetKwFunction{ComputeSingleGPU}{COMPUTE\_SINGLE\_GPU}
\SetKwFunction{Stencil}{STENCIL}
\SetKwFunction{Alloc}{omp\_target\_alloc}
\SetKwFunction{Memcpy}{omp\_target\_memcpy}
\SetKwFunction{Free}{omp\_target\_free}
\SetKwFunction{Wtime}{omp\_get\_wtime}

\begin{algorithm}[htbp]
\DontPrintSemicolon

\caption{\texttt{OpenMP Offloading 1-GPU Baseline}: allocation,
host--device transfers, single-GPU computation, and deallocation.}
\label{alg:1gpu-base}

\tcc{---- Setup ----}
read $n$ and $T$ from command line\;
$h \gets 1/(n-1),\quad
 \kappa_{\max} \gets 0.95,\quad
 \Delta t \gets h^2/(12\kappa_{\max}),\quad
 \beta \gets \Delta t/(2h^2)$\;

allocate host arrays
$u_{\mathrm{old}}, u_{\mathrm{new}}, \mathit{rhs}, \kappa$\;

initialise $u_{\mathrm{old}}, u_{\mathrm{new}}, \mathit{rhs}, \kappa$\;

$\mathit{bytes} \leftarrow n^3 \cdot \texttt{sizeof(double)}$\;

$\mathit{host\_id}
 \leftarrow \texttt{omp\_get\_initial\_device()}$
\tcp*{CPU}

$\mathit{device\_id}
 \leftarrow \texttt{omp\_get\_default\_device()}$
\tcp*{GPU}

\BlankLine

\tcc{---- GPU memory allocation ----}
$d\_u_{\mathrm{old}}
 \leftarrow \Alloc{$\mathit{bytes}$, $\mathit{device\_id}$}$\;

$d\_u_{\mathrm{new}}
 \leftarrow \Alloc{$\mathit{bytes}$, $\mathit{device\_id}$}$\;

$d\_\mathit{rhs}
 \leftarrow \Alloc{$\mathit{bytes}$, $\mathit{device\_id}$}$\;

$d\_\kappa
 \leftarrow \Alloc{$\mathit{bytes}$, $\mathit{device\_id}$}$\;

\BlankLine

\tcc{---- Host $\rightarrow$ device transfer ----}
\Memcpy{
  $d\_u_{\mathrm{old}}$,
  $u_{\mathrm{old}}$,
  $\mathit{bytes}$,
  $0$, $0$,
  $\mathit{device\_id}$,
  $\mathit{host\_id}$
}\;

\Memcpy{
  $d\_u_{\mathrm{new}}$,
  $u_{\mathrm{new}}$,
  $\mathit{bytes}$,
  $0$, $0$,
  $\mathit{device\_id}$,
  $\mathit{host\_id}$
}\;

\Memcpy{
  $d\_\mathit{rhs}$,
  $\mathit{rhs}$,
  $\mathit{bytes}$,
  $0$, $0$,
  $\mathit{device\_id}$,
  $\mathit{host\_id}$
}\;

\Memcpy{
  $d\_\kappa$,
  $\kappa$,
  $\mathit{bytes}$,
  $0$, $0$,
  $\mathit{device\_id}$,
  $\mathit{host\_id}$
}\;

\BlankLine

\tcc{---- Time-stepping loop ----}
$\mathit{start} \leftarrow \Wtime{}$\;

\While{$t < T$}{

  \hyperref[alg:compute-single-gpu]{%
    \ComputeSingleGPU{
      $d\_u_{\mathrm{new}}$,
      $d\_u_{\mathrm{old}}$,
      $d\_\mathit{rhs}$,
      $d\_\kappa$,
      $\mathit{device\_id}$
    }%
  }\;

  swap$(d\_u_{\mathrm{old}},d\_u_{\mathrm{new}})$
    \tcp*{device-pointer swap; no data movement}

  $t \leftarrow t+dt$\;
}

$\mathit{elapsed}
 \leftarrow \Wtime{}-\mathit{start}$\;

\BlankLine

\tcc{---- Device $\rightarrow$ host transfer ----}
\Memcpy{
  $u_{\mathrm{old}}$,
  $d\_u_{\mathrm{old}}$,
  $\mathit{bytes}$,
  $0$, $0$,
  $\mathit{host\_id}$,
  $\mathit{device\_id}$
}\;

\BlankLine

\tcc{---- GPU memory deallocation ----}
\Free{$d\_u_{\mathrm{old}}$, $\mathit{device\_id}$}$\;
\Free{$d\_u_{\mathrm{new}}$, $\mathit{device\_id}$}$\;
\Free{$d\_\mathit{rhs}$,    $\mathit{device\_id}$}$\;
\Free{$d\_\kappa$,          $\mathit{device\_id}$}$\;

free host arrays
$u_{\mathrm{old}},u_{\mathrm{new}},\mathit{rhs},\kappa$\;

\end{algorithm}

\begin{algorithm}[htbp]
\DontPrintSemicolon

\caption{\texttt{COMPUTE\_SINGLE\_GPU}: OpenMP target computation
over the interior grid points.}
\label{alg:compute-single-gpu}

\textbf{procedure} $\mathrm{COMPUTE\_SINLGE\_GPU}
(u_{\mathrm{new}},u_{\mathrm{old}},rhs,\kappa,device\_id)$
\textbf{\{}\;

\Indp

\pragma{target is\_device\_ptr($u_{\mathrm{new}},u_{\mathrm{old}},rhs,\kappa$)
device($device\_id$)}\;

\textbf{\{}\;

\Indp

\pragma{teams distribute parallel for collapse(3)}\;

\For{$k \leftarrow 1$ \KwTo $n_k-2$}{
    \For{$j \leftarrow 1$ \KwTo $n_y-2$}{
        \For{$i \leftarrow 1$ \KwTo $n_x-2$}{
            $u_{\mathrm{new}}[k,j,i]
            \leftarrow
            \Stencil{$u_{\mathrm{old}},rhs,\kappa,k,j,i$}$\;
        }
    }
}

\Indm
\textbf{\}}\;

\Indm
\textbf{\}}\;

\end{algorithm}

\subsubsection{OpenMP Offloading \{2,4\}-GPU Version 1}
In this approach, we partitioned the data blocks when utilizing multiple GPUs. For the case of two GPUs, we implement two distinct kernels: one for interior computations and another for halo computations. In the four-GPU scenario, we use a single kernel for interior computations and two separate kernels for halo computations, designated for the top and bottom layers, respectively, on the 2nd and 3rd GPUs. It is important to note that synchronization of kernel execution and data transfer operations across the entire process can be costly in terms of performance. This synchronization can pose a significant issue in both the 2-GPU and 4-GPU configurations, with the 4-GPU case exhibiting greater performance degradation compared to the 2-GPU setup. In contrast, CUDA inherently supports asynchronous kernel execution, allowing host threads to proceed with subsequent instructions after initiating kernel computations. Unfortunately, the OpenMP Offloading model differs in this respect. To address these limitations, we implemented an asynchronous computation approach for both the interior and halo point kernels using the \texttt{nowait} directive. The pseudocode illustrating this algorithm for a 2-GPU configuration is presented in ~\ref{alg:2gpu-sync}, accompanied by a schematic workflow depicted in Appendix Figure~\ref{fig:2gpu-sync}.
\begin{algorithm}[htbp]
\DontPrintSemicolon
\caption{Common preparation and two-GPU domain decomposition used by
Pseudocodes~\ref{alg:2gpu-sync}--\ref{alg:2gpu-p2p}.}
\label{alg:setup}

read $n$ and $T$; 
$h \gets 1/(n-1),\quad
 \kappa_{\max} \gets 0.95,\quad
 \Delta t \gets h^2/(12\kappa_{\max}),\quad
 \beta \gets \Delta t/(2h^2)$\;
decompose the domain along the $k$-axis into
data blocks $D_0$ and $D_1$, including one ghost plane per interface\;

initialize host arrays
$u_g$, $\mathit{rhs}_g$, and $\kappa_g$
for each GPU $g\in\{0,1\}$\;

\BlankLine

\ForEach{GPU $g\in\{0,1\}$}{

    \tcp{Allocate device-resident arrays}
    $d_g u_{\mathrm{old}}
      \leftarrow
      \texttt{omp\_target\_alloc}(\mathit{size}_g,g)$\;

    $d_g u_{\mathrm{new}}
      \leftarrow
      \texttt{omp\_target\_alloc}(\mathit{size}_g,g)$\;

    $d_g\mathit{rhs}
      \leftarrow
      \texttt{omp\_target\_alloc}(\mathit{size}_g,g)$\;

    $d_g\kappa
      \leftarrow
      \texttt{omp\_target\_alloc}(\mathit{size}_g,g)$\;

    \BlankLine

    \tcp{Transfer initialized data from the host to GPU $g$}
    \texttt{omp\_target\_memcpy}
    $(d_g u_{\mathrm{old}},u_g,\mathit{size}_g,\ldots,
      \mathrm{GPU}_g,\mathit{host})$\;

    \texttt{omp\_target\_memcpy}
    $(d_g u_{\mathrm{new}},u_g,\mathit{size}_g,\ldots,
      \mathrm{GPU}_g,\mathit{host})$\;

    \texttt{omp\_target\_memcpy}
    $(d_g\mathit{rhs},\mathit{rhs}_g,\mathit{size}_g,\ldots,
      \mathrm{GPU}_g,\mathit{host})$\;

    \texttt{omp\_target\_memcpy}
    $(d_g\kappa,\kappa_g,\mathit{size}_g,\ldots,
      \mathrm{GPU}_g,\mathit{host})$\;
}

\end{algorithm}


\begin{algorithm}[htbp]
\DontPrintSemicolon
\caption{\texttt{OpenMP Offloading 2-GPU Version 1}: synchronous two-GPU version with
host-staged halo exchange.}
\label{alg:2gpu-sync}

\SetKwFunction{Interior}{INTERIOR}
\SetKwFunction{Boundary}{BOUNDARY}
\SetKwFunction{Swap}{SWAP}

setup + host initialization as in
Pseudocode~\ref{alg:setup}\;

allocate device arrays and transfer $D_0$ to GPU~0
and $D_1$ to GPU~1\;

allocate host halo buffers $H_0$ and $H_1$\;

\BlankLine

\For{$step \leftarrow 0$ \KwTo $N_{\mathrm{steps}}-1$}{
    
    \tcp{Launch computation on GPU 0}
    \InteriorRef{D_0,\mathrm{GPU}_0}\;
    \BoundaryRef{B_0,\mathrm{GPU}_0}\;

    \BlankLine

    \tcp{Launch computation on GPU 1}
    \InteriorRef{D_1,\mathrm{GPU}_1}\;
    \BoundaryRef{B_1,\mathrm{GPU}_1}\;

    \BlankLine

    \tcp{Complete computation on both GPUs}
    \pragma{taskwait}\;

    \BlankLine

    \tcp{Halo block 0: GPU 0 $\rightarrow$ host $\rightarrow$ GPU 1}
    \texttt{omp\_target\_memcpy}
    $(H_0,B_0,\mathit{halo\_bytes},\ldots,
      \mathit{host},\mathrm{GPU}_0)$\;

    \texttt{omp\_target\_memcpy}
    $(G_1,H_0,\mathit{halo\_bytes},\ldots,
      \mathrm{GPU}_1,\mathit{host})$\;

    \BlankLine

    \tcp{Halo block 1: GPU 1 $\rightarrow$ host $\rightarrow$ GPU 0}
    \texttt{omp\_target\_memcpy}
    $(H_1,B_1,\mathit{halo\_bytes},\ldots,
      \mathit{host},\mathrm{GPU}_1)$\;

    \texttt{omp\_target\_memcpy}
    $(G_0,H_1,\mathit{halo\_bytes},\ldots,
      \mathrm{GPU}_0,\mathit{host})$\;

    \BlankLine

    \tcp{Advance to the next time level}
    \Swap{$u^{old}_0,u^{new}_0$}\;
    \Swap{$u^{old}_1,u^{new}_1$}\;
}

copy the local solutions back to the host
and reconstruct the global domain\;

\end{algorithm}

\begin{algorithm}[htbp]
\DontPrintSemicolon
\caption{OpenMP target kernel for the interior data block.}
\label{alg:interior-kernel}

\SetKwFunction{Stencil}{STENCIL}

\textbf{procedure} $\mathrm{INTERIOR}
(u_{\mathrm{new}},u_{\mathrm{old}},rhs,\kappa,
 k_{\mathrm{start}},k_{\mathrm{end}},device\_id)$
\textbf{\{}\;

\Indp

\pragma{\texttt{target is\_device\_ptr($u_{\mathrm{new}},u_{\mathrm{old}},rhs,\kappa$)
device($device\_id$) nowait}}\;

\textbf{\{}\;

\Indp

\pragma{\texttt{teams distribute parallel for collapse(3)}}\;

\For{$k \leftarrow k_{\mathrm{start}}$ \KwTo $k_{\mathrm{end}}$}{
    \For{$j \leftarrow 1$ \KwTo $n_y-2$}{
        \For{$i \leftarrow 1$ \KwTo $n_x-2$}{
            $u_{\mathrm{new}}[k,j,i]
            \leftarrow
            \Stencil{$u_{\mathrm{old}},rhs,\kappa,k,j,i$}$\;
        }
    }
}

\Indm
\textbf{\}}\;

\Indm
\textbf{\}}\;

\end{algorithm}

\begin{algorithm}[htbp]
\DontPrintSemicolon
\caption{OpenMP target kernel for an interface boundary block.}
\label{alg:boundary-kernel}

\SetKwFunction{Stencil}{STENCIL}

\textbf{procedure} $\mathrm{BOUNDARY}
(u_{\mathrm{new}},u_{\mathrm{old}},rhs,\kappa,
 k_{\mathrm{boundary}},device\_id)$
\textbf{\{}\;

\Indp

\pragma{\texttt{target is\_device\_ptr($u_{\mathrm{new}},u_{\mathrm{old}},rhs,\kappa$)
device($device\_id$) nowait}}\;

\textbf{\{}\;

\Indp

\pragma{\texttt{teams distribute parallel for collapse(2)}}\;

\For{$j \leftarrow 1$ \KwTo $n_y-2$}{
    \For{$i \leftarrow 1$ \KwTo $n_x-2$}{
        $u_{\mathrm{new}}[k_{\mathrm{boundary}},j,i]
        \leftarrow
        \Stencil{$u_{\mathrm{old}},rhs,\kappa,
                  k_{\mathrm{boundary}},j,i$}$\;
    }
}

\Indm
\textbf{\}}\;

\Indm
\textbf{\}}\;

\end{algorithm}

\subsubsection{OpenMP Offloading \{2,4\}-GPU Version 2}
In this scenario, we utilize asynchronous operations for both computation and data transfer, specifically between the Host and Device, as well as between Device and Host. The interior and halo computations are initialized on each GPU block, while the halo computations are linked to a \texttt{depobj ID} to facilitate synchronization for the subsequent Device-to-Host (DtoH) copy operation. Once the DtoH operation is complete, the updated values of \( u \) are swapped. To guarantee that the DtoH operation fully completes before the swapping occurs, we introduce the directive \texttt{\#pragma omp taskwait}. The pseudocode for this algorithm, demonstrated in a 2-GPU scenario, is presented in~\ref{alg:2gpu-stream}, and the corresponding schematic workflow is illustrated in Appendix Figure~\ref{fig:2gpu-stream}.
\newcommand{\BoundaryAsyncRef}[1]{%
  \hyperref[alg:boundary-kernel-async]{%
    \textsc{Boundary\_async}\ensuremath{\left(#1\right)}%
  }%
}

\begin{algorithm}[htbp]
\DontPrintSemicolon
\caption{\texttt{OpenMP Offloading 2-GPU Version 2}: asynchronous two-GPU stream version with host-staged halo exchange.}
\label{alg:2gpu-stream}

\SetKwFunction{Swap}{SWAP}

setup + host initialization as in
Pseudocode~\ref{alg:setup}\;

allocate device arrays and transfer $D_0$ to GPU~0
and $D_1$ to GPU~1\;

allocate host halo buffers $H_0$ and $H_1$\;

\BlankLine

initialize dependency tokens
$tk_{0\rightarrow1}$ and $tk_{1\rightarrow0}$\;

\pragma{\texttt{depobj($dep_{0\rightarrow1}$)
depend(inout:} $tk_{0\rightarrow1}$)}\;

\pragma{\texttt{depobj($dep_{1\rightarrow0}$)
depend(inout:} $tk_{1\rightarrow0}$)}\;

\BlankLine

\For{$step \leftarrow 0$ \KwTo $N_{\mathrm{steps}}-1$}{

    \tcp{Launch computation on GPU 0}
    \InteriorRef{D_0,\mathrm{GPU}_0}\;

    \BoundaryAsyncRef{
      B_0,\mathrm{GPU}_0,tk_{0\rightarrow1}
    }\;

    \BlankLine

    \tcp{Launch computation on GPU 1}
    \InteriorRef{D_1,\mathrm{GPU}_1}\;

    \BoundaryAsyncRef{
      B_1,\mathrm{GPU}_1,tk_{1\rightarrow0}
    }\;

    \BlankLine

    \tcp{Asynchronous halo chain: GPU 0 $\rightarrow$ host $\rightarrow$ GPU 1}
    \texttt{omp\_target\_memcpy\_async}
    $(H_0,B_0,\mathit{halo\_bytes},\ldots,
      \mathit{host},\mathrm{GPU}_0,
      1,\&dep_{0\rightarrow1})$\;

    \texttt{omp\_target\_memcpy\_async}
    $(G_1,H_0,\mathit{halo\_bytes},\ldots,
      \mathrm{GPU}_1,\mathit{host},
      1,\&dep_{0\rightarrow1})$\;

    \BlankLine

    \tcp{Asynchronous halo chain: GPU 1 $\rightarrow$ host $\rightarrow$ GPU 0}
    \texttt{omp\_target\_memcpy\_async}
    $(H_1,B_1,\mathit{halo\_bytes},\ldots,
      \mathit{host},\mathrm{GPU}_1,
      1,\&dep_{1\rightarrow0})$\;

    \texttt{omp\_target\_memcpy\_async}
    $(G_0,H_1,\mathit{halo\_bytes},\ldots,
      \mathrm{GPU}_0,\mathit{host},
      1,\&dep_{1\rightarrow0})$\;

    \BlankLine

    \tcp{Complete both interior kernels and halo chains}
    \pragma{taskwait}\;

    \BlankLine

    \tcp{Advance to the next time level}
    \Swap{$u^{old}_0,u^{new}_0$}\;
    \Swap{$u^{old}_1,u^{new}_1$}\;
}

\BlankLine

\pragma{\texttt{depobj($dep_{0\rightarrow1}$) destroy}}\;
\pragma{\texttt{depobj($dep_{1\rightarrow0}$) destroy}}\;

copy the local solutions back to the host
and reconstruct the global domain\;

\end{algorithm}

\begin{algorithm}[htbp]
\DontPrintSemicolon
\caption{OpenMP asynchronous target kernel for an interface boundary block.}
\label{alg:boundary-kernel-async}

\SetKwFunction{Stencil}{STENCIL}

\textbf{procedure} $\mathrm{BOUNDARY\_ASYNC}
(u_{\mathrm{new}},u_{\mathrm{old}},rhs,\kappa,
 k_{\mathrm{boundary}},device\_id,dep\_token)$
\textbf{\{}\;

\Indp

\pragma{\texttt{target is\_device\_ptr($u_{\mathrm{new}},u_{\mathrm{old}},rhs,\kappa$)
device($device\_id$) nowait
depend(out: $dep\_token[0]$)}}\;

\textbf{\{}\;

\Indp

\pragma{\texttt{teams distribute parallel for collapse(2)}}\;

\For{$j \leftarrow 1$ \KwTo $n_y-2$}{
    \For{$i \leftarrow 1$ \KwTo $n_x-2$}{
        $u_{\mathrm{new}}[k_{\mathrm{boundary}},j,i]
        \leftarrow
        \Stencil{$u_{\mathrm{old}},rhs,\kappa,
                  k_{\mathrm{boundary}},j,i$}$\;
    }
}

\Indm
\textbf{\}}\;

\Indm
\textbf{\}}\;

\end{algorithm}

\subsubsection{OpenMP Offloading \{2,4\}-GPU Version 3}
This version closely resembles the previously described \texttt{OpenMP Offloading \{2,4\}-GPU Version 2}, but it introduces multiple threads for managing GPU control. Each GPU operation, including both kernel execution and memory copying, is handled by its respective host thread. As in the prior version, each kernel execution and data transfer is identified by a unique \texttt{depobj ID}, ensuring that computation and communication can overlap for each device. To mitigate the risk of data races, \texttt{\#pragma omp barrier} constructs are strategically placed to guarantee numerical correctness before any data swapping occurs. The pseudocode for this algorithm, demonstrated in a 2-GPU scenario, is shown in~\ref{alg:2gpu-multithread}, and the corresponding workflow is illustrated in Appendix Figure~\ref{fig:2gpu-multithread}.

\begin{algorithm}[htbp]
\DontPrintSemicolon
\caption{\texttt{OpenMP Offloading 2-GPU Version 3}: asynchronous two-GPU version with
two host threads and host-staged halo exchange.}
\label{alg:2gpu-multithread}

\SetKwFunction{Swap}{SWAP}

setup + host initialization as in
Pseudocode~\ref{alg:setup}\;

allocate device arrays and transfer $D_0$ to GPU~0
and $D_1$ to GPU~1\;

allocate shared host halo buffers $H_0$ and $H_1$\;

\BlankLine

initialize dependency tokens
$tk_{0\rightarrow1}$ and $tk_{1\rightarrow0}$\;

\pragma{\texttt{depobj($dep_{0\rightarrow1}$)
depend(inout:} $tk_{0\rightarrow1}$)}\;

\pragma{\texttt{depobj($dep_{1\rightarrow0}$)
depend(inout:} $tk_{1\rightarrow0}$)}\;

\BlankLine

\pragma{\texttt{parallel num\_threads(2) default(shared)}}
\textbf{\{}\;

\Indp

$tid \leftarrow \texttt{omp\_get\_thread\_num}()$\;
$device\_id \leftarrow tid$\;

\BlankLine

\For{$step \leftarrow 0$ \KwTo $N_{\mathrm{steps}}-1$}{

    \uIf{$tid = 0$}{

        \tcp{Thread 0 controls GPU 0}
        \InteriorRef{D_0,\mathrm{GPU}_0}\;

        \BoundaryAsyncRef{
          B_0,\mathrm{GPU}_0,tk_{0\rightarrow1}
        }\;

        \BlankLine

        \tcp{Push halo block 0: GPU 0 $\rightarrow$ host $\rightarrow$ GPU 1}
        \texttt{omp\_target\_memcpy\_async}
        $(H_0,B_0,\mathit{halo\_bytes},\ldots,
          \mathit{host},\mathrm{GPU}_0,
          1,\&dep_{0\rightarrow1})$\;

        \texttt{omp\_target\_memcpy\_async}
        $(G_1,H_0,\mathit{halo\_bytes},\ldots,
          \mathrm{GPU}_1,\mathit{host},
          1,\&dep_{0\rightarrow1})$\;
    }

    \Else{

        \tcp{Thread 1 controls GPU 1}
        \InteriorRef{D_1,\mathrm{GPU}_1}\;

        \BoundaryAsyncRef{
          B_1,\mathrm{GPU}_1,tk_{1\rightarrow0}
        }\;

        \BlankLine

        \tcp{Push halo block 1: GPU 1 $\rightarrow$ host $\rightarrow$ GPU 0}
        \texttt{omp\_target\_memcpy\_async}
        $(H_1,B_1,\mathit{halo\_bytes},\ldots,
          \mathit{host},\mathrm{GPU}_1,
          1,\&dep_{1\rightarrow0})$\;

        \texttt{omp\_target\_memcpy\_async}
        $(G_0,H_1,\mathit{halo\_bytes},\ldots,
          \mathrm{GPU}_0,\mathit{host},
          1,\&dep_{1\rightarrow0})$\;
    }

    \BlankLine

    \tcp{Complete both compute and communication chains}
    \pragma{\texttt{barrier}}\;

    \BlankLine

    \tcp{One host thread advances both GPU time levels}
    \pragma{\texttt{single}}\;

    \textbf{\{}\;

    \Indp
    \Swap{$u^{old}_0,u^{new}_0$}\;
    \Swap{$u^{old}_1,u^{new}_1$}\;
    \Indm

    \textbf{\}}\;
}

\Indm

\textbf{\}}\;

\BlankLine

\pragma{\texttt{depobj($dep_{0\rightarrow1}$) destroy}}\;
\pragma{\texttt{depobj($dep_{1\rightarrow0}$) destroy}}\;

copy the local solutions back to the host
and reconstruct the global domain\;

\end{algorithm}

\subsubsection{OpenMP Offloading \{2,4\}-GPU Version 4}
This iteration builds upon the previous version, \texttt{OpenMP Offloading \{2,4\}-GPU Version 3}. It eliminates the host-stage data transfer of halo data by leveraging P2P functionality. This approach negates the need for data transfers between the device and host, as detailed in the pseudocode for the 2-GPU scenario presented in~\ref{alg:2gpu-p2p}. The corresponding schematic workflow is illustrated in Appendix Figure~\ref{fig:2gpu-p2p}.

\begin{algorithm}[htbp]
\DontPrintSemicolon
\caption{\texttt{OpenMP Offloading 2-GPU Version 4}: asynchronous two-GPU version with
two host threads and direct P2P halo exchange.}
\label{alg:2gpu-p2p}

\SetKwFunction{Swap}{SWAP}

setup + host initialization as in
Pseudocode~\ref{alg:setup}\;

allocate device arrays and transfer $D_0$ to GPU~0
and $D_1$ to GPU~1\;

\tcp*{No host halo buffers are required}

\BlankLine

initialize dependency tokens
$tk_{0\rightarrow1}$ and $tk_{1\rightarrow0}$\;

\pragma{\texttt{depobj($dep_{0\rightarrow1}$)
depend(inout:} $tk_{0\rightarrow1}$)}\;

\pragma{\texttt{depobj($dep_{1\rightarrow0}$)
depend(inout:} $tk_{1\rightarrow0}$)}\;

\BlankLine

\pragma{\texttt{parallel num\_threads(2) default(shared)}}
\textbf{\{}\;

\Indp

$tid \leftarrow \texttt{omp\_get\_thread\_num}()$\;
$device\_id \leftarrow tid$\;

\BlankLine

\For{$step \leftarrow 0$ \KwTo $N_{\mathrm{steps}}-1$}{

    \uIf{$tid = 0$}{

        \tcp{Thread 0 controls GPU 0}
        \InteriorRef{D_0,\mathrm{GPU}_0}\;

        \BoundaryAsyncRef{
          B_0,\mathrm{GPU}_0,tk_{0\rightarrow1}
        }\;

        \BlankLine

        \tcp{Direct halo push: GPU 0 $\rightarrow$ GPU 1}
        \texttt{omp\_target\_memcpy\_async}
        $(G_1,B_0,\mathit{halo\_bytes},\ldots,
          \mathrm{GPU}_1,\mathrm{GPU}_0,
          1,\&dep_{0\rightarrow1})$\;
    }

    \Else{

        \tcp{Thread 1 controls GPU 1}
        \InteriorRef{D_1,\mathrm{GPU}_1}\;

        \BoundaryAsyncRef{
          B_1,\mathrm{GPU}_1,tk_{1\rightarrow0}
        }\;

        \BlankLine

        \tcp{Direct halo push: GPU 1 $\rightarrow$ GPU 0}
        \texttt{omp\_target\_memcpy\_async}
        $(G_0,B_1,\mathit{halo\_bytes},\ldots,
          \mathrm{GPU}_0,\mathrm{GPU}_1,
          1,\&dep_{1\rightarrow0})$\;
    }

    \BlankLine

    \tcp{Complete both compute and peer-copy chains}
    \pragma{\texttt{barrier}}\;

    \BlankLine

    \tcp{One host thread advances both GPU time levels}
    \pragma{\texttt{single}}\;

    \textbf{\{}\;

    \Indp
    \Swap{$u^{old}_0,u^{new}_0$}\;
    \Swap{$u^{old}_1,u^{new}_1$}\;
    \Indm

    \textbf{\}}\;
}

\Indm

\textbf{\}}\;

\BlankLine

\pragma{\texttt{depobj($dep_{0\rightarrow1}$) destroy}}\;
\pragma{\texttt{depobj($dep_{1\rightarrow0}$) destroy}}\;

copy the local solutions back to the host
and reconstruct the global domain\;

\end{algorithm}

\subsubsection{CUDA, HIP, and SYCL}
In this research, we utilize baseline versions of CUDA, HIP, and SYCL alongside OpenMP Offloading implementations. Our primary objective is to avoid the redundant task of developing three separate versions for CUDA, HIP, and SYCL from the ground up. Instead, we will create the CUDA version first, which can then be converted to HIP using the command \texttt{hipify-perl mycode.cu > mycode.hip} and to SYCL with \texttt{dpct mycode.cu --out-root=sycl\_out}. However, it is important to note that some fine-tuning will be necessary after conversion to achieve optimal performance; otherwise, the performance may fall significantly short compared to the OpenMP Offloading implementations. Leveraging our extensive expertise in CUDA programming, particularly within single-GPU and multi-GPU environments~\cite{Krishnasamy2014HybridCPUGPU, Krishnasamy2015MultiGPU3DSweeping}, we have devised an efficient strategy for both the CUDA single-GPU and multi-GPU implementations that parallels our approach with OpenMP Offloading. This strategy includes effective thread block management, memory coalescing, and enhancements to communication and memory transfer efficiency.
    
\subsubsection{Numerical Correctness}
The numerical double-precision correctness of various implementations of the GPU programming model was compared element-wise with a serial CPU reference using an absolute tolerance of \(10^{-6}\) and a relative tolerance of \(10^{-5}\). A result was accepted when every value satisfied

\[
\left|u_i^{\mathrm{GPU}} - u_i^{\mathrm{CPU}}\right|
\leq
10^{-6} + 10^{-5}\left|u_i^{\mathrm{CPU}}\right|,
\]
with matching array sizes and no \texttt{NaN} or infinite values. All evaluated GPU implementations are satisfied this criterion.

\begin{table}[htbp]
\caption{Comparison of selected low-level API features across CUDA, HIP, SYCL, and OpenMP Offloading.\label{tab:high-low-spec}}
\small
\setlength{\tabcolsep}{4pt}
\renewcommand{\tabularxcolumn}[1]{m{#1}}

\begin{tabularx}{\textwidth}{%
    >{\raggedright\arraybackslash\hsize=0.66\hsize}X
    >{\raggedright\arraybackslash\hsize=1.08\hsize}X
    >{\raggedright\arraybackslash\hsize=1.02\hsize}X
    >{\raggedright\arraybackslash\hsize=1.06\hsize}X
    >{\raggedright\arraybackslash\hsize=1.18\hsize}X
}
\toprule
\textbf{Category} & \textbf{CUDA} & \textbf{HIP} & \textbf{SYCL} & \textbf{OpenMP Offloading} \\
\midrule
\multirow[m]{2}{=}{Memory management}
  & \texttt{cudaMalloc()} & \texttt{hipMalloc()} & \texttt{malloc\_device()} & \texttt{omp\_target\_alloc()} \\
  & \texttt{cudaFree()}   & \texttt{hipFree()}   & \texttt{free()}          & \texttt{omp\_target\_free()} \\
\midrule
Streams/Queues
  & \texttt{cudaStreamCreate()} & \texttt{hipStreamCreate()} & \texttt{queue} & Supported through device tasks/\texttt{nowait} \\
\midrule
Synchronous memory copy
  & \texttt{cudaMemcpy()} & \texttt{hipMemcpy()} & \texttt{queue.memcpy().wait()} & \texttt{omp\_target\_memcpy()} \\
\midrule
Asynchronous memory copy
  & \texttt{cudaMemcpyAsync()} & \texttt{hipMemcpyAsync()} & \texttt{queue.memcpy()} & \texttt{omp\_target\_memcpy\_async()} \\
\midrule
\multirow[m]{2}{=}{Device-to-device memory copy}
  & \texttt{cudaMemcpyPeer()}      & \texttt{hipMemcpyPeer()}      & \texttt{queue.memcpy().wait()} & \texttt{omp\_target\_memcpy()} with device IDs \\
  & \texttt{cudaMemcpyPeerAsync()} & \texttt{hipMemcpyPeerAsync()} & \texttt{queue.memcpy()}        & \texttt{omp\_target\_memcpy\_async()} with device IDs \\
\bottomrule
\end{tabularx}
\end{table}
\begin{figure}[htbp]
\centering

\subfloat[\centering NVIDIA H100, 2 GPUs.\label{fig:h100_2gpu}]{
    \includegraphics[width=0.48\linewidth]
    {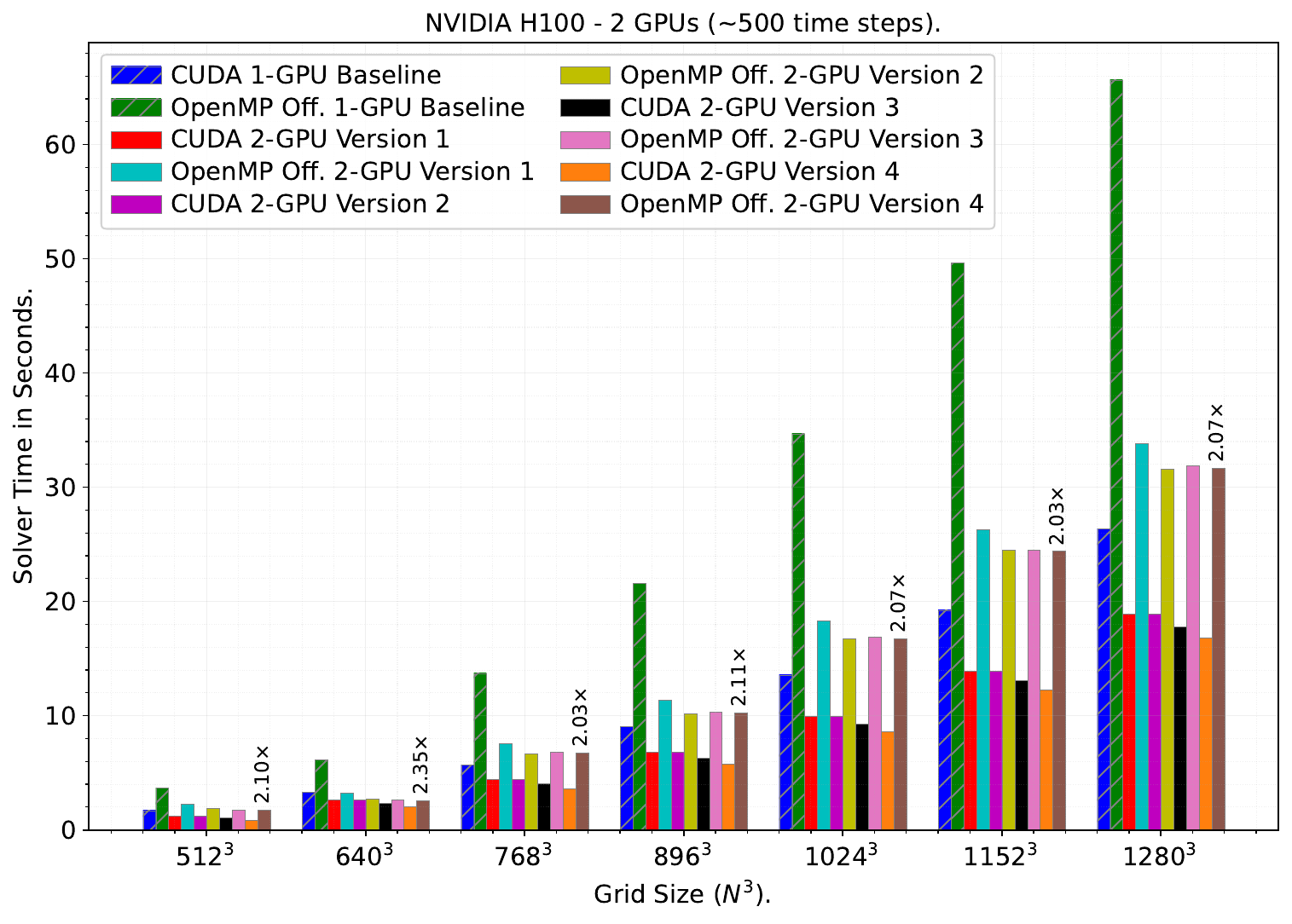}
}
\hfill
\subfloat[\centering NVIDIA H100, 4 GPUs.\label{fig:h100_4gpu}]{
    \includegraphics[width=0.48\linewidth]
    {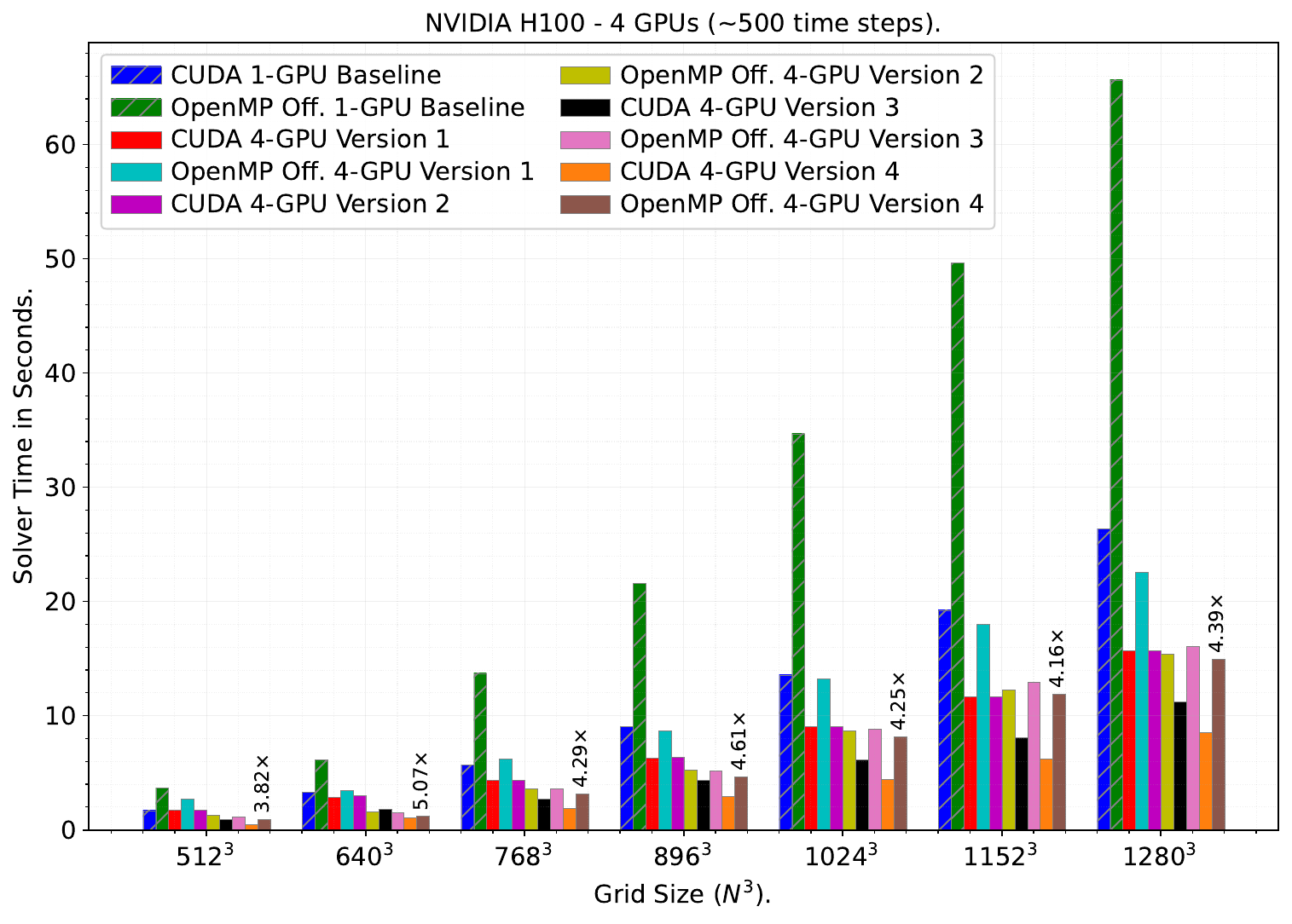}
}

\medskip

\subfloat[\centering AMD MI250X, 2 GPUs.\label{fig:amd_2gpu}]{
    \includegraphics[width=0.48\linewidth]
    {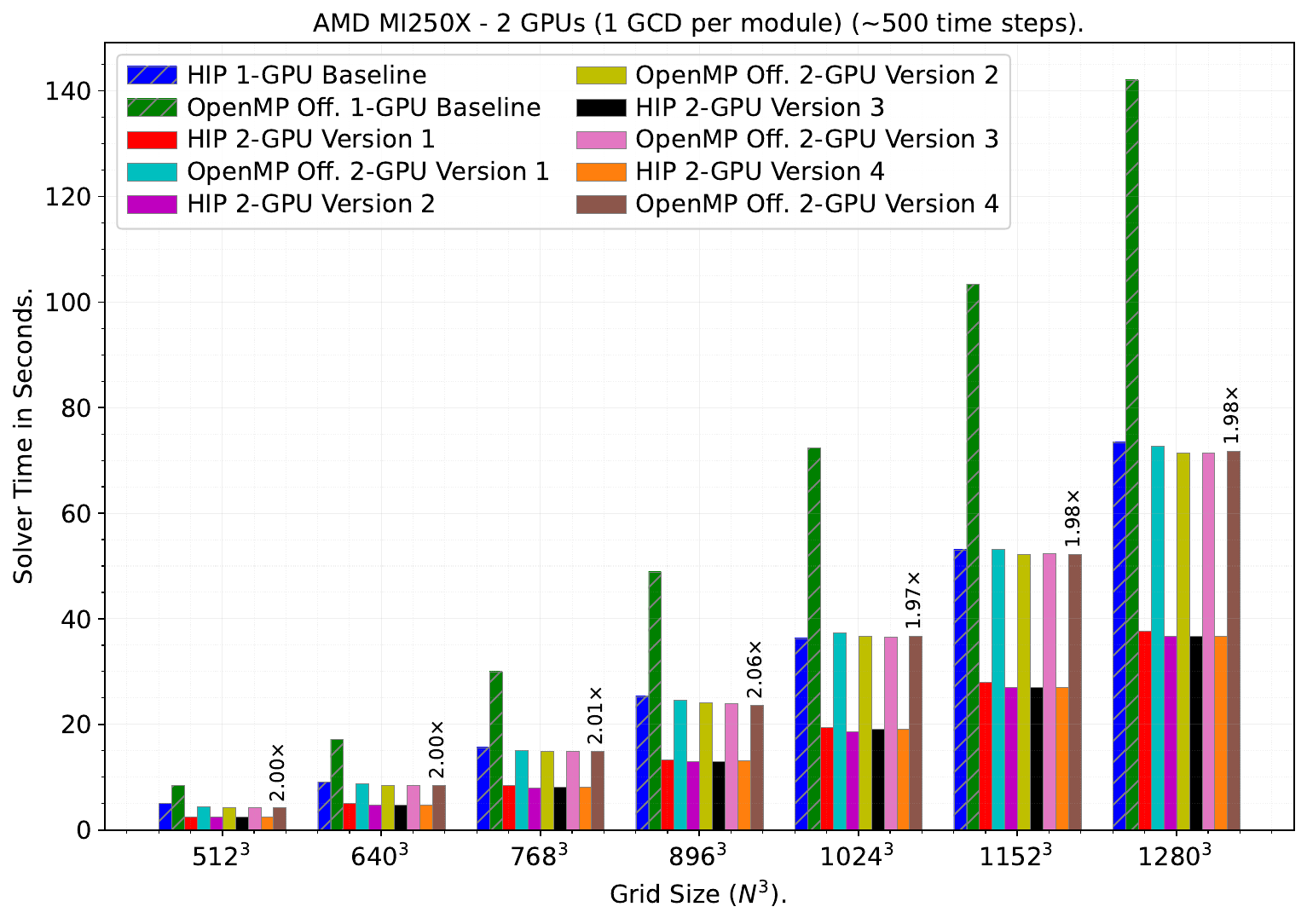}
}
\hfill
\subfloat[\centering AMD MI250X, 4 GPUs.\label{fig:amd_4gpu}]{
    \includegraphics[width=0.48\linewidth]
    {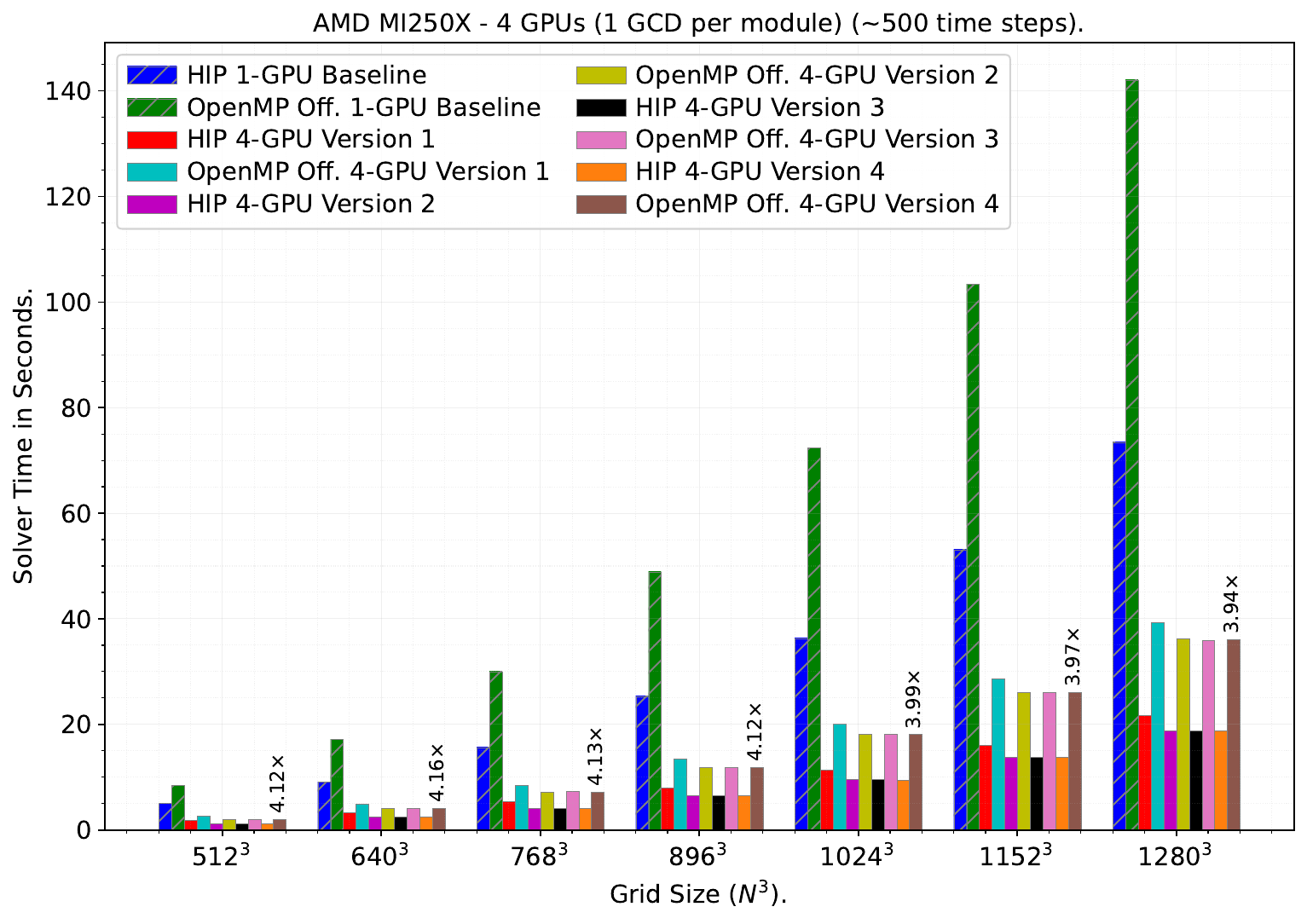}
}

\medskip

\subfloat[\centering Intel 1550, 2 GPUs.\label{fig:intel_2gpu}]{
    \includegraphics[width=0.48\linewidth]
    {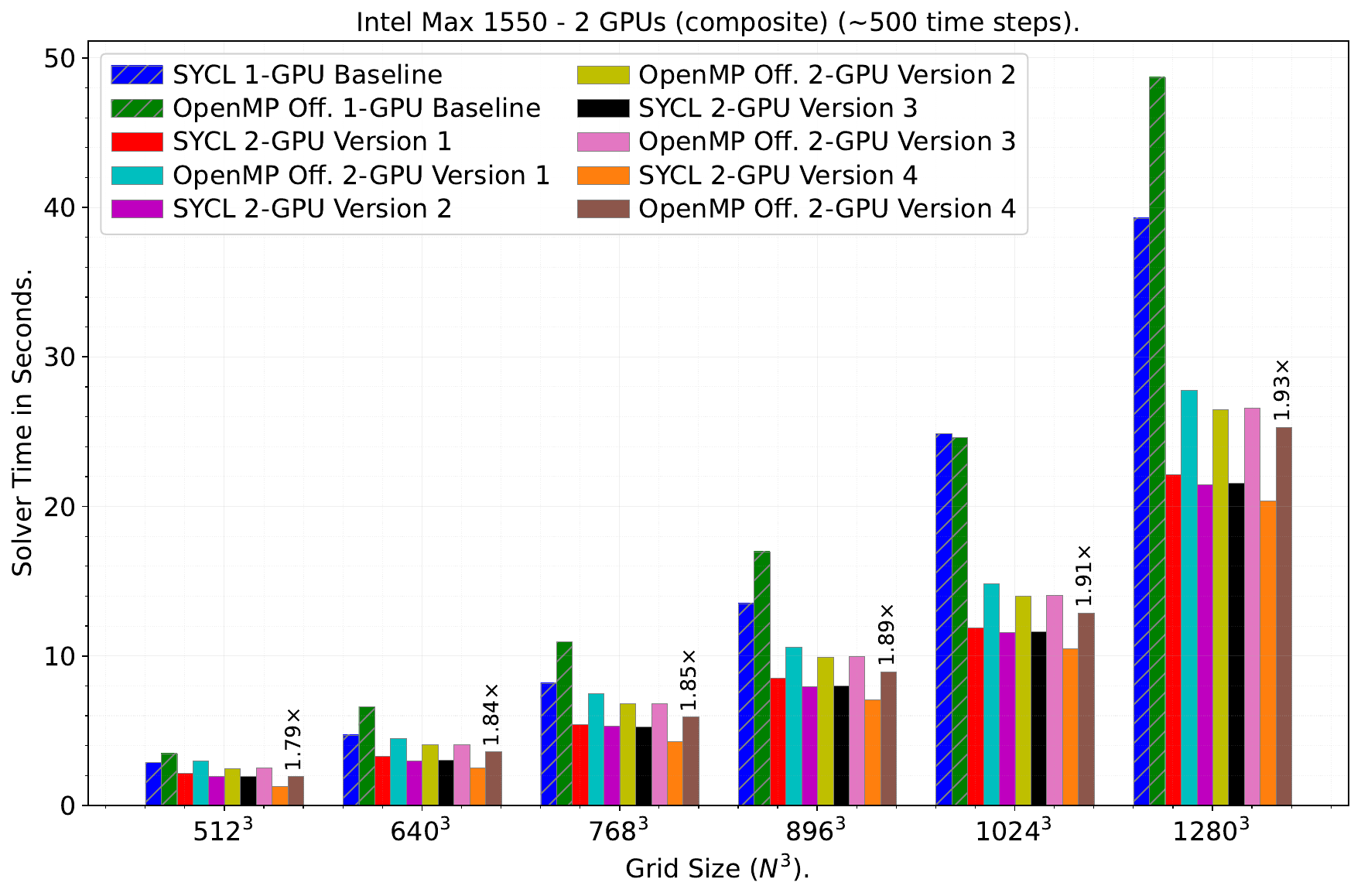}
}
\hfill
\subfloat[\centering Intel 1550, 4 GPUs.\label{fig:intel_4gpu}]{
    \includegraphics[width=0.48\linewidth]
    {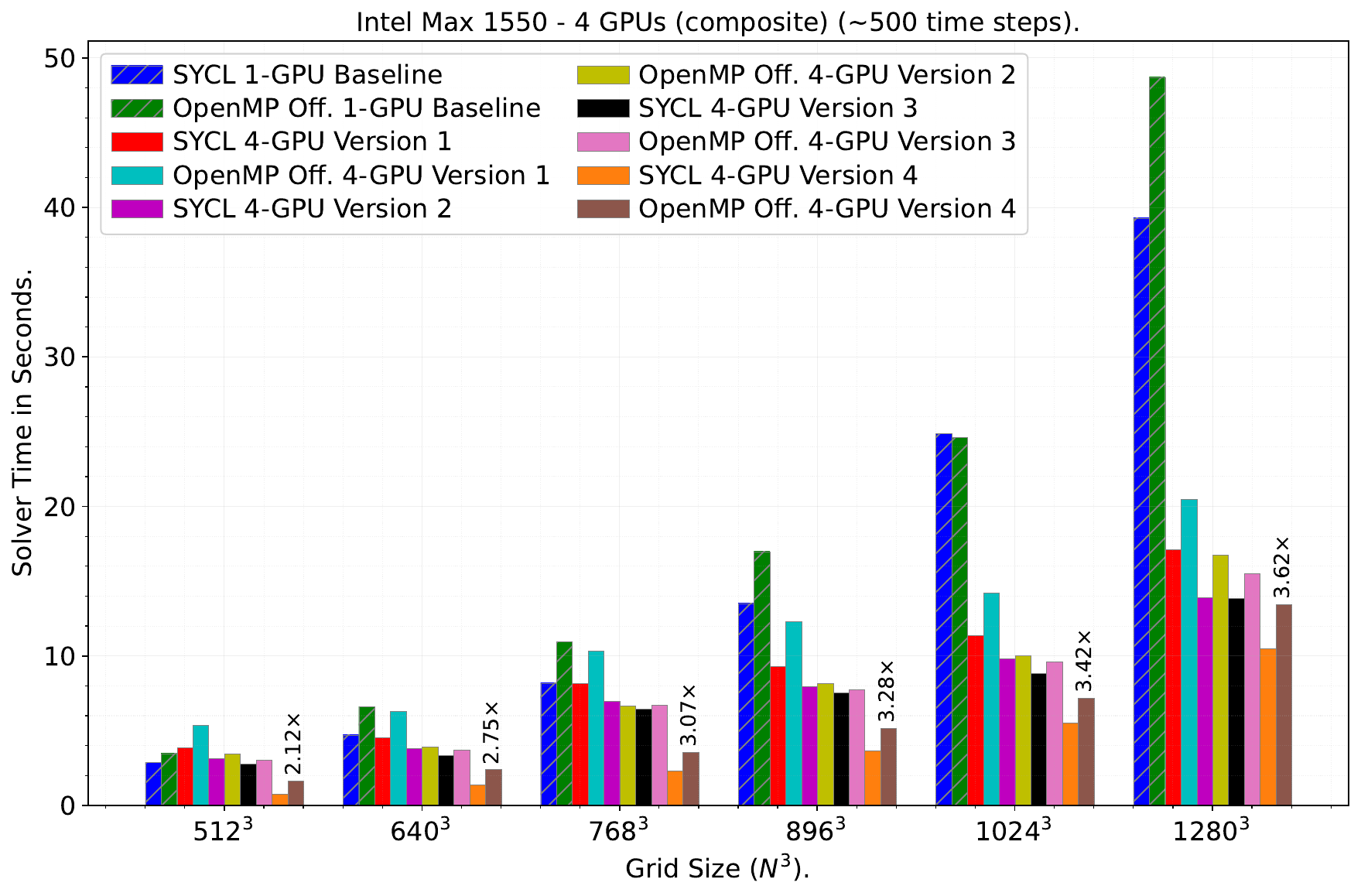}
}


\caption{
Comparison of runtime performance across different GPU programming models along with various GPU architectures for different grid sizes \( N^3 \), utilizing configurations with 2 and 4 GPUs. The speedup values indicate the performance of \texttt{OpenMP Offloading \{2,4\}-GPU version 4} relative to the \texttt{OpenMP Offloading 1-GPU Baseline}.
}
\label{fig:diffusion_comparison}
\end{figure}


\newcommand{\bwpanel}[2]{%
    \begin{minipage}[t]{0.325\linewidth}
        \centering
        \includegraphics[width=\linewidth]{#1}\\[-2pt]
        {\scriptsize (#2)}
    \end{minipage}%
}

\begin{figure}[p]
\centering

\bwpanel{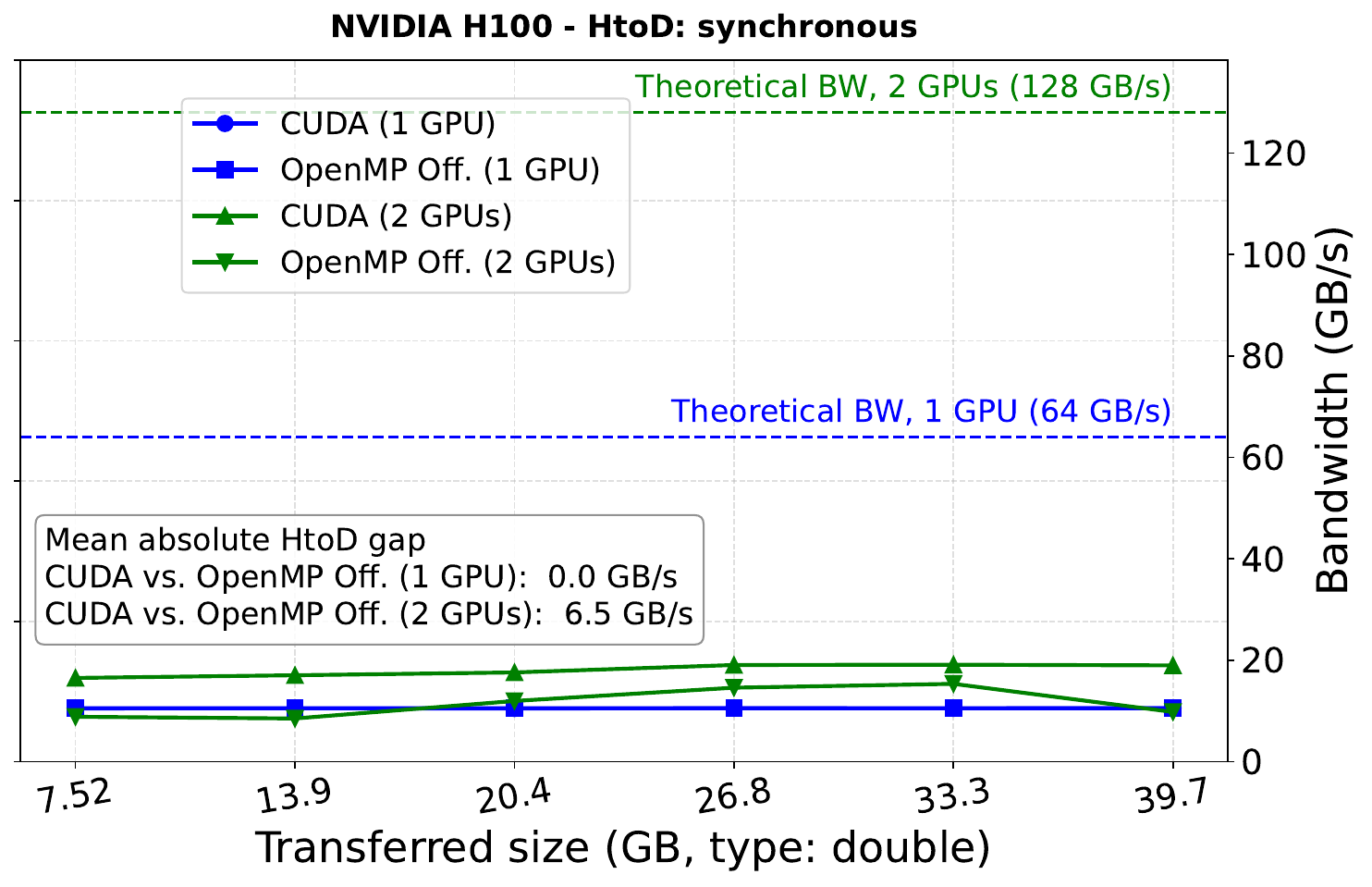}{a}\hfill
\bwpanel{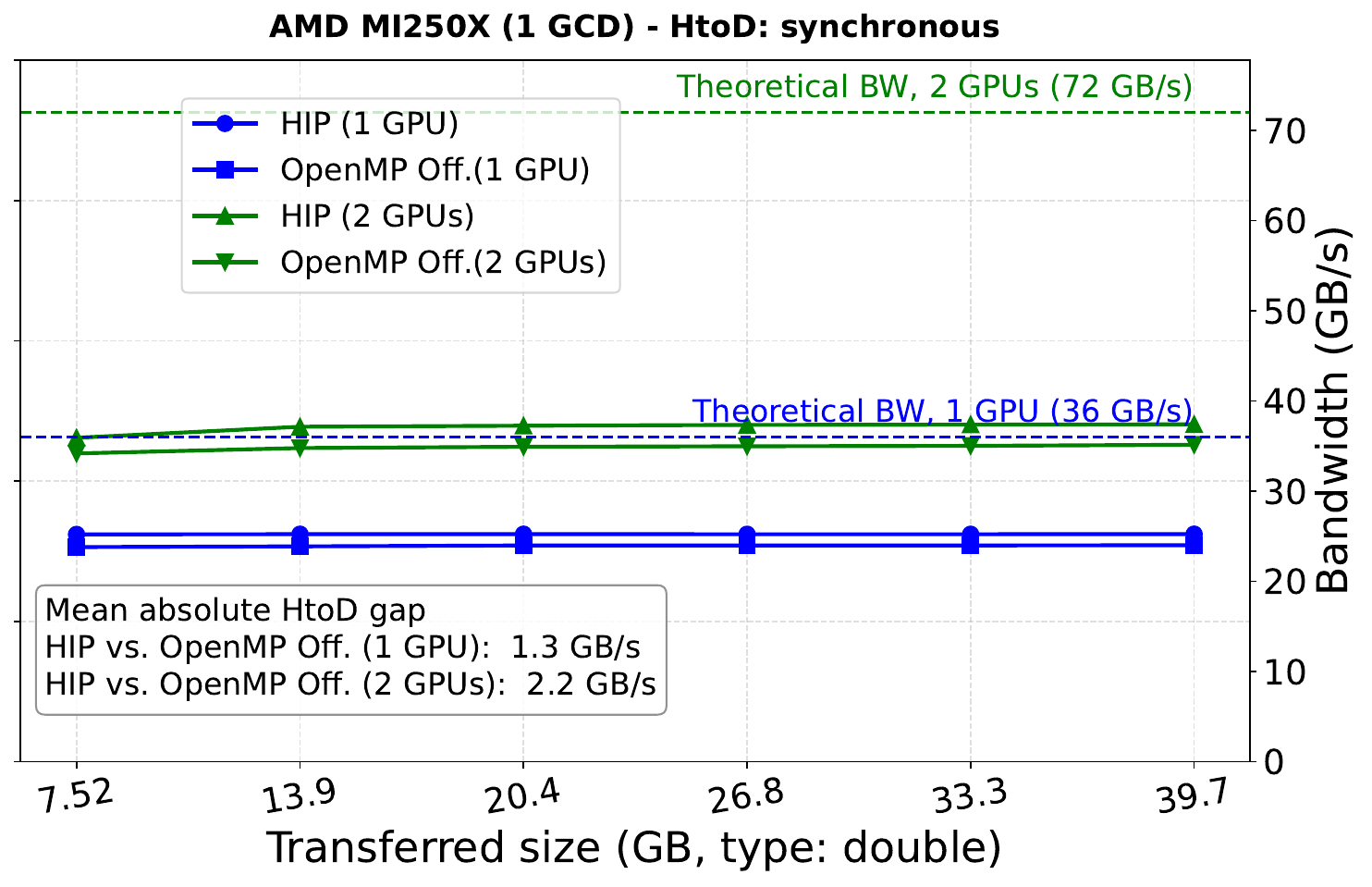}{b}\hfill
\bwpanel{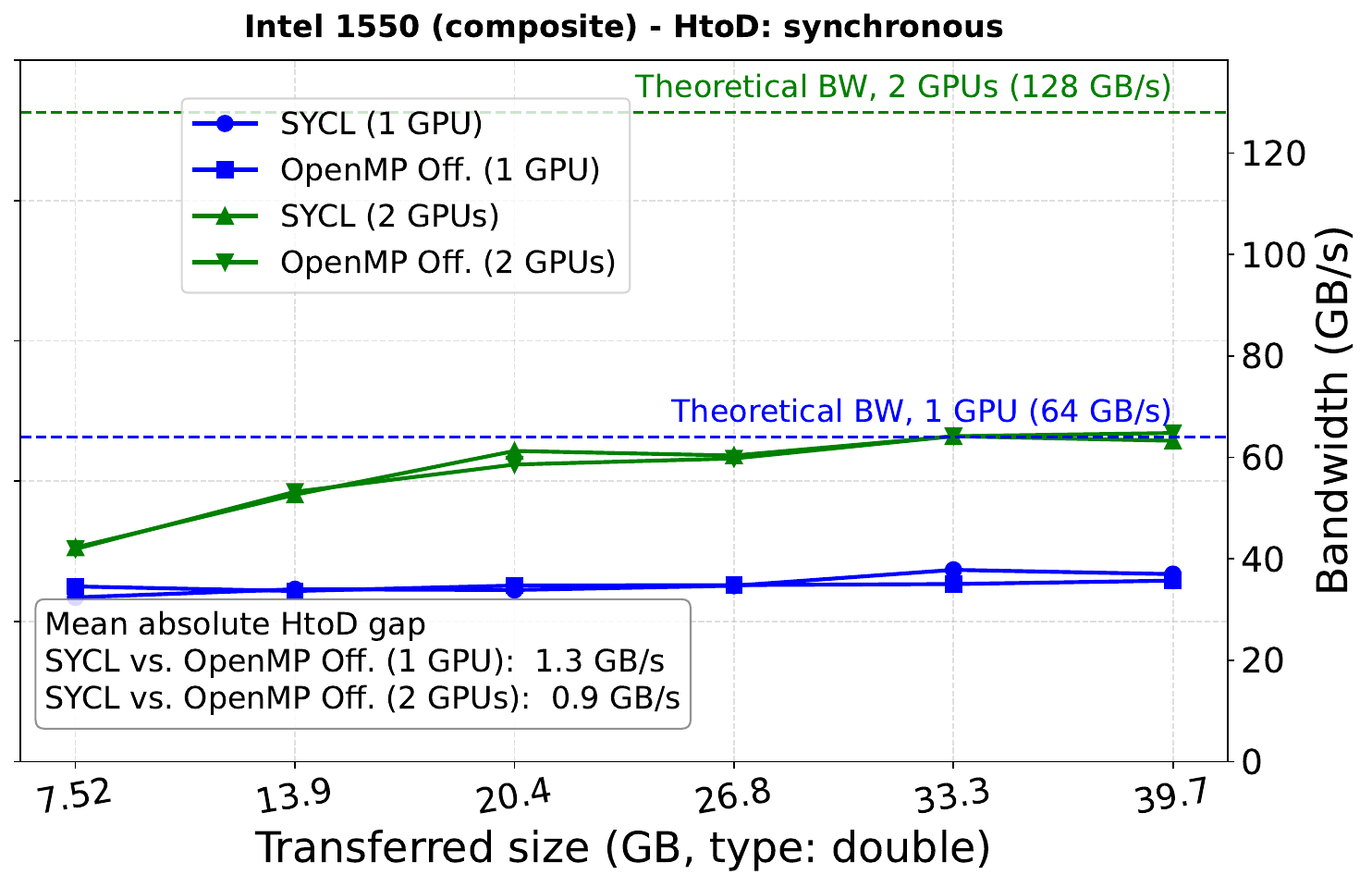}{c}

\par\medskip

\bwpanel{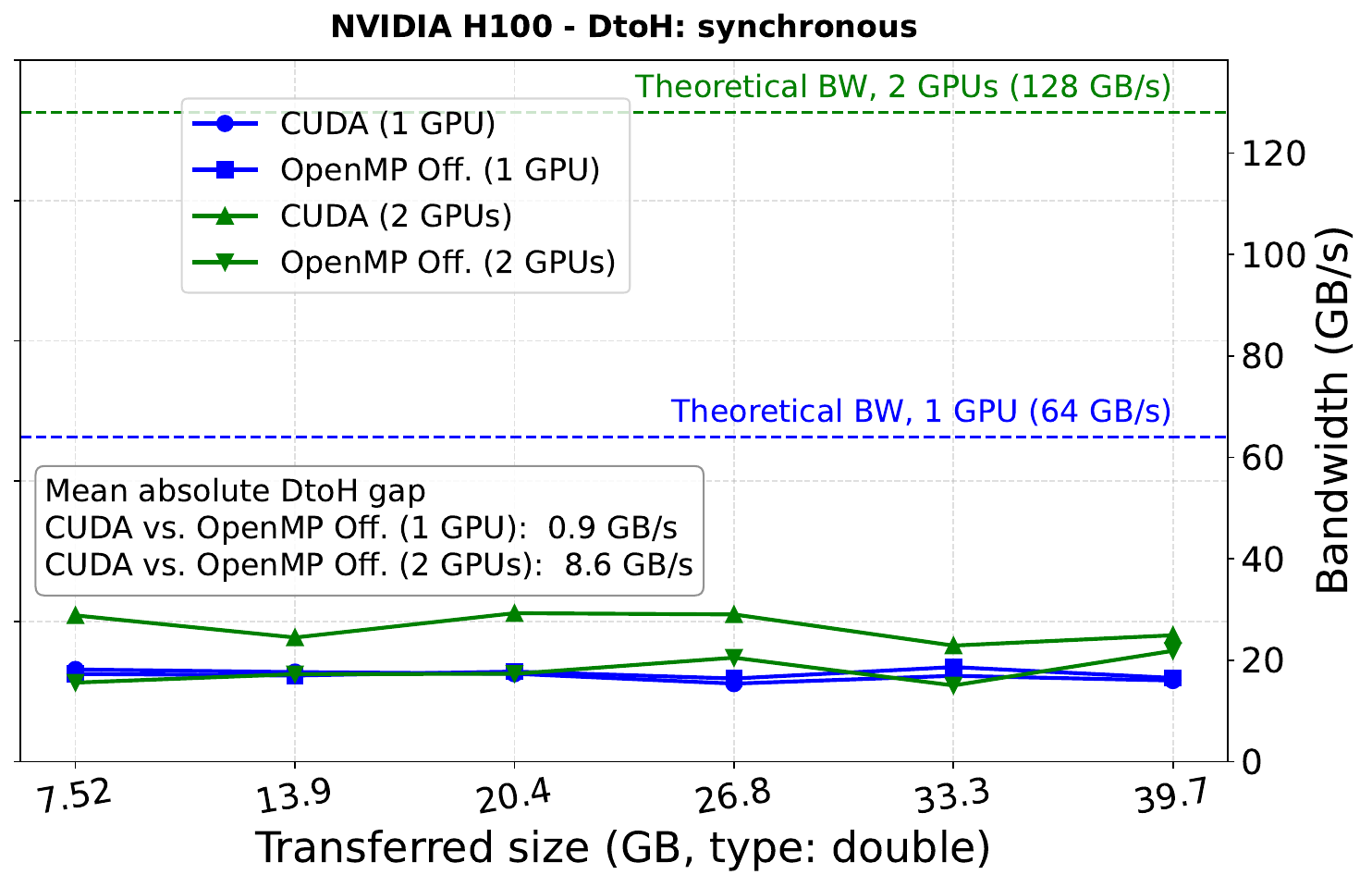}{d}\hfill
\bwpanel{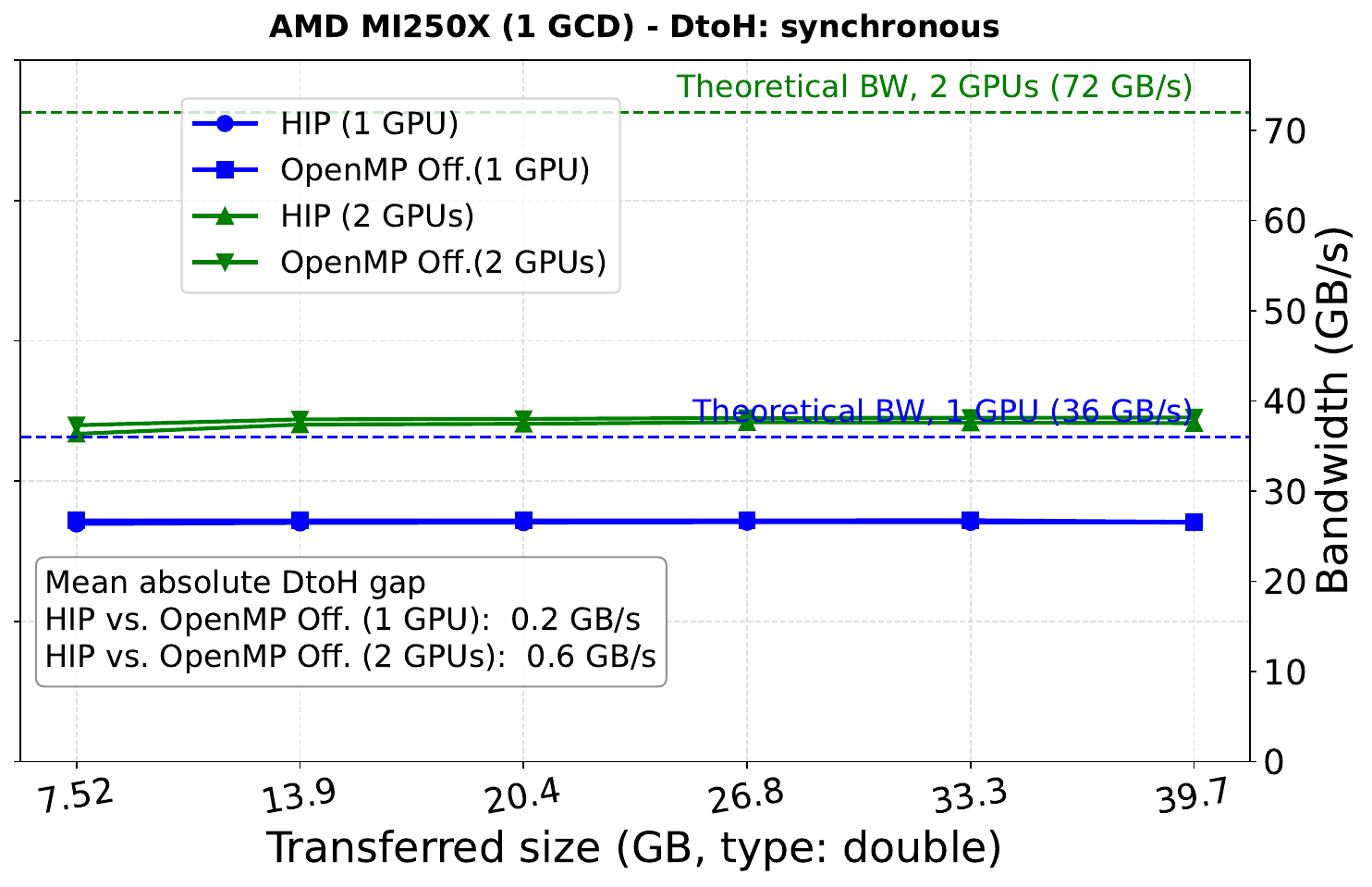}{e}\hfill
\bwpanel{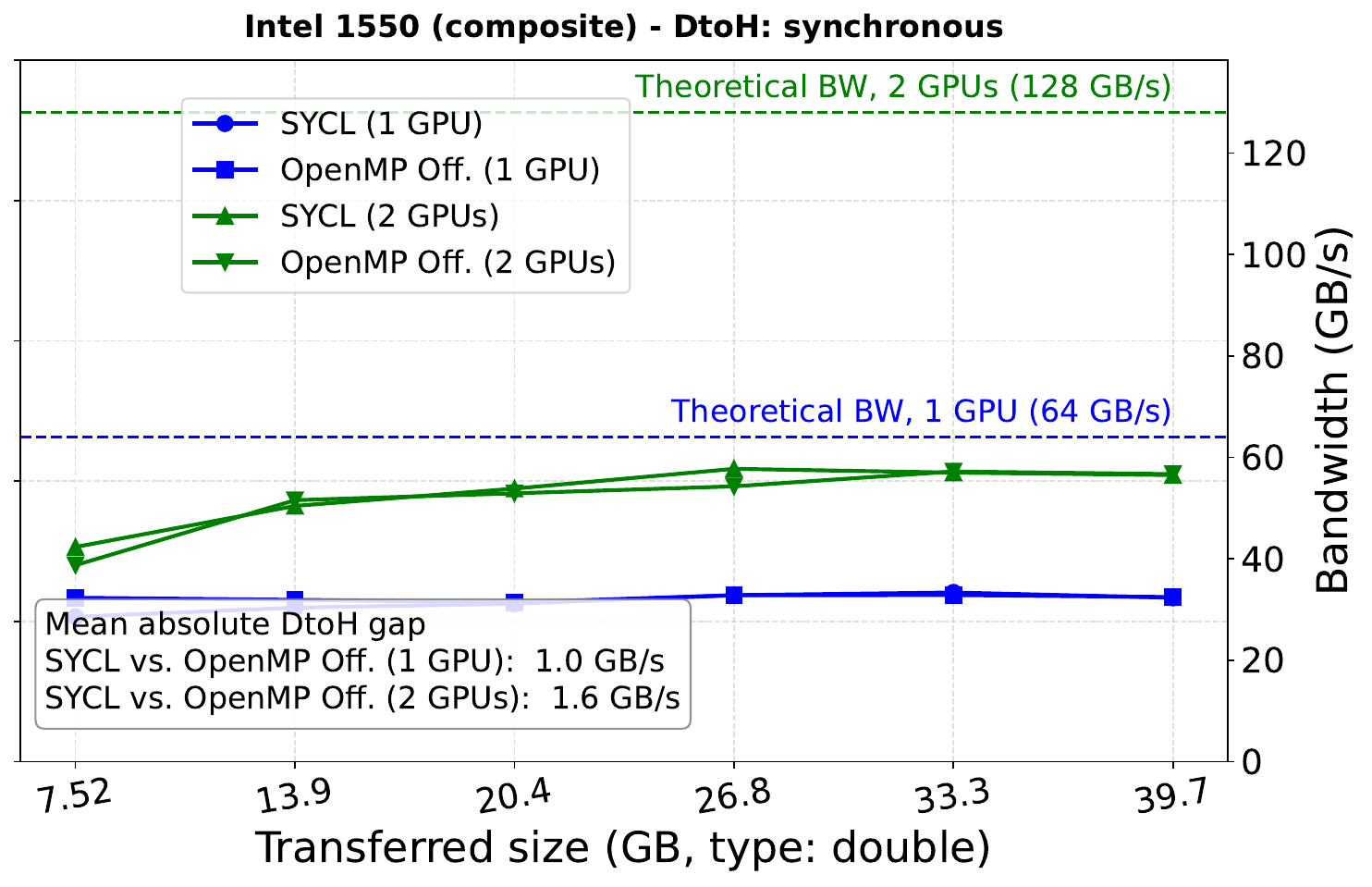}{f}

\par\medskip

\bwpanel{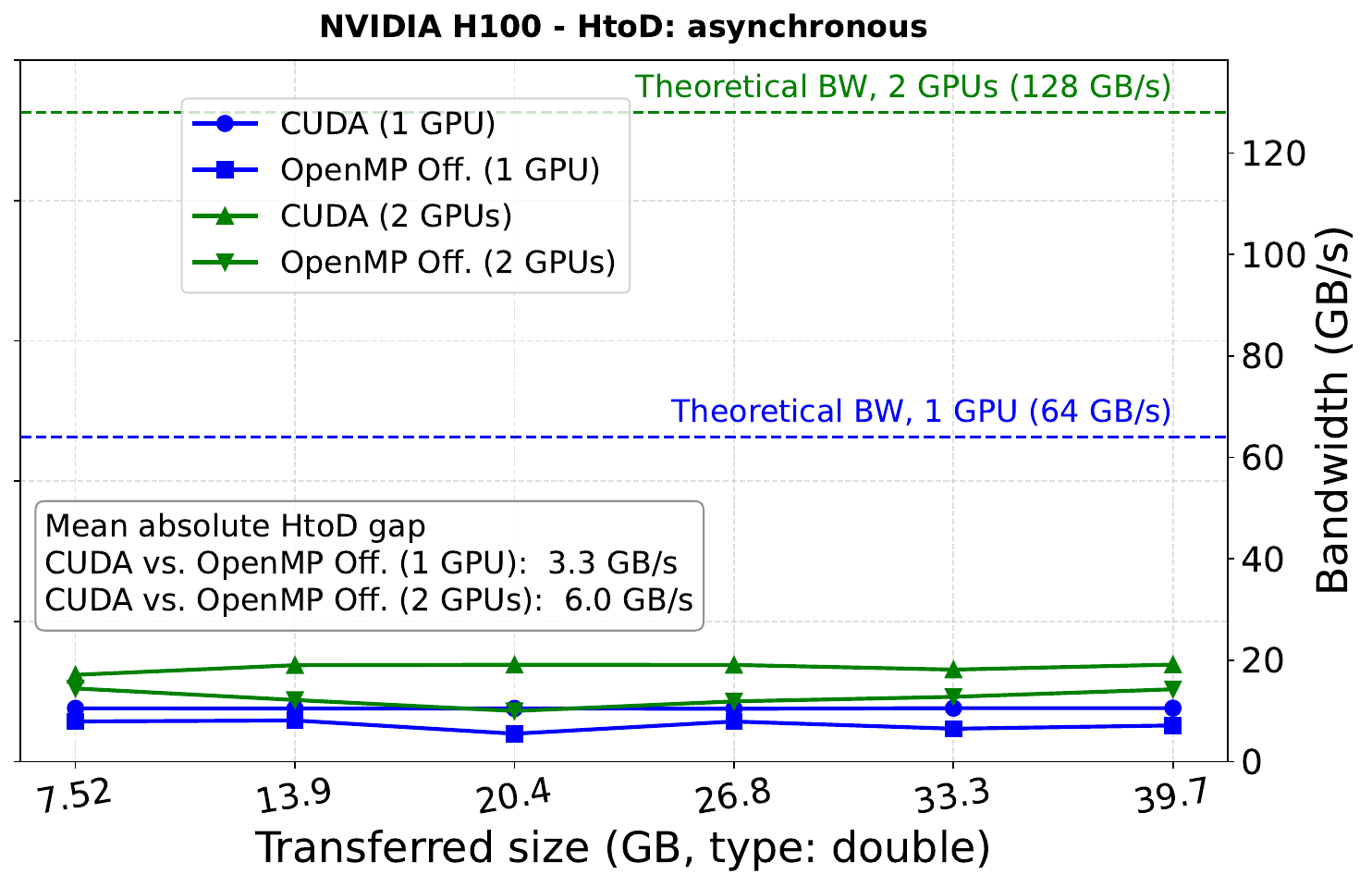}{g}\hfill
\bwpanel{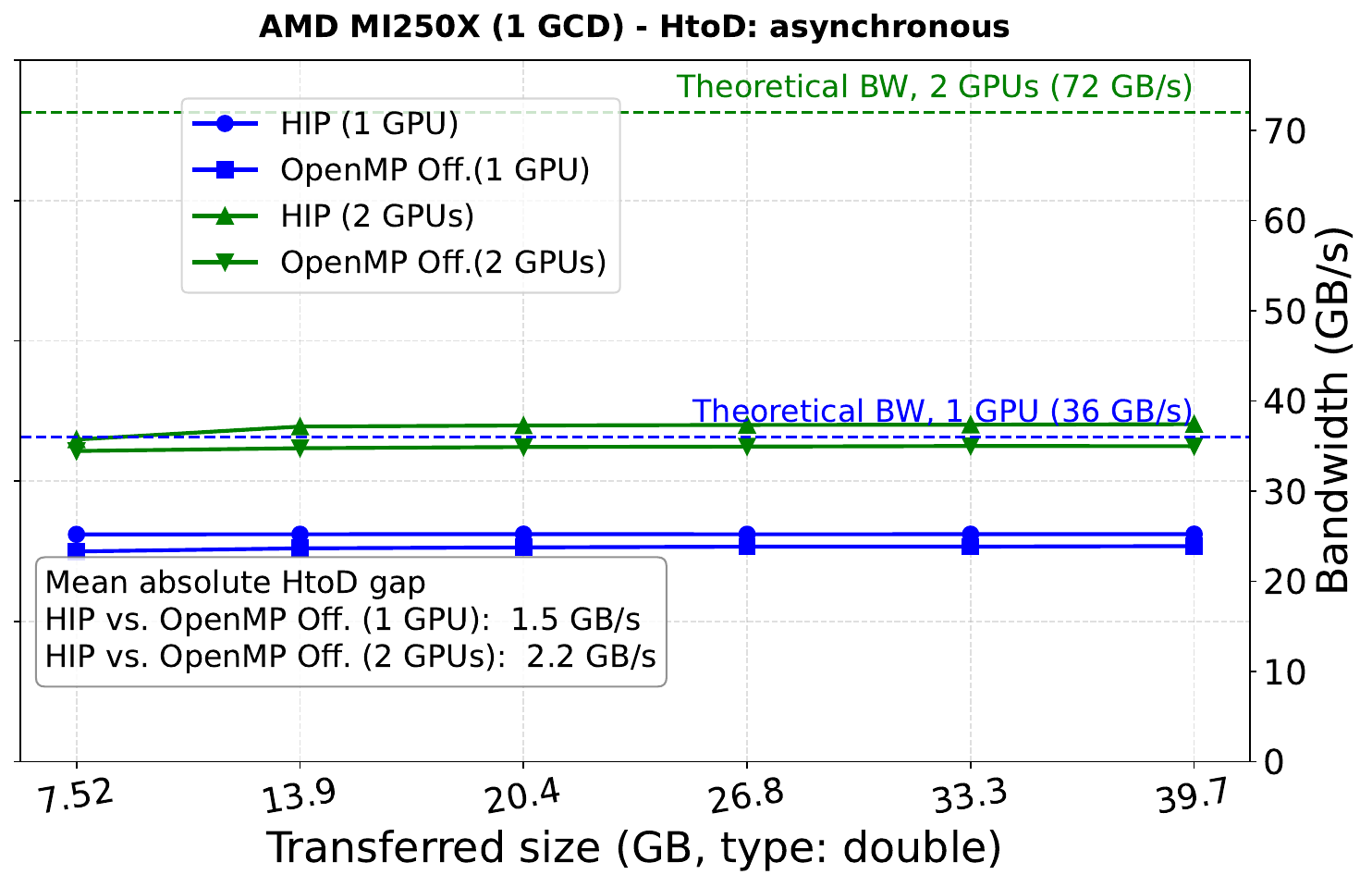}{h}\hfill
\bwpanel{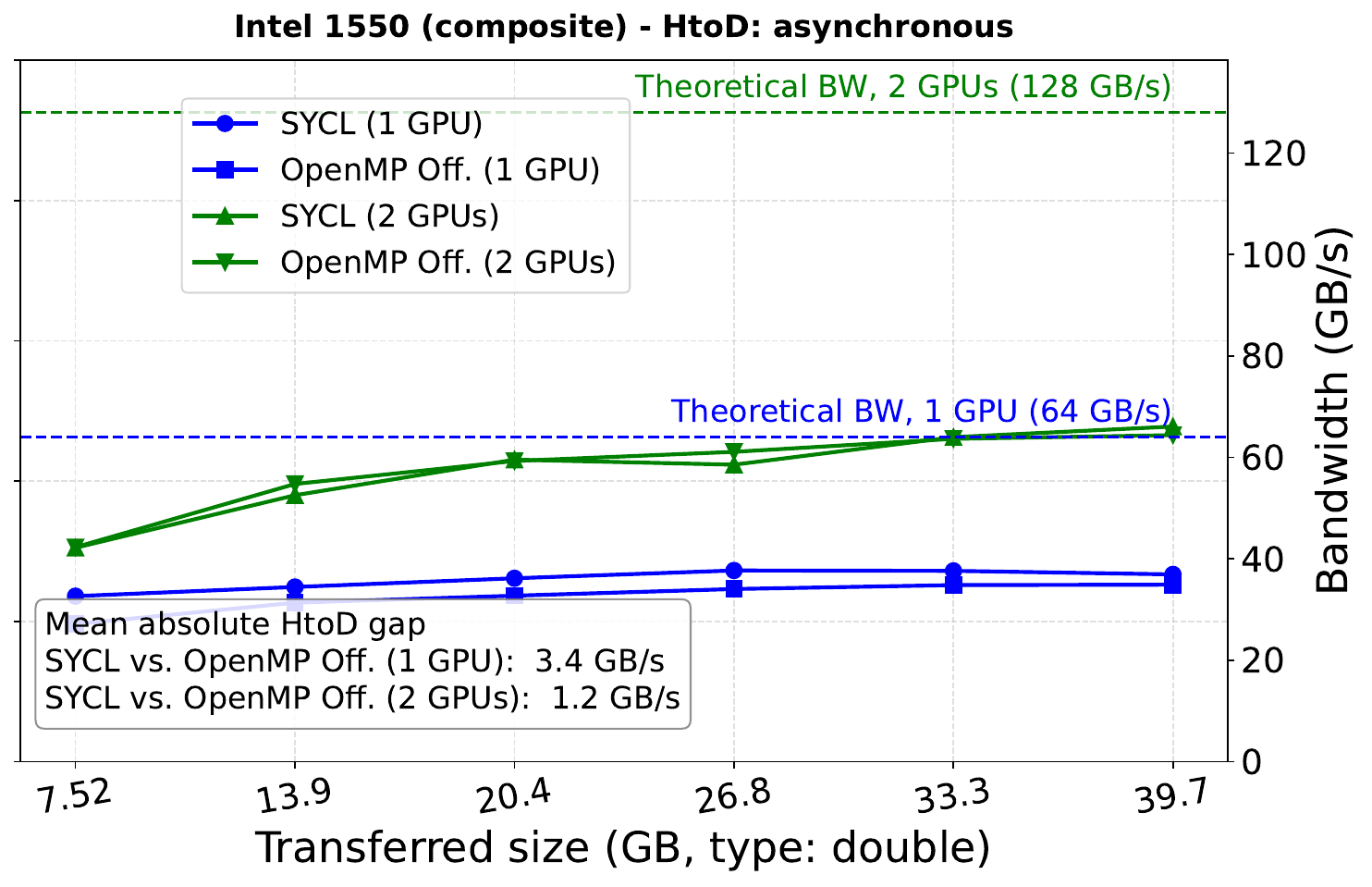}{i}

\par\medskip

\bwpanel{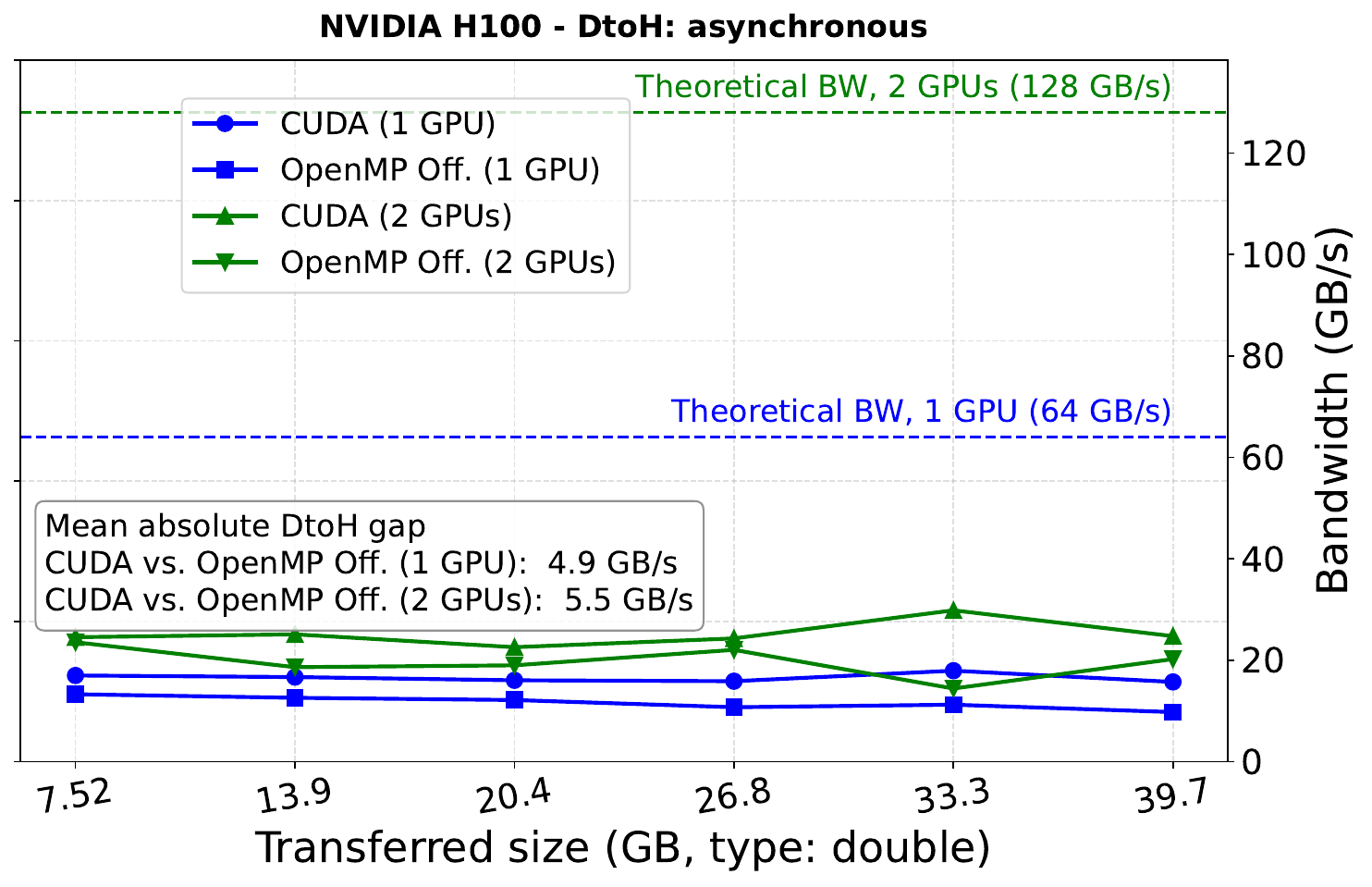}{j}\hfill
\bwpanel{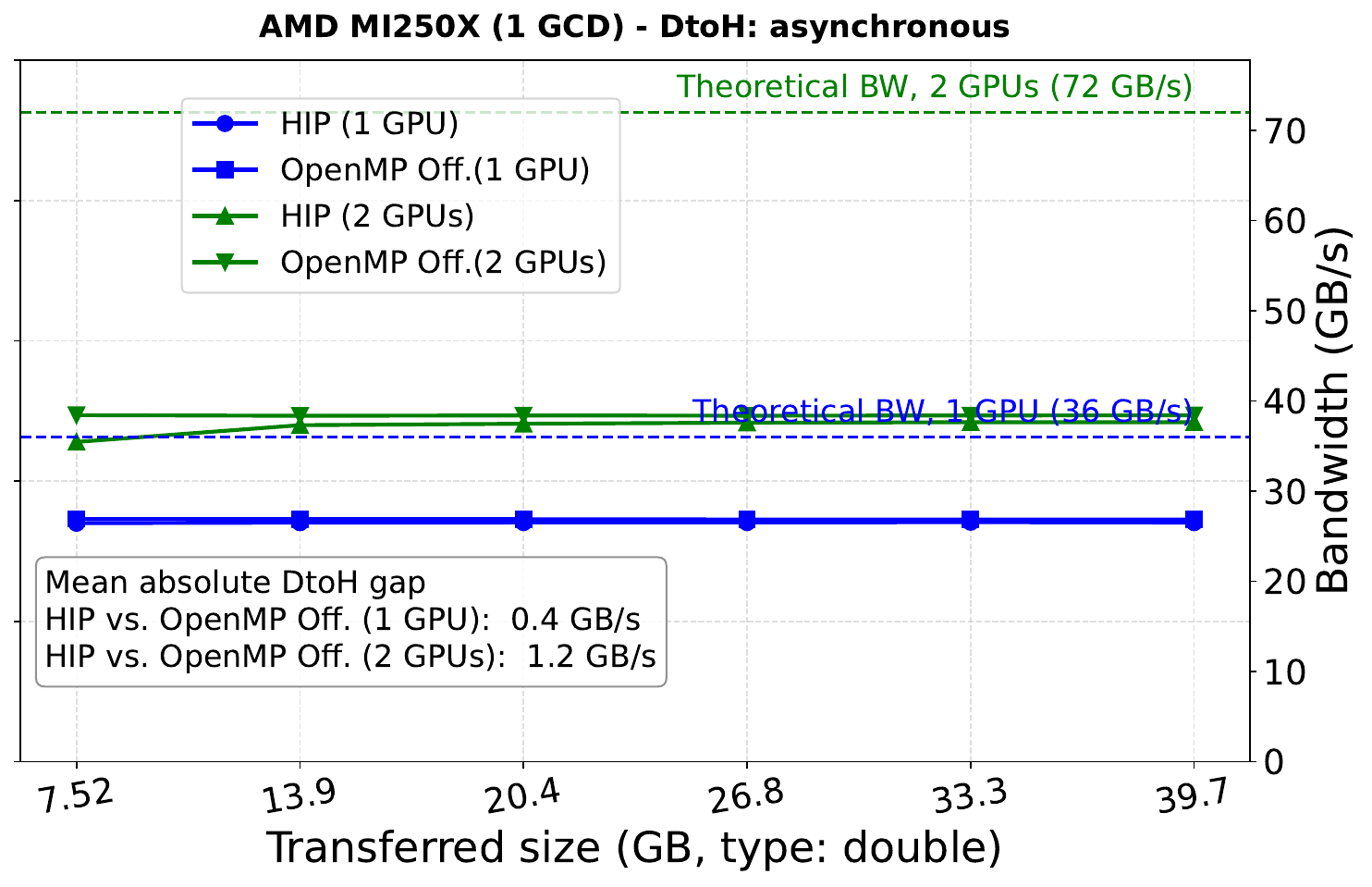}{k}\hfill
\bwpanel{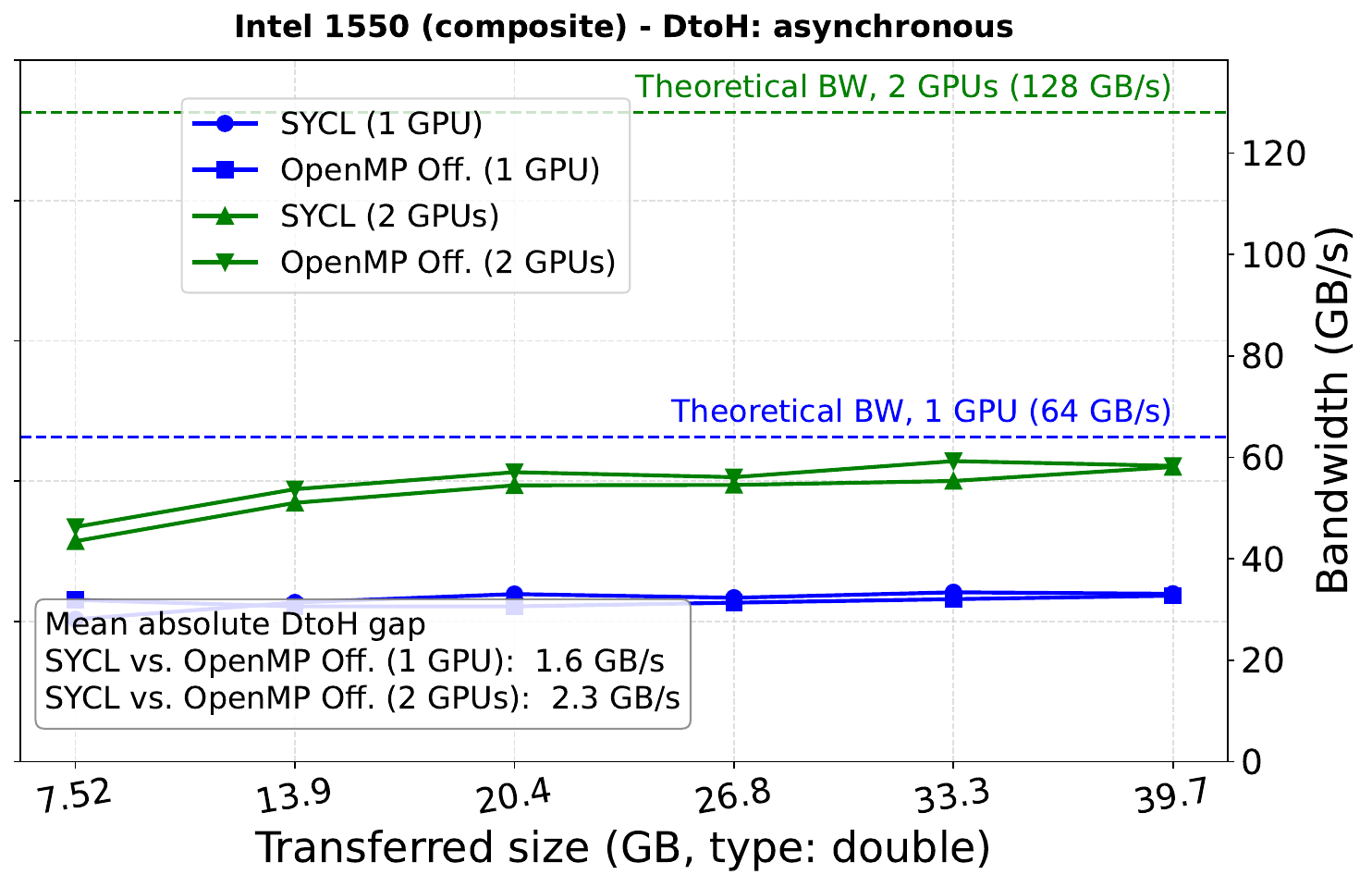}{l}

\par\medskip

\bwpanel{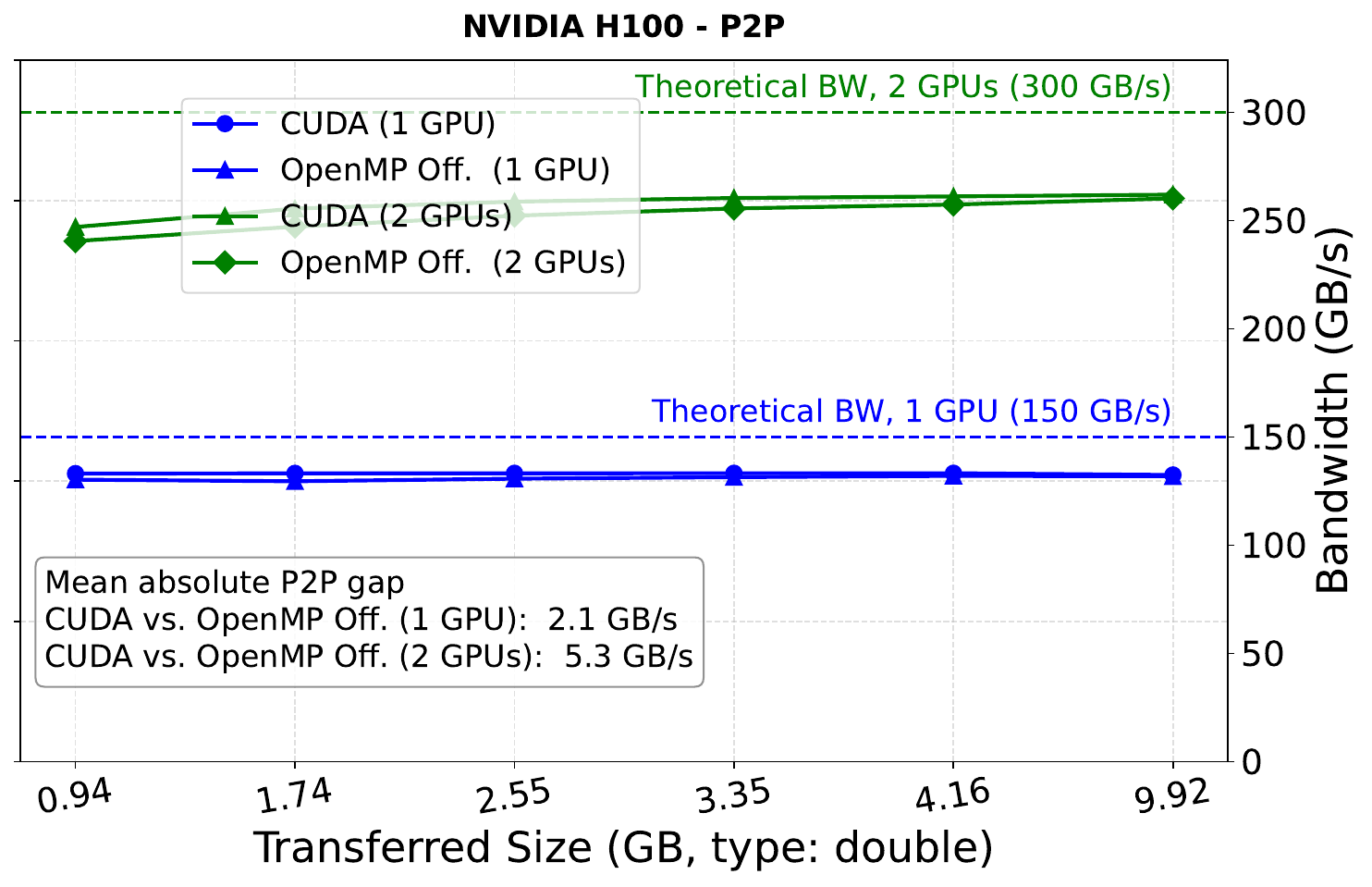}{m}\hfill
\bwpanel{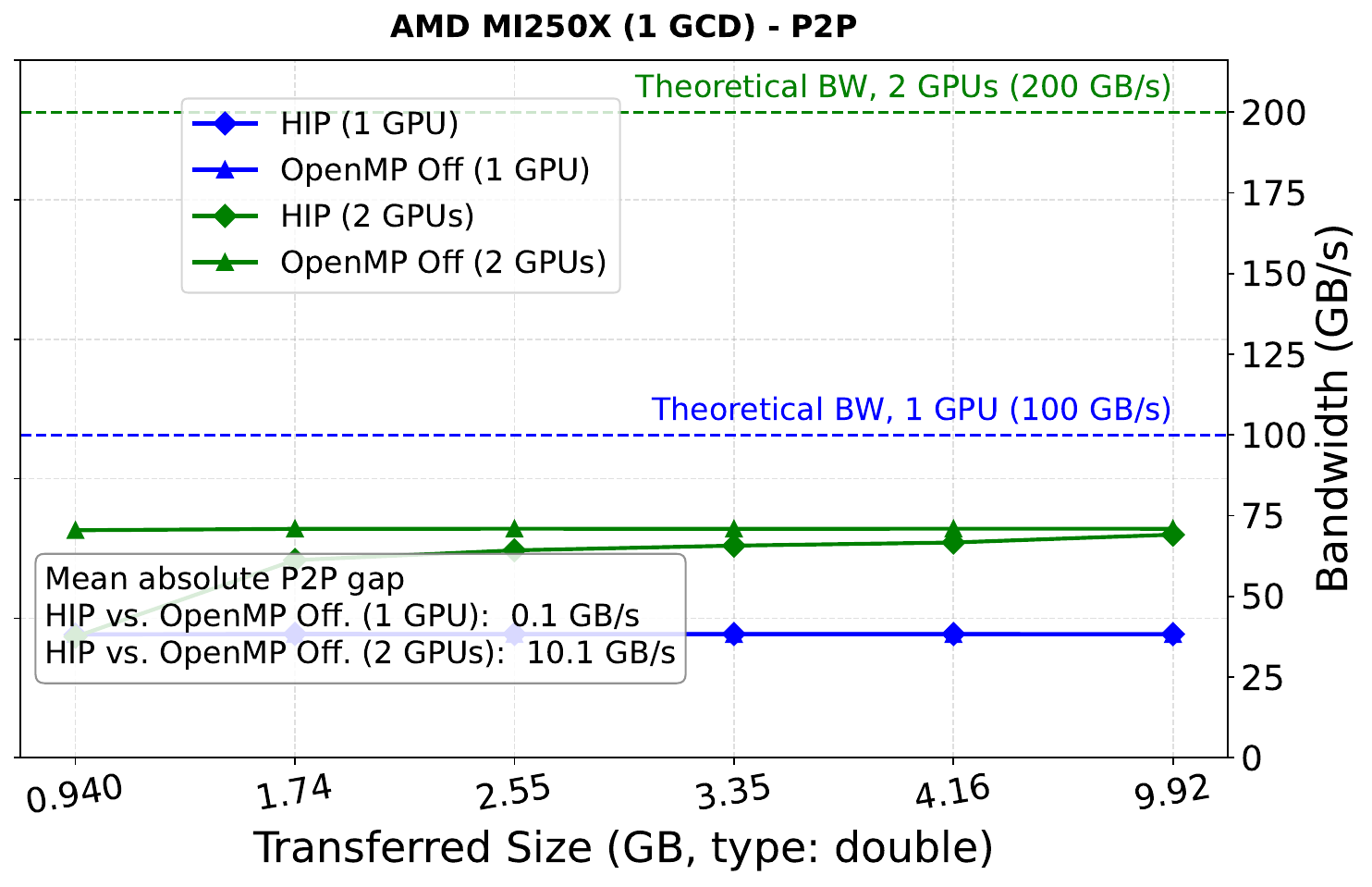}{n}\hfill
\bwpanel{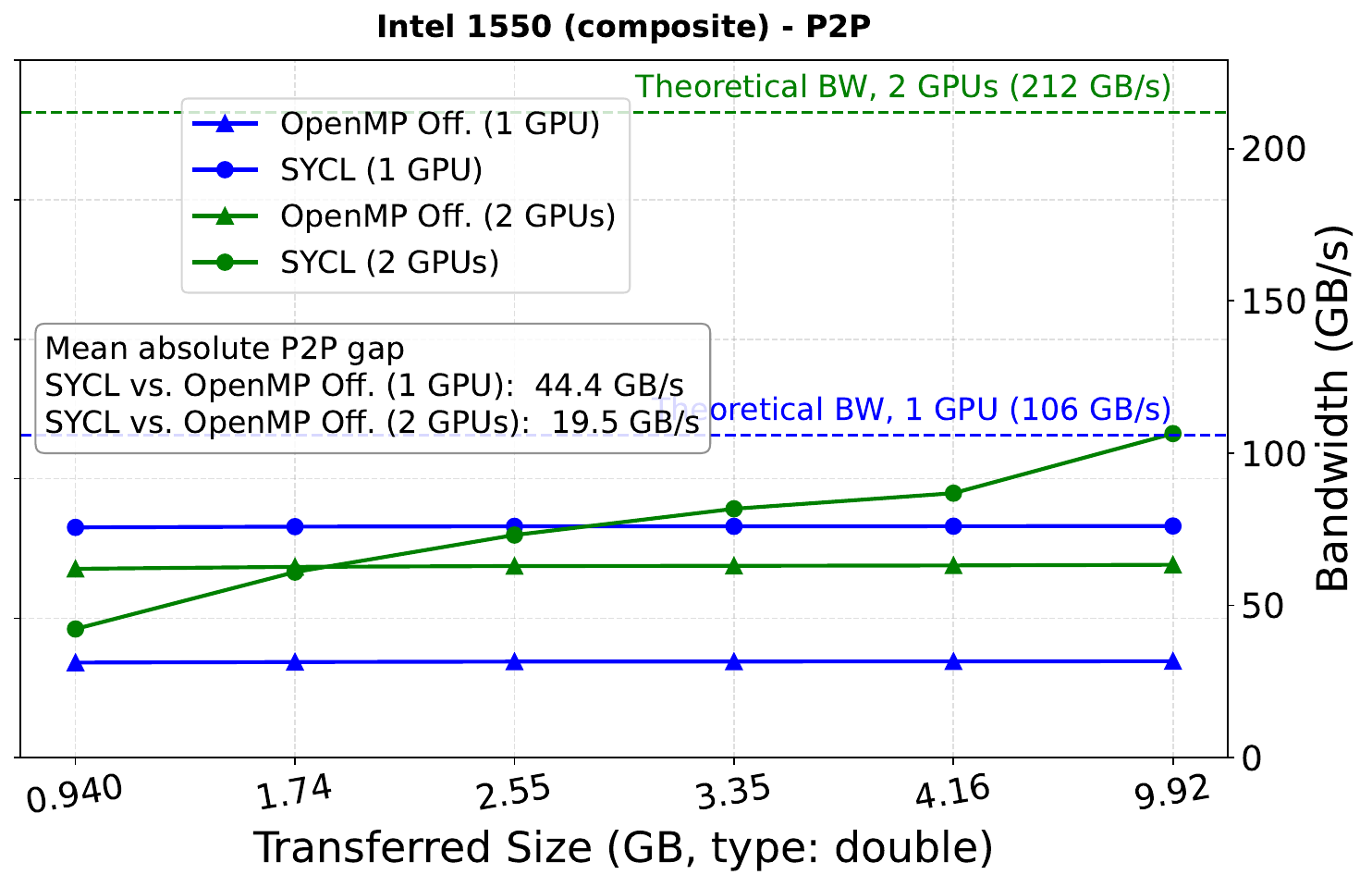}{o}
\caption{Theoretical and measured intra-node transfer bandwidth on three GPU architectures (columns: NVIDIA~H100, AMD~MI250X, Intel~Max~1550) for five transfer modes (rows: synchronous HtoD and DtoH, asynchronous HtoD and DtoH, and P2P). In the host-transfer rows, a single GPU is driven by one host thread and two GPUs by two host threads using OpenMP; the P2P row uses two GPUs and two host threads throughout.}
\label{fig:p2p}
\end{figure}

\subsection{Evaluation of Hidden Helper Thread Configurations} 
OpenMP provides a mechanism to utilize hidden helper threads, which can be controlled via the environment variable \texttt{LIBOMP\_NUM\_HIDDEN\_HELPER\_THREADS=N}. These threads enhance the asynchronous behavior of compute kernels that utilize the \texttt{nowait} configuration during GPU Offloading and various data copy operations. By default, eight hidden helper threads are employed. Therefore, it is crucial to evaluate the impact of these threads within the context of Multi-GPU OpenMP Offloading and their overall effect on performance. To conduct this investigation, we established three categories of hidden helper thread configurations, as detailed in Table~\ref{tab:helper-thread-configurations-main}.  
\begin{table}[htbp]
\caption{Host-thread and hidden-helper-thread configurations used for the
evaluting \texttt{OpenMP Offloading \{2,4\}-GPU Versions}.}
\label{tab:helper-thread-configurations-main}
\centering

\small
\setlength{\tabcolsep}{2pt}
\renewcommand{\arraystretch}{1.08}

\begin{tabularx}{\columnwidth}{
@{}
>{\hsize=1.30\hsize\raggedright\arraybackslash}X
>{\hsize=0.70\hsize\centering\arraybackslash}X
c@{\hspace{12pt}}
c@{\hspace{4pt}}
c
@{}
}
\toprule

\multirow{2}{*}{\textbf{Implementation}}
&
\multirow{2}{*}{%
  \shortstack{\textbf{Host}\\\textbf{threads}}}
&
\multicolumn{3}{c}{\textbf{Hidden Helper Threads (HHT)}} \\

\cmidrule(lr){3-5}

& &
\textbf{HHT-0} &
\textbf{HHT-GPU} &
\textbf{HHT-8} \\
\midrule

\texttt{OpenMP Offloading 2-GPU Version 1} & 1 & 0 & 2 & 8 \\
\texttt{OpenMP Offloading 2-GPU Version 2} & 1 & 0 & 2 & 8 \\
\texttt{OpenMP Offloading 2-GPU Version 3} & 2 & 0 & 2 & 8 \\
\texttt{OpenMP Offloading 2-GPU Version 4} & 2 & 0 & 2 & 8 \\
\midrule
\texttt{OpenMP Offloading 4-GPU Version 1} & 1 & 0 & 4 & 8 \\
\texttt{OpenMP Offloading 4-GPU Version 2} & 1 & 0 & 4 & 8 \\
\texttt{OpenMP Offloading 4-GPU Version 3} & 4 & 0 & 4 & 8 \\
\texttt{OpenMP Offloading 4-GPU Version 4} & 4 & 0 & 4 & 8 \\
\bottomrule
\end{tabularx}

\medskip
\noindent\footnotesize
HHT-0 disables hidden helper threads; HHT-GPU uses one hidden helper
thread per GPU; and HHT-8 uses a fixed total of eight hidden helper threads per GPU.
\end{table}

\subsection{Bandwidth Measurement on NVIDIA H100, AMD MI250, and Intel Max 1550}  
To fully understand the capabilities of the GPU bandwidth network between GPUs and the CPU, it is essential to analyze the performance of memory transfer APIs across various GPU programming models and architectures. We have established a set of benchmarks designed to measure the performance of GPU programming model APIs for data transfer between the CPU and GPU, as well as between GPUs (P2P). For instance, we utilized all the functionalities of the memory transfer APIs outlined in Table~\ref{tab:high-low-spec}. This methodology expands upon the framework we established in our earlier work on OpenACC~\cite{prior2026}. 

In the context of OpenMP Offloading, we developed several benchmark cases to evaluate the actual bandwidth offered by the OpenMP Offloading API for communication between the CPU and GPU, as well as between multiple GPUs.

\begin{figure}[htbp]
\centering
\subfloat[]{\includegraphics[width=0.48\linewidth]{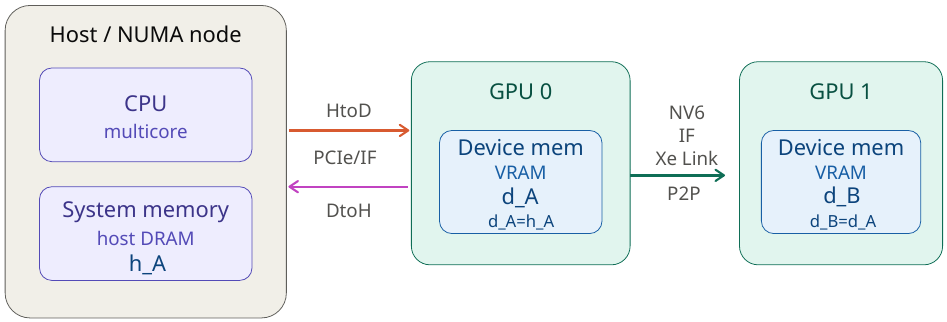}\label{fig:comparison-1}}
\hfill
\subfloat[]{\includegraphics[width=0.48\linewidth]{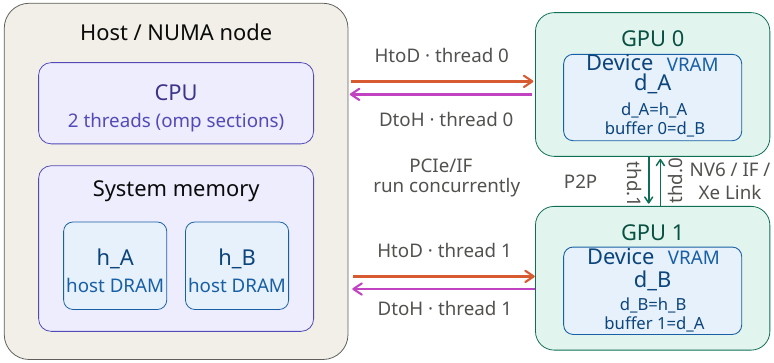}\label{fig:comparison-2}}
\caption{Comparison of bandwidth measurement scenarios: (\textbf{a}) 1 GPU: Schematic overview of measurements for HtoD, DtoH, and P2P between the host and GPU, as well as between GPUs using 1 host thread; (\textbf{b}) 2 GPUs: Schematic overview of measurements for HtoD, DtoH, and P2P between the host and GPU, and between GPUs using 2 host threads concurrently.}
\label{fig:comparison}
\end{figure}

\begin{itemize}
\item HtoD (Host to Device):
\begin{itemize}
    \item 1 GPU: Using \texttt{omp\_target\_memcpy()} and \texttt{omp\_target\_memcpy\_async()} with a single host thread, data is transferred from the host to the device, as depicted in Figure~\ref{fig:comparison-1}.
    \item 2 GPUs: Using \texttt{omp\_target\_memcpy()} and \texttt{omp\_target\_memcpy\_async()} with two host threads, two sets of data can be sent concurrently from the host to both devices, as illustrated in Figure~\ref{fig:comparison-2}.
\end{itemize}
\item DtoH (Device to Host):
\begin{itemize}
    \item 1 GPU: Using \texttt{omp\_target\_memcpy()} and \texttt{omp\_target\_memcpy\_async()} with a single host thread, data is transferred from the device to the host, as shown in Figure~\ref{fig:comparison-1}.
    \item 2 GPUs: Using \texttt{omp\_target\_memcpy()} and \texttt{omp\_target\_memcpy\_async()} with two host threads, two sets of data can be sent concurrently from the device to the host, as illustrated in Figure~\ref{fig:comparison-2}.
\end{itemize}
\item P2P (Peer-to-Peer):
\begin{itemize}
    \item 1 GPU: Using \texttt{omp\_target\_memcpy\_async()} with a single host thread, data is transferred from one GPU to another, as depicted in Figure~\ref{fig:comparison-1}.
    \item 2 GPUs: Using \texttt{omp\_target\_memcpy\_async()} with two host threads, data can be transferred concurrently between the two GPUs, as illustrated in Figure~\ref{fig:comparison-2}.
\end{itemize}
\end{itemize}

In this research on solving the 3D heat equation, the estimates for halo data movement are as follows: For double-precision floating-point values (\(b=8\) bytes) and a one-layer halo (\(h=1\)), each GPU is responsible for sending and receiving one \(N^2\) plane. Consequently, the communication volume per GPU can be calculated as:

\[
V_{\mathrm{GPU}} = 2hN^2b = 2 \times 1 \times N^2 \times 8 = 16N^2\,\text{bytes}. 
\]

For global domain sizes \(N^3\) in the set \(\{512^3, 640^3, 768^3, 896^3, 1024^3, 1152^3, 1280^3\}\), the corresponding send-and-receive communication volumes per GPU are \(0.004194\), \(0.006554\), \(0.009437\), \(0.012845\), \(0.016777\), \(0.021234\), and \(0.026214\,\text{GB}\), respectively. This volume of data is essential for developing the performance model for multi-GPU systems, which is further discussed in Subsections~\ref{sec:perf-model} and ~\ref{sec:perf-results}.

\subsection{Performance Model}
\label{sec:perf-model}
Here, we derive the performance model to gain a better understanding of the parallel implementations of OpenMP Offloading, along with hardware bandwidth measurements. This will help predetermine the expected runtime cost and, overall, draw the performance model. For this, we assume that the computational cost is determined by the number of points that need to be calculated in the 3D domain. Specifically, \((N-2)^3\) points need to be computed, where \(N\) represents the total number of points in the grid, excluding the boundary layers. Therefore, the computation time per step can be modeled as:

\begin{equation}
t_{\text{step}} = k (N-2)^3,
\label{eq:tstep}
\end{equation}

where \(k\) denotes the overhead cost, which may arise from latency or other machine-related factors, and we assume it remains constant.

Let \( n_{\text{steps}} \) denote the total number of time steps. In our measurements, the reported runtime \( T_{\text{total}} \) refers to the execution of \( n_{\text{steps}} \, t_{\text{step}} \). We can express this relationship as follows:

\[
T_{\text{total}} = n_{\text{steps}} \, t_{\text{step}}.
\]

Substituting the expression for \(t_{\text{step}}\) from Eq.~\eqref{eq:tstep}, we obtain:

\[
T_{\text{total}} = n_{\text{steps}} \, k (N-2)^3.
\]

Rearranging this equation to solve for \(k\), we arrive at:

\[
k = \frac{T_{\text{total}}}{n_{\text{steps}}} \, (N-2)^3.
\]

Since the grid size changes, we need to take the average of \(k\). To do this, we model \(\bar{k}\) as follows:

\begin{equation}
\bar{k} = \frac{1}{M} \sum_{j=1}^{M} k(N_j).  
\label{eq:kbar}
\end{equation}

Here, \( N_j \) represents the \( j \)-th grid size, and \( M \) is the total number of grid sizes considered. For instance, \( k(N_j) \) is evaluated at \( N_j = 512, 640, \ldots, 1280 \).

Now we can model the estimated total computation time for the multi-GPU version. For the multi-GPU setup, we must consider the data movement involving HtoD, DtoH, and P2P transfers. Each operation is modeled as follows:

\[
\text{HtoD} = \frac{N^2 \text{ bytes}}{\text{Meas.BW}_{\text{HtoD}}}, \quad
\text{DtoH} = \frac{N^2 \text{ bytes}}{\text{Meas.BW}_{\text{DtoH}}}, \quad
\text{P2P} = \frac{N^2 \text{ bytes}}{\text{Meas.BW}_{\text{P2P}}}.
\]

Since we have two approaches for data transfer, the total time for data movement via CPU and between GPUs through P2P can be represented as:

\[
T_{\text{stage}} = n_{\text{steps}} \left( \text{HtoD} + \text{DtoH} \right)
\]

\[
T_{\text{P2P}} = n_{\text{steps}} \left( \text{P2P} \right)
\]

The total computation time for the multi-GPU setup can be calculated as follows:

For overlapping (concurrent) transfers:
\begin{equation}
T_{\text{total\_G}}^{\text{stage}} = n_{\text{steps}} \left[ \frac{\bar{k} (N-2)^{3}}{G} + T_{\text{stage}} \right], 
\label{eq:stage}
\end{equation}
where \( G \) represents the number of GPUs.

For P2P transfers:
\begin{equation}
T_{\text{total\_G}}^{\text{P2P}} = n_{\text{steps}} \left[ \frac{\bar{k} (N-2)^{3}}{G} + T_{\text{P2P}} \right].
\label{eq:p2p}
\end{equation}

If communication is fully overlapped with computation, the equations referenced in Eq.~\ref{eq:stage} and Eq.~\ref{eq:p2p} become:

\begin{equation}
T_{\mathrm{total},G}^{\mathrm{stage,overlap}} = n_{\mathrm{steps}} \max \left[ \frac{\bar{k}(N-2)^3}{G}, T_{\mathrm{stage}} \right],
\label{eq:stage-overlap}
\end{equation}
and
\begin{equation}
T_{\mathrm{total},G}^{\mathrm{P2P,overlap}} = n_{\mathrm{steps}} \max \left[ \frac{\bar{k}(N-2)^3}{G}, T_{\mathrm{P2P}} \right].
\label{eq:p2p-overlap}
\end{equation}

These models assume ideal computational load balancing and neglect additional costs such as kernel launch overhead, synchronization, communication latency, and bandwidth contention.

\section{Results and Discussion}  
\label{discussion}
In this section, we present a discussion of the results obtained from the experiment based on the methodology described in Section~\ref{methodology}. This discussion includes an examination of various versions of multi-GPU OpenMP Offloading, along with comparisons to CUDA, HIP, and SYCL. We will also analyze the impact of HHT and discuss the proposed performance model using the measured bandwidth. The remainder of this section will elaborate on these findings in detail.

\subsection{Multi-GPU intra-node OpenMP Offloading Performance Analysis}
This subsection focuses on the performance analysis of solving 3D heat transfer problems using various versions of OpenMP Offloading in an intra-node setting. Figure~\ref{fig:diffusion_comparison} illustrates the overall runtime for different grid sizes of \(N\), comparing the performance of CUDA, HIP, and SYCL. We measure variability using the formula \((\max - \min)/\mathrm{median}\) for each configuration. We apply this method consistently across OpenMP Offloading and native implementations (CUDA, HIP, and SYCL) on their respective GPU platforms. We observe median variability values of 0.6\% (CUDA) and 0.9\% (OpenMP Off.) on the H100; 2.7\% (HIP) and 2.9\% (OpenMP Off.) on the MI250X; and 4.4\% (SYCL) and 5.9\% (OpenMP Off.) on the Max 1550, with maxima of 7.4\%, 10.9\%, and 21.6\%, respectively. We report run-to-run variations on each platform and its corresponding programming model (with different approaches), including OpenMP Offloading (with different approaches). Nevertheless, the speedups discussed below are substantial. We will further analyze these approaches based on the results we obtained.

\subsubsection{NVIDIA H100}  
In our tests on the NVIDIA H100, we examined the implementations of CUDA and OpenMP Offloading using single, 2, and 4 GPU configurations. As shown in Figures~\ref{fig:h100_2gpu} and ~\ref{fig:h100_4gpu}, the performance of \texttt{OpenMP Offloading \{2,4\}-GPU Version 4} in both the 2 (2x) and 4 (4x) GPU scenarios is competitive with the single GPU OpenMP Offloading implementation. Notably, the 4 GPU implementations (Versions 3 and 4) outperformed the single GPU implementation of the CUDA version. Below, we summarize the key findings from our analysis of the NVIDIA H100 regarding the 4 GPU implementation:

\begin{itemize}  
\item The 2 GPU implementation is less efficient than the 4 GPU implementation primarily because, while the workload can be distributed across two GPUs, it remains insufficient for optimal performance. Additionally, the overhead from communication contributes to the reduced efficiency of the 2 GPU implementations compared to their 4 GPU counterparts.
\item In contrast, all 4 GPU implementations allow for a more effectively distributed workload, making computations less expensive compared to those in single and 2 GPU implementations. Although the increased number of GPUs results in more communication, the overall computation cost is still lower. This combination leads to a significant speedup, particularly when compared to all versions of the 2 GPU implementations and the single GPU implementation.
\item The \texttt{OpenMP Offloading \{2,4\}-GPU Version 4} not only surpasses the single GPU CUDA implementations but also outperforms other variations of OpenMP Offloading. Its enhanced speed is primarily achieved by concurrently launching GPU kernels with multiple host threads. Furthermore, the H100's P2P bandwidth, as illustrated in Figure~\ref{fig:h100_case}, provides higher bandwidth, which significantly contributes to the speedup by avoiding data transfer routing through the CPU.
\end{itemize}

\subsubsection{AMD MI250X}  
Within the AMD MI250X, we evaluate the HIP implementations as a reference to compare them against OpenMP Offloading implementations. As illustrated in Figure~\ref{fig:amd_case}, the bandwidth between the Graphics Compute Dies (GCDs) of the GPUs is modest, resulting in only a slight difference in bandwidth between the CPU and GPU. However, the AMD MI250X features a unique design in which each module contains two GCDs. The memory bandwidth between these two GCDs is significantly higher than the data transfer rates between the modules of the MI250X. Notably, all multi-GPU implementations of HIP and OpenMP Offloading utilize a single GCD from each module of the MI250X, a design choice that impacts the communication dynamics between multiple GPUs compared to single-GPU implementations.

Despite this characteristic, the overarching conclusion remains valid: the implementation utilizing four GPUs in OpenMP Offloading performs competitively against the two-GPU implementations and the single-GPU implementation of the HIP version. Below, we summarize key insights regarding the multi-GPU implementations of HIP and OpenMP Offloading.

\begin{itemize}
    \item Ideally, the \texttt{OpenMP Offloading \{2,4\}-GPU Version 4} is expected to outperform all other variants. However, due to the limited computational power of the AMD MI250X (1 GCD), we do not observe a significant difference between the \texttt{OpenMP Offloading 2-GPU Version 3} and \texttt{OpenMP Offloading 2-GPU Version 4}. In scenarios where computational time is extensive, the communication costs become negligible, leading to minimal differences in performance between Versions 3 and 4. Interestingly, 2-GPU Version 2 performs slightly better than Versions 3 and 4, indicating that the latter two achieve a better overlap of computation compared to 2-GPU Versions 1 and 2.

    \item Additionally, in the context of the 4-GPU implementations, both \texttt{OpenMP Offloading \{2,4\}-GPU Version 3} and \texttt{OpenMP Offloading \{2,4\}-GPU Version 4} demonstrate significant speedup compared to the other versions within the 4-GPU category. As the workload is distributed across the four GPUs, we observe a remarkable speedup in 2- and 4-GPU Version 4, achieving nearly 2x and 4x the performance compared to a single GPU implementation.
\end{itemize}

\subsubsection{Intel Max 1550}
Concerning Intel 1550, we do the comparison of the reference GPU programming model SYCL against OpenMP Offloading on Intel GPUs. Figures~\ref{fig:intel_2gpu} and \ref{fig:intel_4gpu} illustrate the runtime implementations of SYCL and OpenMP Offloading with 2 GPUs and 4 GPUs, respectively. The performance characteristics of the Intel 1550 GPU closely resemble those of the NVIDIA H100. Below, we summarize the key insights derived from the OpenMP Offloading implementation:

\begin{itemize}
    \item The data transfer between GPUs is slightly higher than the data transfer via the CPU, making \texttt{OpenMP Offloading \{2,4\}-GPU Version 4} an improvement over \texttt{OpenMP Offloading \{2,4\}-GPU Version 3}. 
    \item As an end user, programmers can opt to use either a single tile or the entire composite GPU of the Intel Max 1550. This flexibility minimizes overall computational time compared to the other GPUs evaluated in this research. For specifications of the Intel GPU, please refer to Table~\ref{tab:h100-mi250x-max1550}.
    \item The P2P data transfer performance between SYCL and OpenMP Offloading does not appear particularly impressive. The measured bandwidth gap is slightly greater than what we observed with the NVIDIA H100 and the AMD MI250X. This discrepancy is primarily due to OpenMP Offloading not delivering performance on par with SYCL when the Intel 1550 is considered as a composite.
    \item The ideal speedup of 4x is not achieved when using 4 GPUs with the \texttt{OpenMP Offloading 4-GPU Version 4}. This shortfall is attributed to an increase in communication overhead, kernel launches, synchronization, and other factors, which ultimately raise the overall computational time. As a result, employing 4 GPUs only yields approximately a 3.5x speedup with the OpenMP Offloading 4-GPU version. In contrast, utilizing 2 GPUs achieves nearly a 2x speedup with the \texttt{OpenMP Offloading 2-GPU Version 4}. This is due to the substantial computational workload, which is sufficient to mask the communication overhead.
\end{itemize}

\subsection{Performance Impact of Hidden Helper Threads}  
\label{sec:Hidden-Helpers}  
\begin{figure}[htbp]
\centering
\includegraphics[
    width=\textwidth,
    height=0.9\textheight,
    keepaspectratio
]{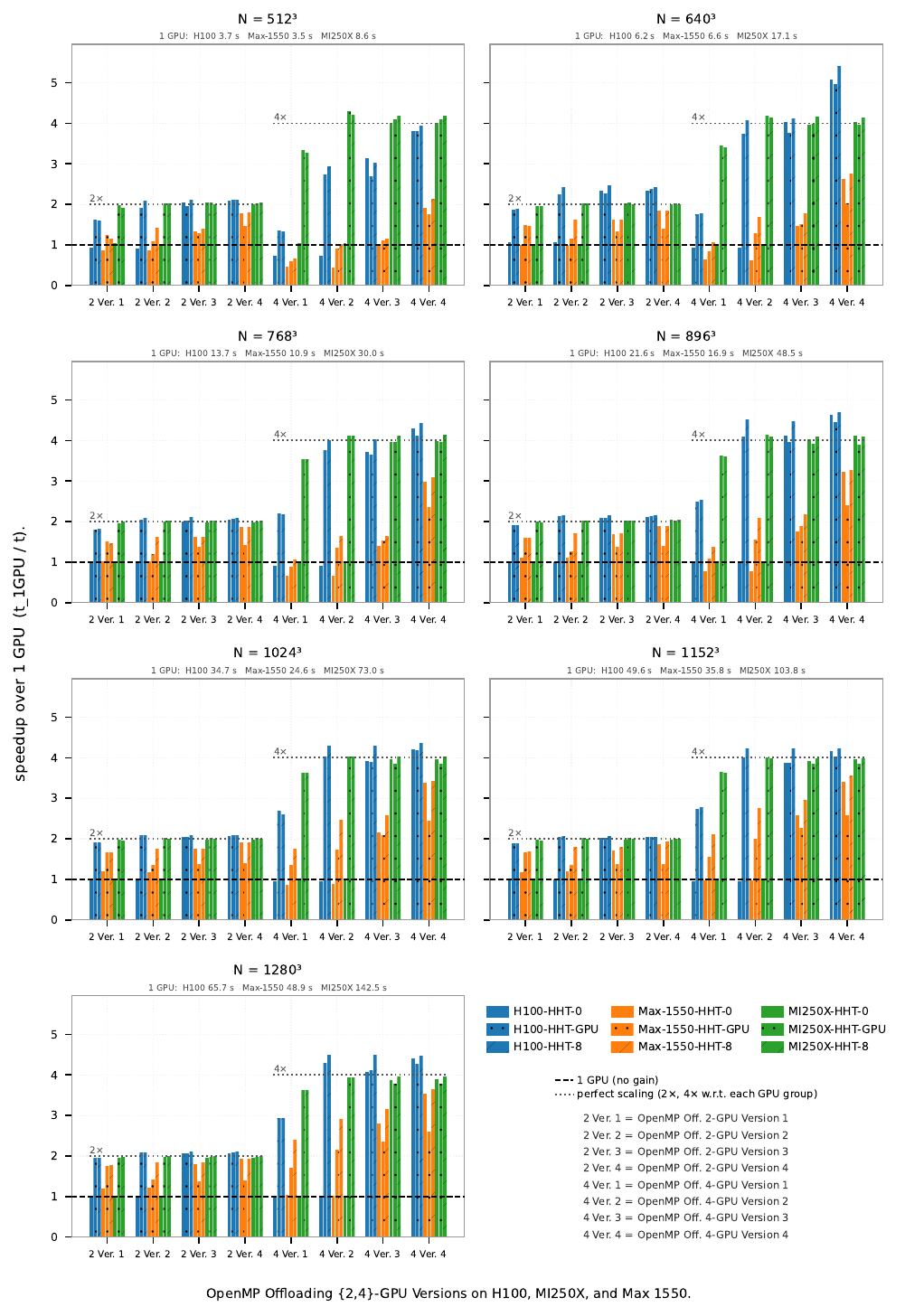}
\caption{Speedup of the \texttt{OpenMP Offloading \{2,4\}-GPU Versions} implementations over each machine's own
\texttt{OpenMP Offloading 1-GPU Baseline} single-GPU run, for problem sizes $N = 512$ to $N = 1280$. Variant names are
abbreviated: 2~Ver.~1 = OpenMP Offloading 2-GPU Version 1; and so on.}
\label{fig:gains_openmp}
\end{figure}

The impact of Hidden Helper Threads (HHT) has shown a notable performance gain across the three GPUs tested in this research. Figure~\ref{fig:gains_openmp} provides an overview of the speedup achieved through the three methodologies considered, as summarized in Table~\ref{tab:helper-thread-configurations-main}. In every case, HHT-8 demonstrated superior performance compared to both HTT-0 and HTT-GPU. It is evident that utilizing a greater number of HHTs significantly enhances the concurrency of asynchronous operations within the application.  

\begin{figure}[htbp]
\centering
\includegraphics[width=\textwidth]{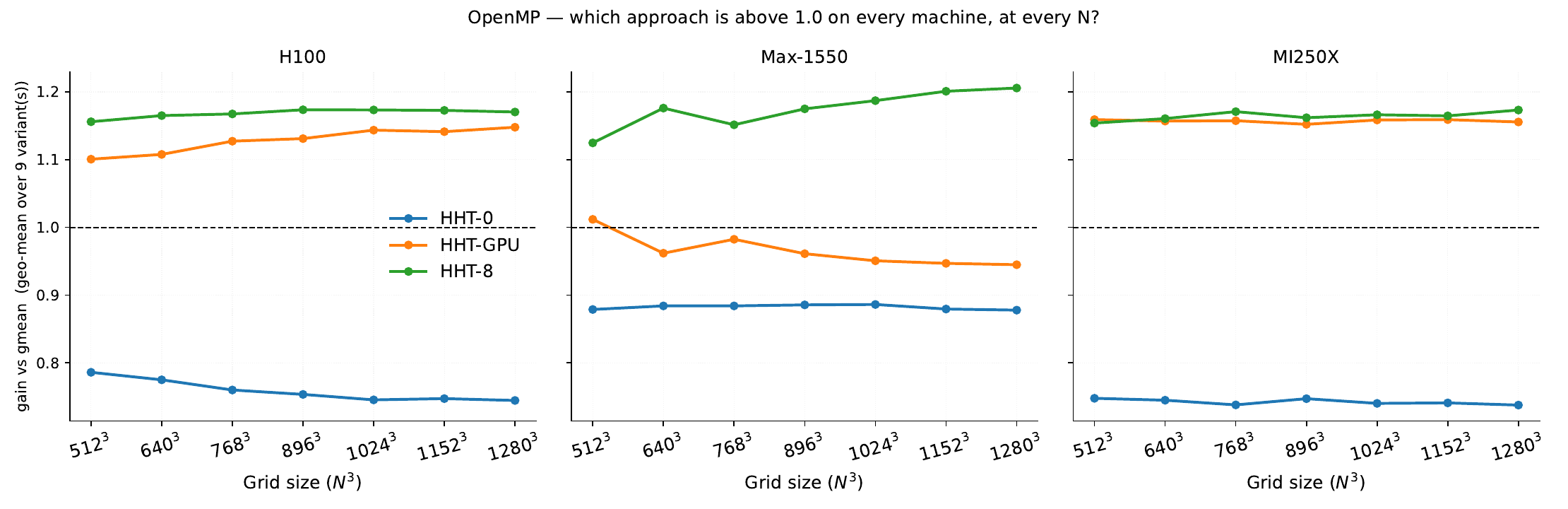}
\caption{Geometric-mean performance gain of the HHT-0, HHT-GPU, and HHT-8
approaches for the \texttt{OpenMP Offloading \{2,4\}-GPU Versions} on the NVIDIA H100,
Intel-1550, and AMD MI250X. The dashed horizontal line represents a gain of
one relative to the geometric-mean reference. HHT-8 remains above unity for
all evaluated problem sizes and GPU architectures.}
\label{fig:decision-openmp}
\end{figure}

To further illustrate this scenario, we plotted another model using the geometric mean (gmean). As shown in Figure~\ref{fig:decision-openmp}, we divided the execution time by the number of time steps (approximately 500) for each GPU architecture, implementation version, and grid size. The geometric mean of the normalized times for HHT-0, HHT-GPU, and HHT-8 served as a neutral reference. The gain was then calculated by dividing this reference time by the normalized time of each configuration. A gain above 1 indicates better performance, a gain of 1 indicates equal performance, and a gain below 1 indicates lower performance. For the overall comparison, gains across the selected implementation versions were also combined using the geometric mean.  

We did not experiment with more than 8 threads because, in the current configuration involving four GPU cases, there are approximately 10 asynchronous operations (comprising 4 interior kernels and 6 boundary kernels) that must be managed by the host. Additionally, each GPU is assigned a corresponding OpenMP thread to facilitate the streaming of operations. We hypothesize that increasing the number of threads beyond 8 would not yield further speedups within this problem setting.  

In terms of performance efficiency among the three GPUs—NVIDIA H100, AMD MI250X, and Intel 1550—the Intel 1550 exhibited slightly lower speedup compared to the NVIDIA H100 and AMD MI250X. This is attributed to the Intel 1550's superior computational power, which allows it to complete computational tasks more quickly but incurs a penalty in communication time. Conversely, when computation takes longer, as observed with the other two GPUs (NVIDIA H100 and AMD MI250X), there is sufficient time for halo computation and communication to be completed. Consequently, we observe nearly perfect speedups with the NVIDIA H100 and AMD MI250X. This observation is further supported by the lower computational power of the NVIDIA H100 and AMD MI250X compared to the Intel 1550. The production analysis plots in Figure~\ref{fig:diffusion_comparison} are based on the HHT-8 configuration (with HHT set to 8).  

\subsection{Performance Model Analysis}
\label{sec:perf-results}
\begin{figure}[htbp]
\centering
\subfloat[\centering NVIDIA H100, 2 GPUs.\label{fig:model_h100_2gpu}]{%
    \includegraphics[width=0.48\linewidth]{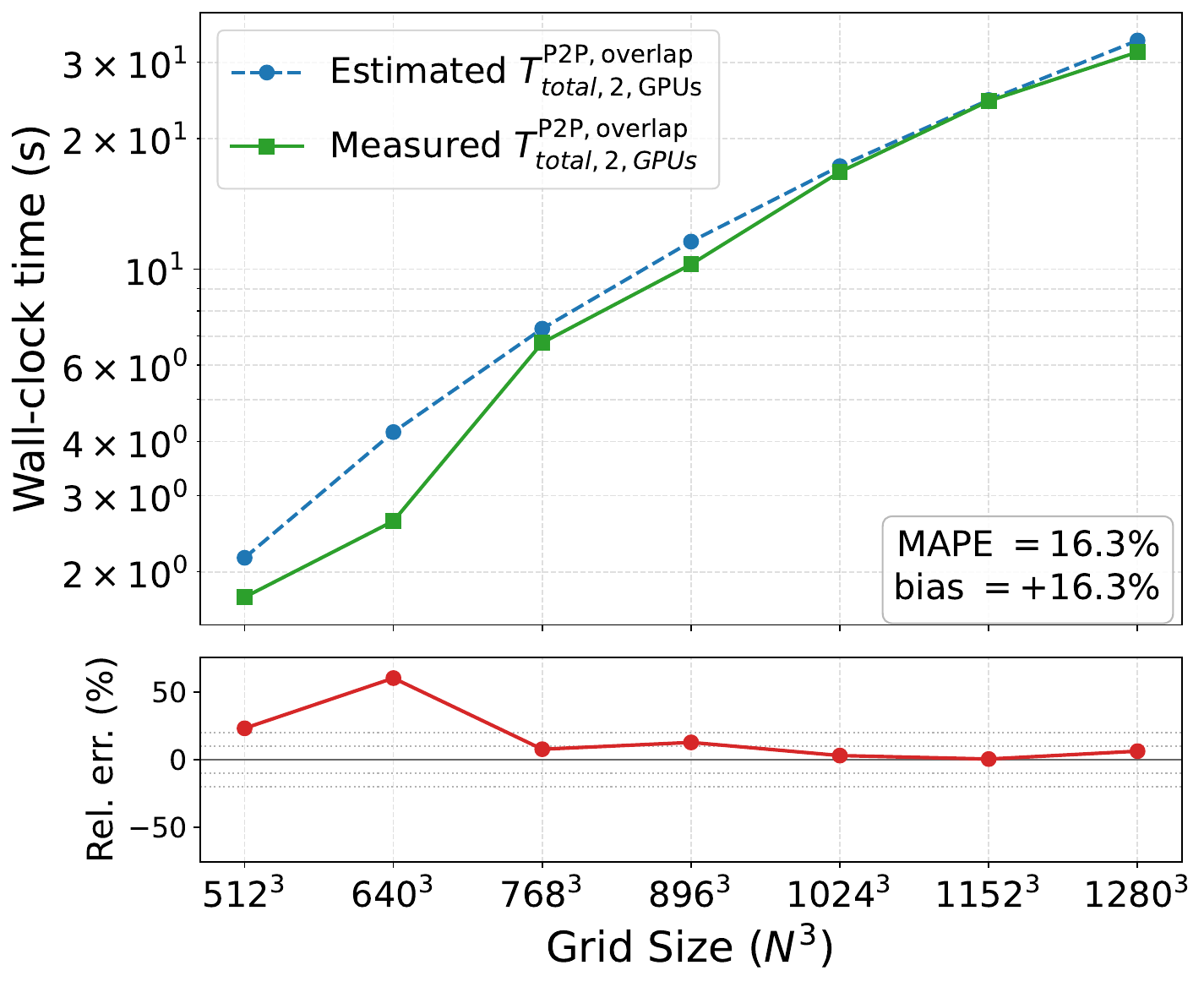}%
}\hfill
\subfloat[\centering NVIDIA H100, 4 GPUs.\label{fig:model_h100_4gpu}]{%
    \includegraphics[width=0.48\linewidth]{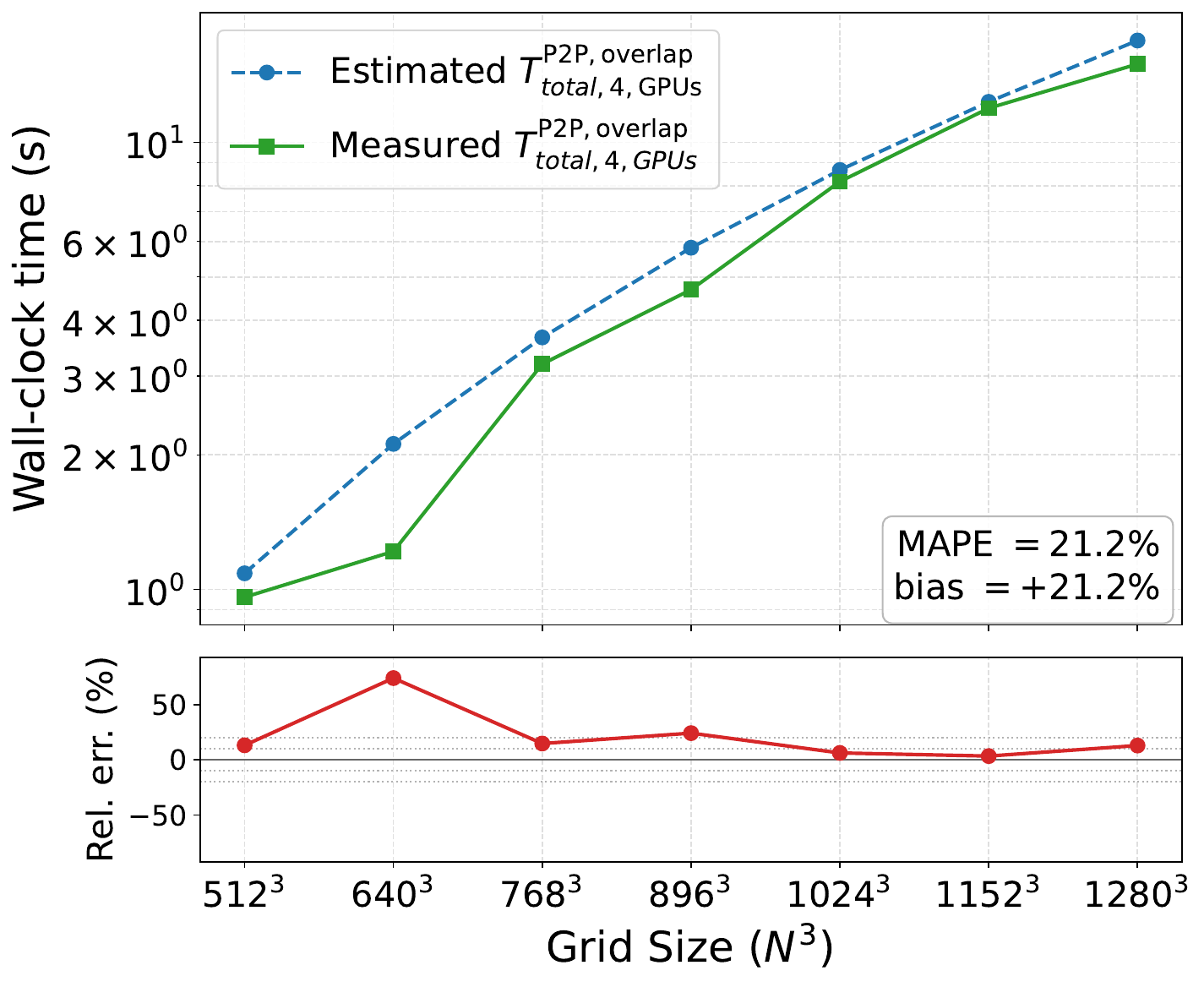}%
}\\[\medskipamount]
\subfloat[\centering AMD MI250X (1 GCD), 2 GPUs.\label{fig:model_amd_2gpu}]{%
    \includegraphics[width=0.48\linewidth]{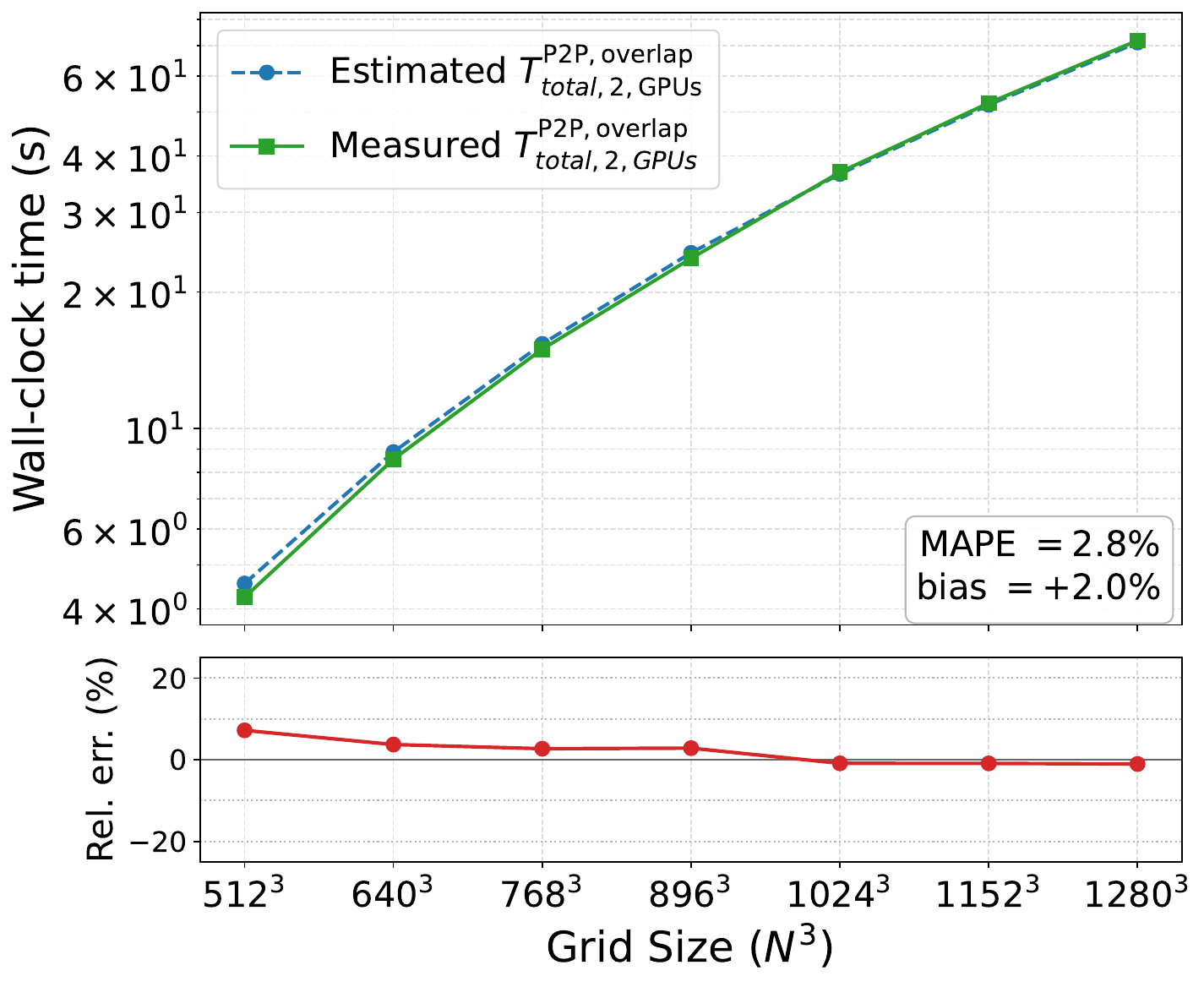}%
}\hfill
\subfloat[\centering AMD MI250X (1 GCD), 4 GPUs.\label{fig:model_amd_4gpu}]{%
    \includegraphics[width=0.48\linewidth]{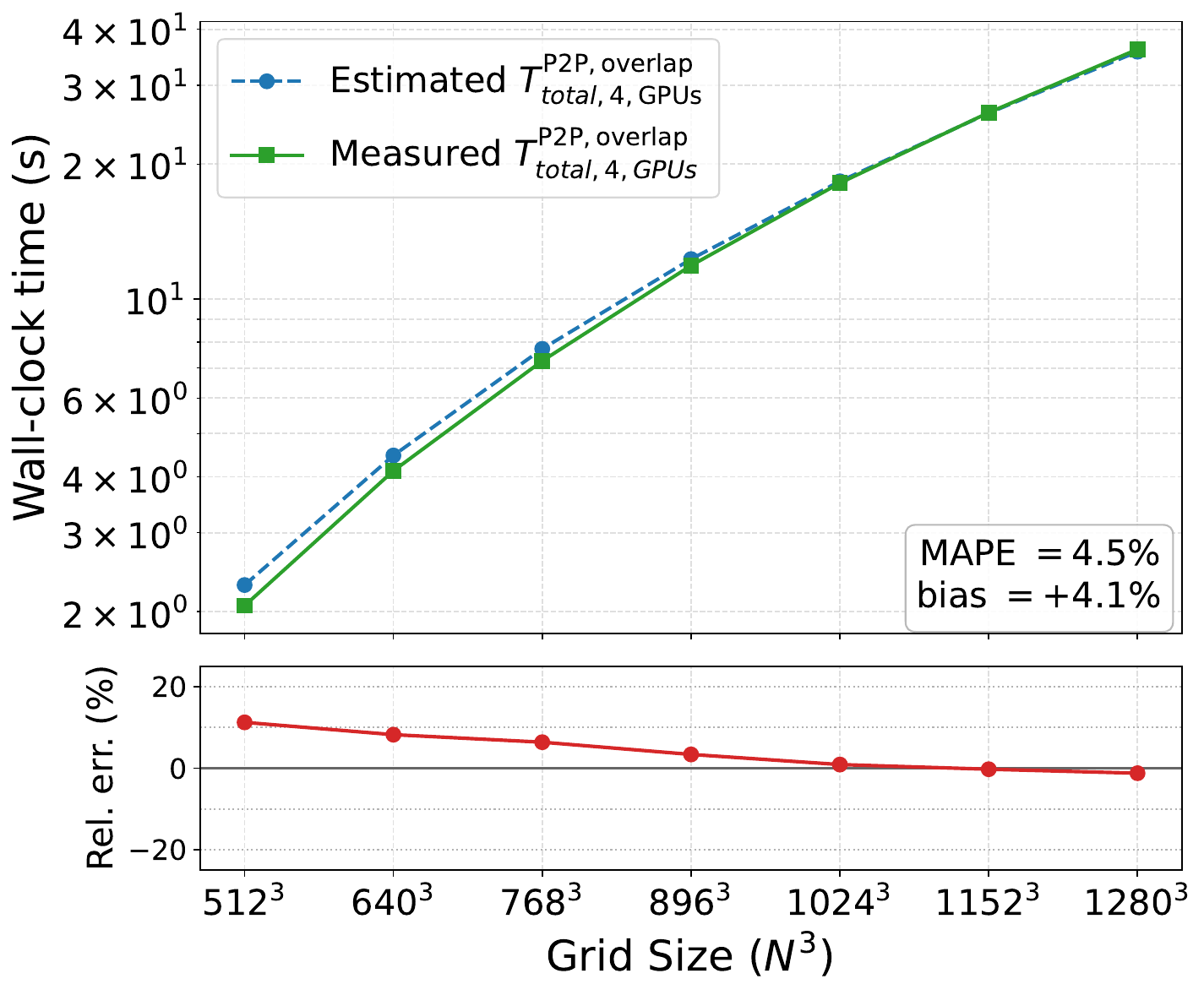}%
}\\[\medskipamount]
\subfloat[\centering Intel {\color{blue}Max} 1550 (composite), 2 GPUs.\label{fig:model_intel_2gpu}]{%
    \includegraphics[width=0.48\linewidth]{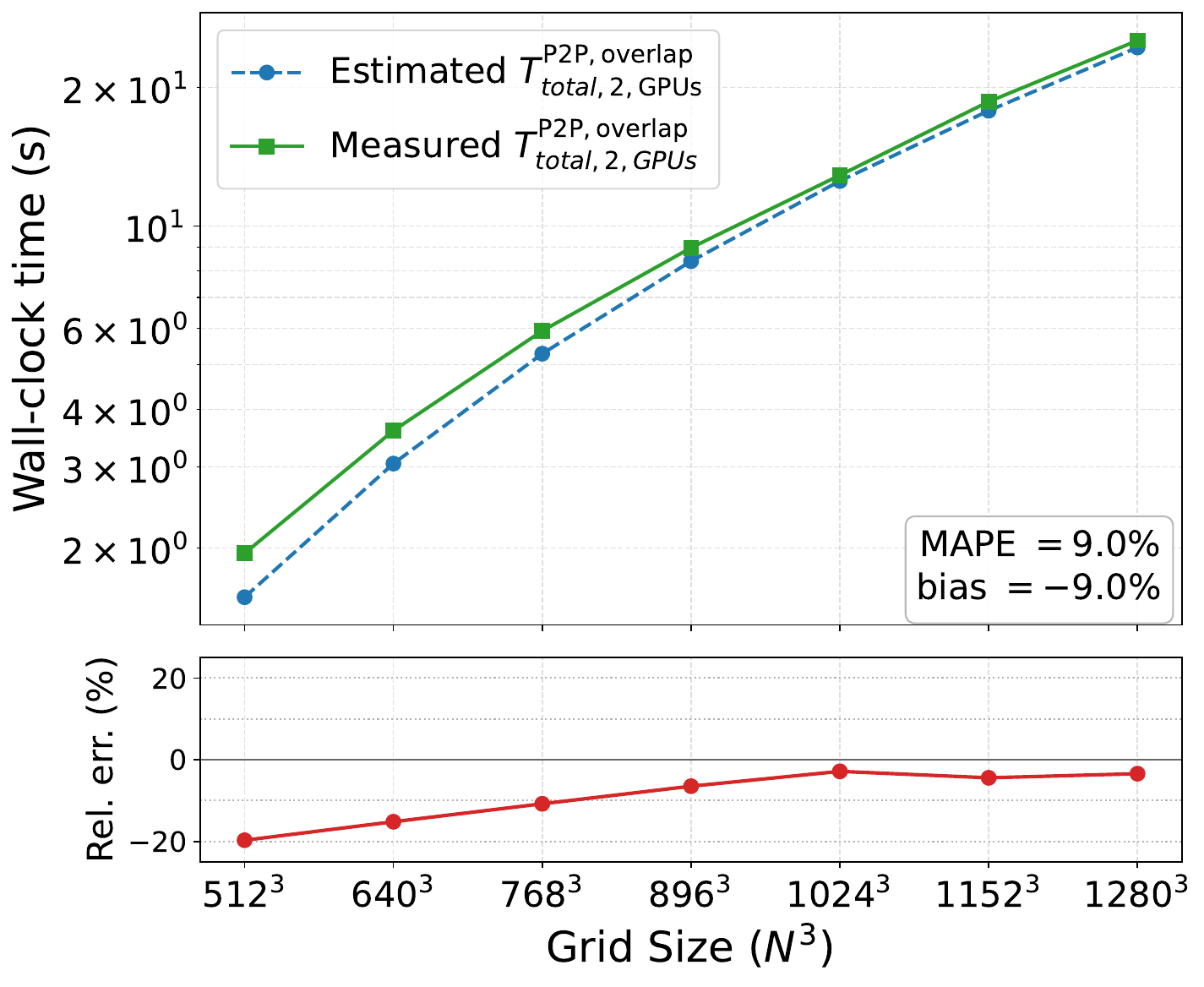}%
}\hfill
\subfloat[\centering Intel {\color{blue}Max} 1550 (composite), 4 GPUs.\label{fig:model_intel_4gpu}]{%
    \includegraphics[width=0.48\linewidth]{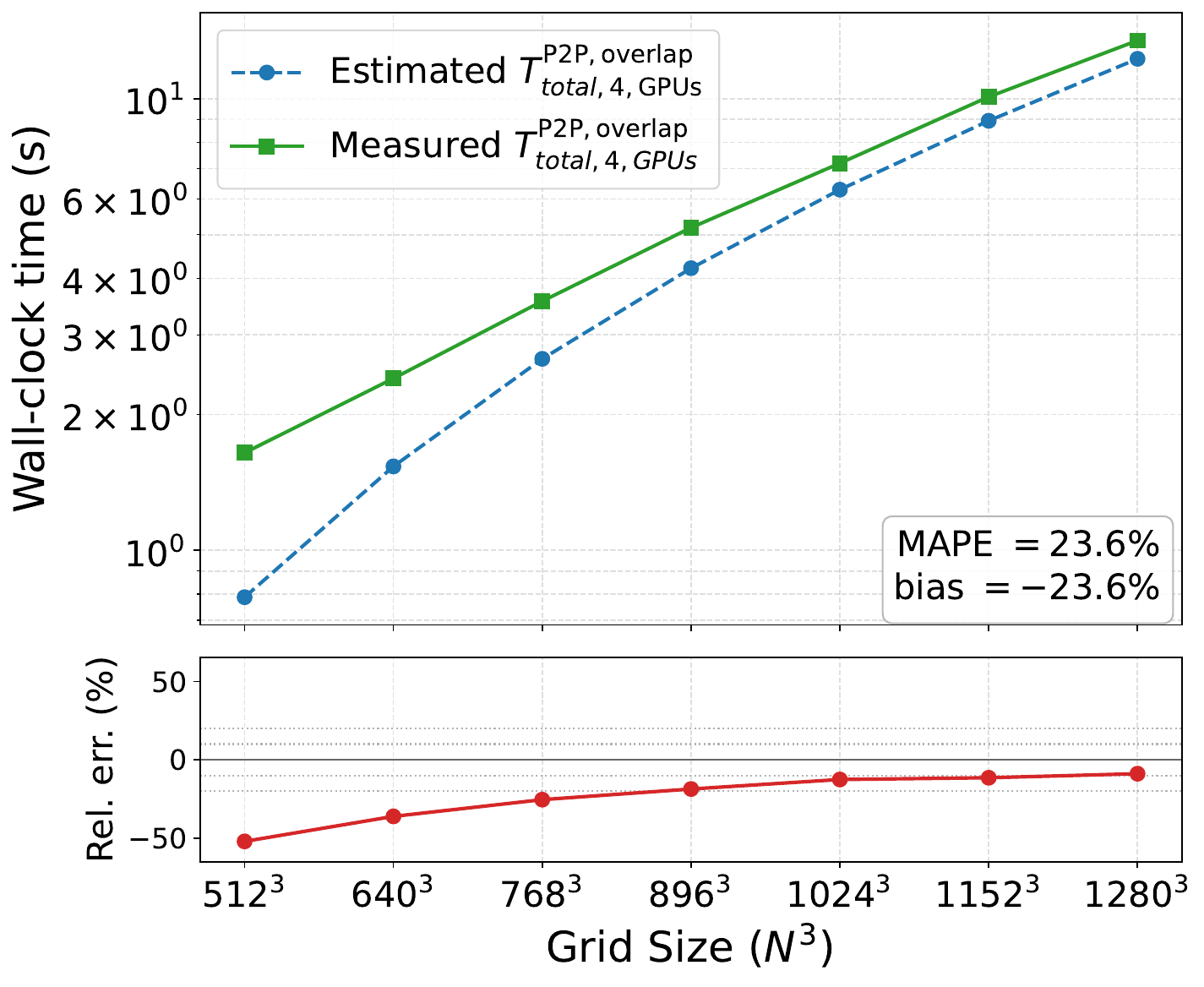}%
}
\caption{
Comparison of measured vs.\ estimated wall-clock time for \texttt{OpenMP Offloading \{2,4\}-GPU Version 4} across three GPU architectures: NVIDIA H100, AMD MI250X, and Intel {\color{blue}Max} 1550: model estimation (dashed)
against measurement (solid). Lower panels: relative error
$(T_{\mathrm{esti}}-T_{\mathrm{meas}})/T_{\mathrm{meas}}$ per grid size,
with $\pm10\%$ and $\pm20\%$ guides; note the wider vertical scale in
panel~(f). Mean absolute percentage error and signed bias are annotated
in each upper panel.}
\label{fig:model}
\end{figure}

To evaluate the performance efficiency of our parallelized OpenMP implementations, we compared the estimated times against the measured times. Specifically, we utilized the equation from Eq.~\ref{eq:p2p} in Subsection~\ref{sec:perf-model}. While we proposed several estimated time models, we are primarily focusing on the P2P concept (\texttt{OpenMP Offloading \{2,4\}-GPU Version 4}), which demonstrates the most effective parallelization among the other OpenMP Offloading models. To analyze \texttt{OpenMP Offloading \{2,4\}-GPU Version 4}, we calculated the average value, denoted as $\bar{k}$, using Eq.~\ref{eq:kbar}, along with P2P values gleaned from Figure~\ref{fig:p2p}. These data are summarized in Table~\ref{tab:used-measurements}. By substituting these values into Eq.~\ref{eq:p2p}, we generated the model plot illustrated in Figure~\ref{fig:model}. Upon examining Figure~\ref{fig:model}, it is evident that our estimated values closely align with the measured values across all scenarios for \texttt{OpenMP Offloading \{2,4\}-GPU Version 4}. However, we noted some fluctuations in the results for the NVIDIA H100 and Intel 1550 GPUs. To enhance clarity, we also plotted the mean absolute error (MAE), including bias, alongside the relative error. Notably, for smaller grid sizes on the NVIDIA H100, we observed significant deviations. In contrast, the performance measurements for the AMD MI250X closely corresponded with our estimated values, although some discrepancies also appeared with the Intel 1550.
\begin{table}[htbp]
\caption{Measured values of $\beta_{\mathrm{P2P}}$ and the average computational constant ratio $\bar{k}$ (see Eq.~\ref{eq:kbar}) for the evaluated GPU architectures, which are used to model performance (see Figure~\ref{fig:model}). \label{tab:used-measurements}} 
\begin{tabularx}{\textwidth}{lCCC}
\toprule
\textbf{Architecture} & \textbf{GPUs} & \boldmath$\bar{k}$ \textbf{(\boldmath$\times10^{-11}$)} & \textbf{\boldmath$\beta_{\mathrm{P2P}}$ (GB/s)} \\
\midrule
NVIDIA H100  & 2, 4 & 6.4042 & 253.11 \\
AMD MI250X   & 2, 4 & 13.584 & 61.28  \\
Intel 1550   & 2, 4 & 4.6676 & 62.85  \\
\bottomrule
\end{tabularx}
\end{table}
6.4042e-11

On the AMD MI250X, the deviation between estimated and measured values was minimal. This is primarily because we utilized only 1-GCD on the AMD MI250X, which offers lower computational power compared to the other GPUs in this research. Consequently, we experienced higher overall computational times, effectively masking the communication overhead in the \texttt{OpenMP Offloading \{2,4\}-GPU Versions}, resulting in a mean absolute percentage error (MAPE) of less than 3\%. Similarly, while the NVIDIA H100 showed slightly higher deviations than the AMD MI250X, these deviations were most pronounced with smaller problem sizes. As the problem size increased, the estimated and measured values demonstrated a closer agreement. Overall, for both GPU cases, the NVIDIA H100 and AMD MI250X, we must consider the overhead costs associated with kernel launches, synchronization, and communication contention.

On the Intel 1550 GPU, our estimated model tended to underestimate the measured wall-clock time. For example, in the 2-GPU case, the MAPE was around 9.0\%, indicating an almost optimal scaling model. However, in the 4-GPU wall-clock scenario involving the Intel 1550, we found a mean absolute error of 23.6\%, coupled with a slight negative bias of -23.6\%. This negative bias reflects inherent execution overheads, including kernel launch delays, synchronization delays, communication latencies, and bandwidth contention, which our idealized estimated overlap model did not account for. It is important to note that the Intel 1550 (composite stack of GPUs) possesses greater computational power than the other two GPUs examined in this research. This increased power results in reduced computation time, consequently creating additional communication overhead. Therefore, we observe larger deviations at smaller grid sizes, such as N=512. However, this gap tends to decrease as the computational domain expands, indicating that as computation time increases, the significance of communication overhead diminishes. To determine whether the performance model is predictive, we conducted tests using two additional grid sizes: \( 1408^3 \) and \( 1536^3 \). For these tests, we used the averaged computational constant \( \bar{k} \) over grid sizes from \( 1024^3 \) to \( 1280^3 \). Table~\ref{tab:holdout} shows that the model closely predicts the measured times; the estimated and measured values agree within \( 2.5\% \) for H100 and MI250X, and within \( 10\% \) for Max~1550. In summary, we conclude that our proposed model and methodology provide an optimized framework for multi-GPU computations on a single compute node. This model can be confidently applied to similar methodologies and to various problem-domain sizes to test parallel efficiency.

\begin{table}[H]
\caption{Validation of the performance model for {\textit{OpenMP Offloading {2,4}-GPU Version 4}} across grid sizes ($1408^3$ to $1536^3$). The computational constant $\bar{k}$, averaged over grid sizes from $1024^3$ to $1280^3$, is used.}
\label{tab:holdout}
\begin{tabularx}{\textwidth}{llCcCc}
\toprule
\multirow{2.5}{*}{GPU} & \multirow{2.5}{*}{$N$} & \multicolumn{2}{c}{2 GPUs} & \multicolumn{2}{c}{4 GPUs} \\
\cmidrule(lr){3-4}\cmidrule(lr){5-6}
& & Meas.\,/\,Est.\ (s) & Dev.\ (\%) & Meas.\,/\,Est.\ (s) & Dev.\ (\%) \\
\midrule
H100     & $1408^3$ & 45.09\,/\,44.72   & $-0.8$ & 21.88\,/\,22.42 & $+2.5$ \\
H100     & $1536^3$ & 59.27\,/\,57.94   & $-2.2$ & 29.05\,/\,29.05 & $\phantom{+}0.0$ \\
MI250X   & $1408^3$ & 96.22\,/\,94.85   & $-1.4$ & 48.27\,/\,47.56 & $-1.5$ \\
MI250X   & $1536^3$ & 124.01\,/\,122.91 & $-0.9$ & 61.38\,/\,61.74 & $+0.6$ \\
Max 1550 & $1408^3$ & 34.24\,/\,32.59   & $-4.8$ & 18.14\,/\,16.34 & $-9.9$ \\
Max 1550 & $1536^3$ & 44.55\,/\,42.23   & $-5.2$ & 23.10\,/\,21.17 & $-8.3$ \\
\bottomrule
\end{tabularx}
\end{table}

\subsection{Kernel-Level Comparison of OpenMP Offloading and Native Programming Models}
\label{sec:kernel-test}
From our previous analysis, we have established that OpenMP Offloading exhibits suboptimal performance relative to architecture-specific programming models. While we could hypothesize that OpenMP Offloading behaves similarly to Single Instruction Multiple Data (SIMD) approaches, further investigation is warranted. To this end, we examined three distinct problem sets: vector addition, dense matrix-vector multiplication, and dense matrix-matrix multiplication. Each problem showcases a distinct computation methodology concerning loop index access and floating-point operations. The kernel code syntax is detailed in Appendix Pseudocodes~\ref{alg:vecadd}, ~\ref{alg:matvec}, and ~\ref{alg:matmul}. 

Our experimental tests were conducted on these problem sets with memory footprints of 30, for vector addition with \(N\) set at \(1,250,000,000\); for matrix-vector multiplication with \(N\) values of \(61,184\); and for matrix-matrix multiplication with \(N\) values of \(35,328\). The tests were performed on the H100, MI250X, and Max 1550 GPUs, employing profiling tools such as Nsight Compute (NVIDIA), rocprofv3 (AMD) and VTune (Intel) to evaluate performance metrics via the command line. The performance metrics for these problem sets are summarized in Tables~\ref{tab:all-kernel}. From Table~\ref{tab:all-kernel}, it is evident that the majority of the decline in computational efficiency is primarily observed within the kernel computational blocks. This degradation is influenced by several factors, including the runtime mapping of loop iterations onto threads, data reuse in matrix multiplication, memory coalescing in matrix-vector multiplication, and the degree of parallelism inherent in vector addition.

Although OpenMP Offloading facilitates GPU execution through the use of blocks and threads, it adheres to a similar methodology as CPU (multicore) OpenMP but utilizes an increased number of threads. Despite this increased threading in OpenMP Offloading, each GPU thread typically manages more elements compared to CUDA, HIP, or SYCL implementations. This observation is further elucidated by the thread-per-row and elements-per-thread metrics shown in Table~\ref{tab:all-kernel}. Additionally, minor contributors, for example, less than 1\% of the performance degradation in OpenMP Offloading, come from kernel launch overhead and other factors. To draw more comprehensive conclusions, we also evaluated a memory footprint of 45 GB, which exhibited similar performance trends consistent with those observed at 30 GB. The results of 45GB are presented in Appendix Table~\ref{tab:additional45}.

\begin{table}[htbp]
\centering
\caption{Measurement metrics utilizing a 30 GB memory footprint for various operations: This analysis includes native implementations (CUDA, HIP, SYCL) of matrix-matrix multiplication using $32 \times 32$ shared-memory tiling, matrix-vector multiplication performed with one warp, wavefront, or sub-group per row, and vector addition executed with one thread or work item per element. We compare these results against OpenMP Offloading with H100, MI250X (1 GCD), and Max 1550 (1 tile). A summary of the code syntax is provided in Table~\ref{tab:mapping}. All kernels use double-precision (FP64) data, and the reported GB/s and GFLOP/s are based on this precision. The reported times are the median over five timed kernel launches, following one untimed warm-up launch.}
\label{tab:all-kernel}

\setlength{\tabcolsep}{3pt}
\small

\begin{tabularx}{\textwidth}{>{\raggedright\arraybackslash}p{4.2cm}CCCCCC}
\toprule
 & \multicolumn{2}{c}{\textbf{NVIDIA H100}}
 & \multicolumn{2}{c}{\textbf{AMD MI250X}}
 & \multicolumn{2}{c}{\textbf{Intel Max 1550}} \\
\cmidrule(lr){2-3}
\cmidrule(lr){4-5}
\cmidrule(lr){6-7}

\textbf{Metrics}
 & \textbf{CUDA} & \textbf{OpenMP}
 & \textbf{HIP} & \textbf{OpenMP}
 & \textbf{SYCL} & \textbf{OpenMP} \\
\midrule

\multicolumn{7}{c}{
\textbf{Vector addition $C=A+B$ ($A,B,C\in\mathbb{R}^{N}$, $N=1{,}250{,}000{,}000$)}
} \\
\midrule

Time (ms)
& 19.68 & 21.35 & 22.88 & 31.30 & 28.77 & 28.90 \\

GB/s (\% of DRAM peak)
& 1524 (93) & 1405 (86) & 1311 (80) & 958 (59) & 1043 (64) & 1038 (63) \\

OpenMP slower by
& 1.08$\times$ & -- & 1.37$\times$ & -- & 1.00$\times$ & -- \\

Groups $\times$ work items
& 4,882,813\newline $\times$ 256
& 3200\newline $\times$ 128
& 4,882,813\newline $\times$ 256
& 440\newline $\times$ 256
& 4,882,813\newline $\times$ 256
& 19,531,256\newline $\times$ 64 \\

Elements per thread
& 1 & 3052 & 1 & 11,097 & 1 & 1 \\

Occupancy, theor. (\%)
& 100 & 75 & 100 & 100 & n/a & n/a \\

Occupancy, achieved (\%)
& 82.0 & 39.5 & 87.9 & 41.2 & 91.9 & 96.2 \\

Memory throughput (\%)
& 93.0 & 85.5 & 80.1 & 58.5 & 63.6 & 63.1 \\

Launch overhead ($\mu$s)
& 7.5 & 28.3 & 12.6 & 15.9 & 6.5 & 6.4 \\

\midrule

\multicolumn{7}{c}{
\textbf{Dense matrix-vector multiplication $y=Ax$ 
($A\in\mathbb{R}^{N\times N}$, $x,y\in\mathbb{R}^{N}$, $N=61{,}184$)}
} \\
\midrule

Time (ms)
& 19.20 & 34.71 & 22.91 & 23.47 & 31.10 & 31.46 \\

GB/s (\% of DRAM peak)
& 1560 (96) & 863 (53) & 1307 (80) & 1276 (78) & 963 (59) & 952 (58) \\

OpenMP slower by
& 1.81$\times$ & -- & 1.02$\times$ & -- & 1.01$\times$ & -- \\

Groups $\times$ work items
& 7648\newline $\times$ 256
& 3200\newline $\times$ 128
& 15,296\newline $\times$ 256
& 660\newline $\times$ 256
& 7648\newline $\times$ 256
& 512\newline $\times$ 1024 \\

Threads per row
& 32 & 128 & 64 & 256 & 32 & 1024 \\

Occupancy, theor. (\%)
& 100 & 50 & 100 & 100 & n/a & n/a \\

Occupancy, achieved (\%)
& 96.1 & 22.2 & 96.8 & 72.7 & 97.2 & 98.5 \\

Memory throughput (\%)
& 96.9 & 53.4 & 79.9 & 77.9 & 58.5 & 58.1 \\

Launch overhead ($\mu$s)
& 7.5 & 28.2 & 11.9 & 14.9 & 6.0 & 6.4 \\

\midrule

\multicolumn{7}{c}{
\textbf{Dense matrix-matrix multiplication $C=AB$
($A,B,C\in\mathbb{R}^{N\times N}$, $N=35{,}328$)}
} \\
\midrule

Time (s)
& 18.09 & 209.39 & 38.51 & 279.28 & 48.07 & 394.14 \\

GFLOP/s
& 4875 & 421 & 2290 & 316 & 1835 & 224 \\

OpenMP slower by
& 11.58$\times$ & -- & 7.25$\times$ & -- & 8.20$\times$ & -- \\

Groups $\times$ work items
& 1,218,816\newline $\times$ 1024
& 3200\newline $\times$ 128
& 1,218,816\newline $\times$ 1024
& 440\newline $\times$ 256
& 1,218,816\newline $\times$ 1024
& 552\newline $\times$ 64 \\

Elements per thread
& 1 & 3047 & 1 & 11,080 & 1 & 35,328 \\

Occupancy, theor. (\%)
& 100 & 25 & 100 & 100 & n/a & n/a \\

Occupancy, achieved (\%)
& 99.99 & 24.49 & 99.3 & 48.5 & 100.0 & 99.9 \\

Memory throughput (\%)
& 38.8 & 97.4 & n/a & n/a & 14.1 & 60.3 \\

Launch overhead ($\mu$s)
& 7.5 & 28.0 & 11.5 & 15.2 & 6.3 & 6.5 \\

\bottomrule
\end{tabularx}

\end{table}
\begin{table}[htbp]
\caption{Comparison of OpenMP Offloading and native programming models (blocks and threads), tested on H100, MI250X, and Max 1550. Performance metrics are presented in Table~\ref{tab:all-kernel}, with complete Pseudocodes of kernels shown in Appendix: Pseudocode~\ref{alg:vecadd} (Vector Addition), ~\ref{alg:matvec} (Dense Matrix-Vector Multiplication), and ~\ref{alg:matmul} (Dense Matrix Multiplication).}
\label{tab:mapping}
\setlength{\tabcolsep}{4pt}
\begin{tabularx}{\textwidth}{>{\raggedright\arraybackslash}m{2.3cm}CCC}
\toprule
\textbf{Programming Models} & \textbf{Vector addition} & \textbf{Matrix-vector product} & \textbf{Matrix multiplication} \\
\midrule
OpenMP Offloading
  & \texttt{teams}\newline\texttt{distribute}\newline\texttt{parallel for}
  & \texttt{teams distribute} over rows,\newline \texttt{parallel for reduction(+:s)}\newline over columns
  & \texttt{teams}\newline\texttt{distribute}\newline\texttt{parallel for}\newline\texttt{collapse(2)} \\
\addlinespace[3pt]
Native\newline (CUDA / HIP / SYCL)
  & one thread per element
  & one warp (32), wavefront (64) or sub-group (32) per row, with a shuffle reduction
  & $32 \times 32$ tiles staged in shared memory, one element of $C$ per thread \\
\bottomrule
\end{tabularx}
\end{table}

\subsection{OpenMP Offloading Scalability: Conjugate Gradient Case Study}
Although the 3D stencil application demonstrates good scalability across diverse GPU architectures, asserting its effectiveness across all problem classes remains challenging. This difficulty arises primarily because the 3D stencil approach operates within a structured grid, where memory access is contiguous. Therefore, it is essential to validate and assess the general behavior of OpenMP offloading on irregular grids that exhibit non-uniform memory access. To illustrate this, we consider the conjugate gradient method~\cite{saad2003}—an iterative solver widely employed in various applications for solving sparse matrices, typically resulting from the finite element, finite difference, and finite volume methods used to tackle both time-dependent and time-independent partial differential equations (PDEs)~\cite{heath2018}. Table~\ref{tab:stencil-vs-cg} further shows differences between these two approaches. Examining the performance of this method is crucial because, in the context of PDEs, we often encounter sparse matrices characterized by both zero and non-zero entries. Storing the entire matrix can be redundant and inefficient in terms of memory usage. A common and effective option for storing such matrices is the Compressed Sparse Row (CSR)~\cite{tinney1967} format, which can lead to non-uniform memory access patterns when solving sparse matrices. Moreover, the conjugate gradient method involves several key operations, including Sparse Matrix-Vector multiplication (SpMV)~\cite{williams2007}, dot products, and various vector operations. In the context of OpenMP Offloading, testing this problem and determining whether scalability can be demonstrated is important, as it helps draw general conclusions about the scalability of OpenMP Offloading.

The following were considered for the scalability study of OpenMP Offloading on CG computation:
In a multi-GPU setup, the CSR matrix is partitioned row-wise, ensuring that each GPU manages approximately the same number of nonzero entries. This approach promotes an even distribution of the workload across the GPUs, with the associated vectors adhering to the same row distribution. To facilitate this process, a halo exchange is necessary before each Sparse Matrix-Vector multiplication (SpMV). This exchange consists of several steps: first, the boundary values required by neighboring GPUs are packed; next, these values are transferred via P2P data transfer between GPUs; afterward, the received halo values are unpacked; and finally, the SpMV computation for the boundary rows is conducted. Notably, the SpMV operations for the interior rows are overlapped with the P2P halo transfer, optimizing overall efficiency. A schematic overview of the multi-GPU OpenMP Offloading implementation of this workflow is illustrated in Appendix Figure~\ref{fig:2gpu-packedhalo}.

\begin{itemize}
    \item For benchmarking, we use double precision and the natural order of the CSR matrix. The convergence tolerance is set to \(10^{-8}\).
    \item Pseudocode~\ref{alg:spmv} shows three different kernel computations of SpMV on multicore CPUs, a single GPU, and multiple GPUs. 
    \item We tested matrix-vector sizes ranging from 5 GB to 50 GB to stay within single-GPU memory.
\end{itemize}

\begin{figure}[htbp]
\centering
\subfloat[NVIDIA H100]{\includegraphics[width=0.32\linewidth]{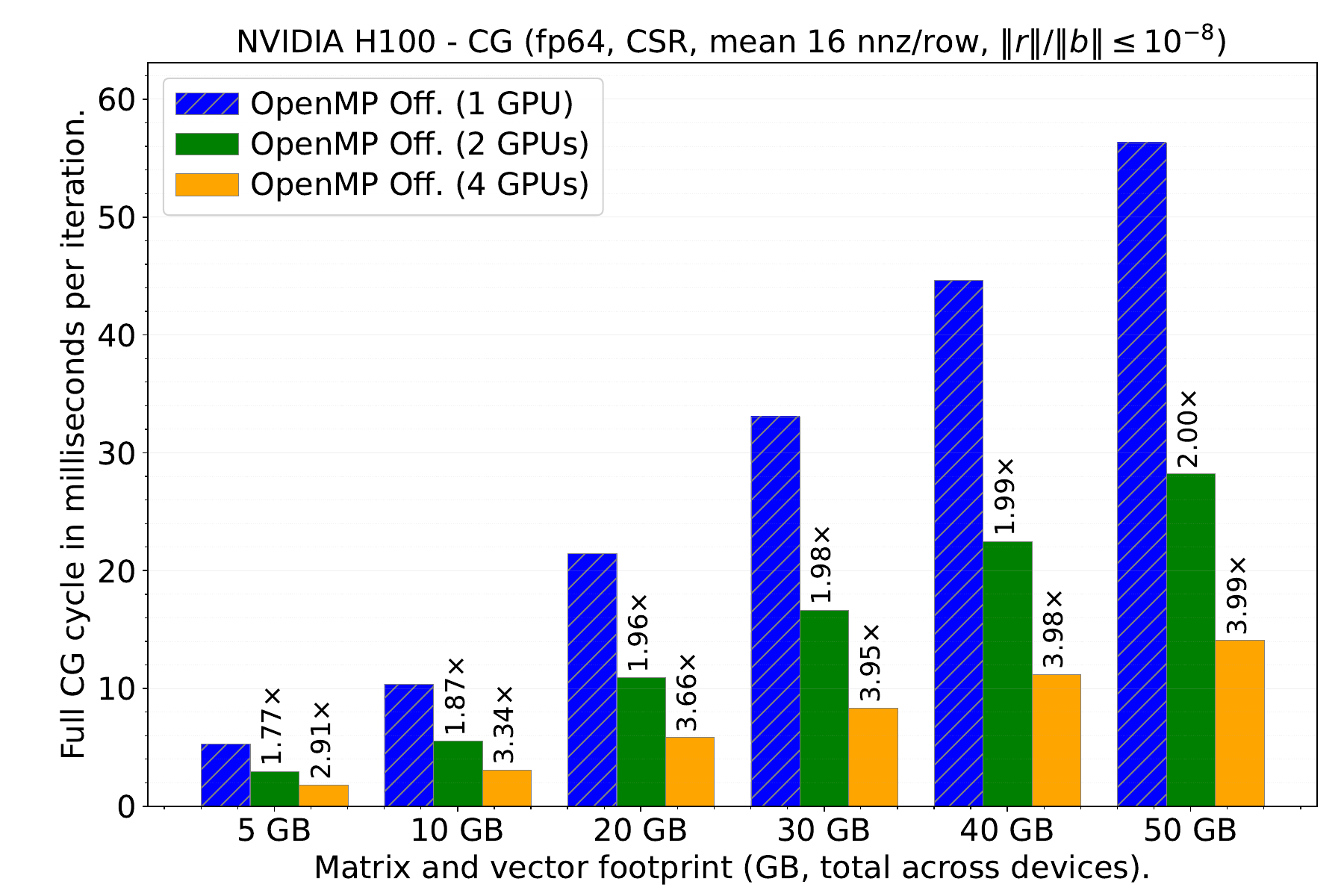}}%
\hspace{1mm}%
\subfloat[AMD MI250X]{\includegraphics[width=0.32\linewidth]{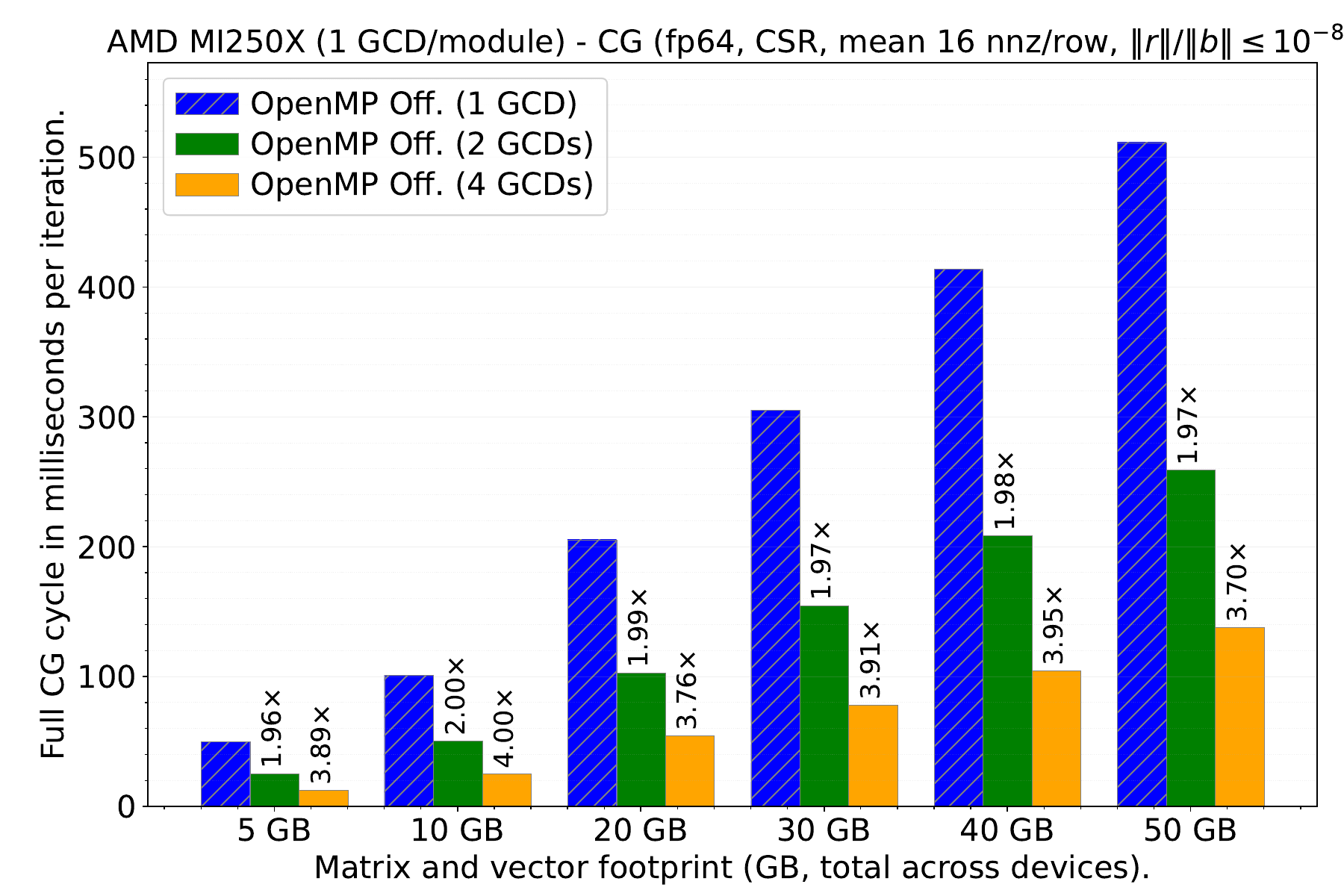}}%
\hspace{1mm}%
\subfloat[Intel Max 1550]{\includegraphics[width=0.32\linewidth]{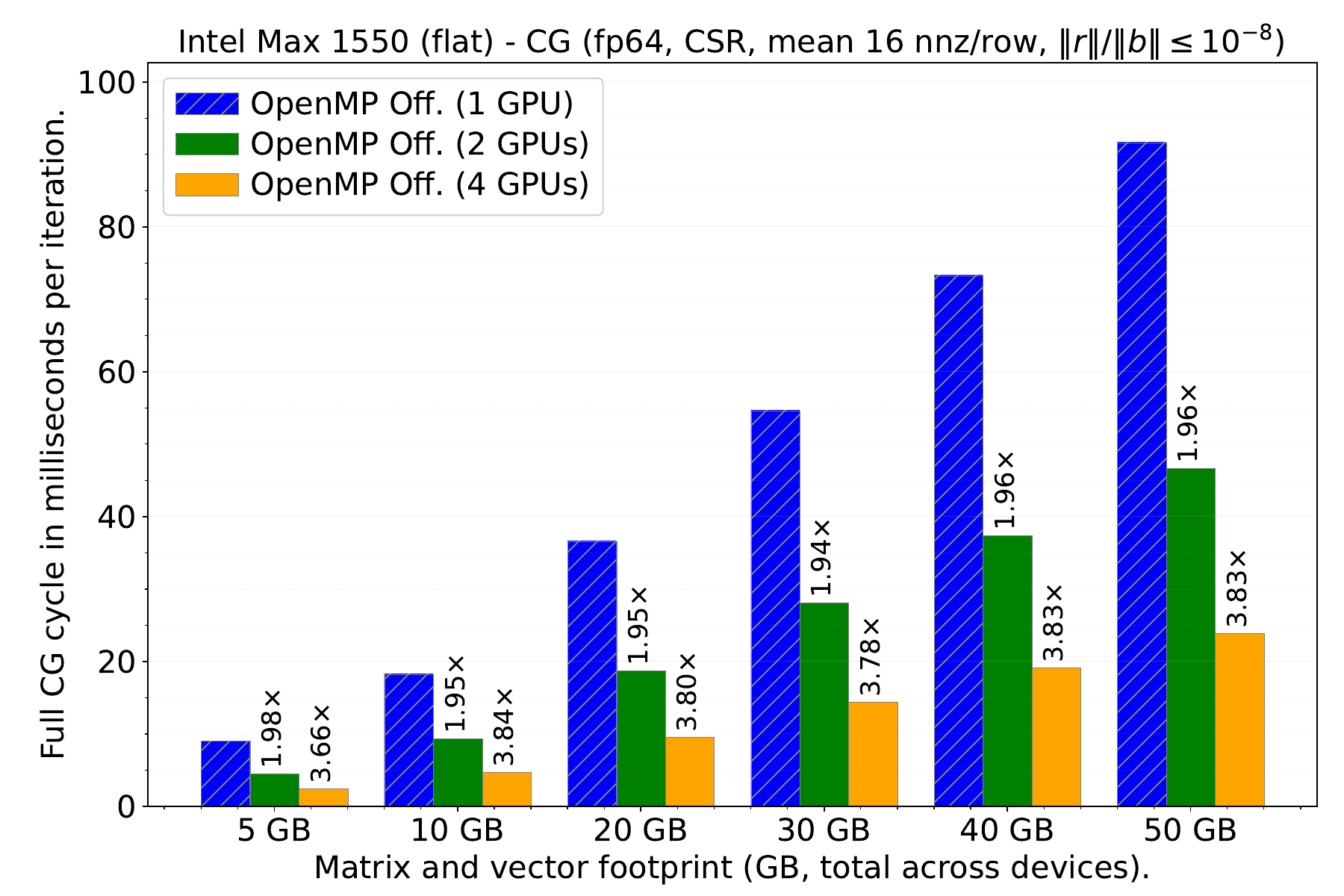}}%
\caption{Performance of the CG cycle per iteration as problem size goes from 5GB to 50 GB, utilizing 1, 2, and 4 GPU devices. The experiments are conducted using double precision (fp64) and CSR format, with an average of 16 non-zeros per row and natural ordering, achieving a relative residual below \(10^{-8}\). Results represent the median of three repetitions.}
\label{fig:cg_strong}
\end{figure}

Figure~\ref{fig:cg_strong} presents a comparison of single GPU, 2 GPUs, and 4 GPUs using OpenMP Offloading in computing CG on NVIDIA, AMD, and Intel GPUs. From this analysis, we observe that for 2 and 4 GPUs, there is nearly a 2x and 4x speedup, respectively, similar to the results seen with the 3D heat stencil. This speedup arises from distributing the computation across additional resources; specifically, using 2 and 4 GPUs with optimized communication significantly reduces overall computational time. As previously discussed, OpenMP Offloading provides a P2P memcpy API that allows direct memory access between GPUs. Although performance may vary depending on different GPU node configurations, the machines we tested demonstrated adequate bandwidth that aligns well with the capabilities of the OpenMP API.

It is important to highlight that when communication is effectively hidden or overlapped during computation, we can achieve nearly perfect speedup. However, one might wonder if such performance gains can be sustained consistently. The answer is no; this holds true only as long as the computational load exceeds the communication cost. For instance, handling very small sparse matrix-vector operations or small problem sizes would not yield performance scaling when using 2 or 4 GPUs. In such cases, it is advisable to utilize a single GPU.

\begin{algorithm}[htbp]
\DontPrintSemicolon
\LinesNotNumbered
\caption{Compute block of SpMV ($Ap$ in CSR) in CG for
multicore CPU, one GPU, and across $N$ GPUs.}
\label{alg:spmv}
\SetKwFunction{SpMVrows}{SpMV\_ROWS}
\SetKwProg{Fn}{Procedure}{:}{}

\KwIn{CSR arrays \texttt{rowptr}, \texttt{col}, \texttt{val}; vector $p$}
\KwOut{$Ap$}
\BlankLine
\BlankLine
\tcp{(a) Compute block - SpMV: multicore CPU, $T$ threads}
\pragma{parallel for schedule(static)}\;
\For{$r \leftarrow 0$ \KwTo $n-1$}{
    $s \leftarrow 0$\;
    \For{$k \leftarrow \texttt{rowptr}[r]$ \KwTo $\texttt{rowptr}[r+1]-1$}{
        $s \leftarrow s + \texttt{val}[k]\cdot p[\texttt{col}[k]]$\;
    }
    $Ap[r] \leftarrow s$\;
}
\BlankLine
\hrulefill\;
\BlankLine

\tcp{(b) Compute block - SpMV: one GPU OpenMP Offloading}
\pragma{target teams distribute parallel for schedule(static,1)}\;
\Indp\texttt{device(0) is\_device\_ptr(rowptr,col,val,p,Ap)}\;\Indm
\For{$r \leftarrow 0$ \KwTo $n-1$}{
    $s \leftarrow 0$\;
    \For{$k \leftarrow \texttt{rowptr}[r]$ \KwTo $\texttt{rowptr}[r+1]-1$}{
        $s \leftarrow s + \texttt{val}[k]\cdot p[\texttt{col}[k]]$\;
    }
    $Ap[r] \leftarrow s$\;
}
\BlankLine
\hrulefill\;
\BlankLine

\tcp{(c) Compute block - SpMV: $N$ GPUs on device $d$, over a row list $R$}
\pragma{target teams distribute parallel for schedule(static,1)}\;
\Indp\texttt{device($d$) is\_device\_ptr(rowptr$_d$,col$_d$,val$_d$,$p_d$,Ap$_d$,R)}\;\Indm    
\For{$i \leftarrow 0$ \KwTo $|R|-1$}{
        $r \leftarrow R[i]$\;
        $s \leftarrow 0$\;
        \For{$k \leftarrow \texttt{rowptr}_d[r]$ \KwTo $\texttt{rowptr}_d[r+1]-1$}{
            $s \leftarrow s + \texttt{val}_d[k]\cdot p_d[\texttt{col}_d[k]]$\;
        }
        $Ap_d[r] \leftarrow s$\;
    }
\BlankLine
\end{algorithm}

\begin{table}[htbp]
\caption{Computational characteristics of 3D stencils (structured grids) versus sparse matrices (unstructured grids).}
\label{tab:stencil-vs-cg}
\begin{tabularx}{\textwidth}{lXX}
\toprule
& \textbf{3D Stencil (structured)} & \textbf{CSR SpMV in CG (unstructured)} \\
\midrule
Index formation
& loop counters held in registers; address computed as \((i \cdot n_y + j) \cdot n_z + k\)
& \texttt{col[k]} loaded from memory, then used to compute the address in \(p\) \\
\addlinespace

Access pattern
& two fixed-stride streams for the \(j\) and \(i\)
neighbours along with stride-1 of \(k\); coalesced and prefetchable
& data-dependent gather from \(p\); adjacent threads access unrelated
addresses \\
\addlinespace

Halo
& a face of the subdomain, which is determined from the domain decomposition
& remote columns identified at setup define the set requiring explicit
gather and scatter index lists \\

\bottomrule
\end{tabularx}
\end{table}

\subsection{Summary of Key Findings}  
We summarize the key findings of this research as follows:

\begin{itemize}
\item By default, OpenMP Offloading for kernel computations is generally less efficient than native GPU programming models, which are specifically designed for their respective GPU architectures. While OpenMP Offloading provides specialized clauses and constructs similar to those found in CUDA—such as thread blocks and grids—the performance does not match that of CUDA on NVIDIA, HIP on AMD, or SYCL on Intel. This performance discrepancy can be attributed to several factors:
\begin{itemize}
    \item CUDA, HIP, and SYCL effectively utilize the concept of SIMT, which optimizes GPU resource usage.
    \item In contrast, OpenMP Offloading behaves more like a SIMD~\cite{flynn1972some} approach; a small verification experiment (see Subsection~\ref{sec:kernel-test}) supports this, a method that may not achieve optimal performance on GPUs. Although there is an option to manually set the thread grid and block in OpenMP Offloading, the current compilers associated with the OpenMP Offloading API do not execute this task as efficiently as those in CUDA, HIP, or SYCL for GPUs.
\end{itemize}

\item This performance gap is clearly illustrated in Figure~\ref{fig:diffusion_comparison}, which shows that CUDA, HIP, and SYCL demonstrate significantly superior performance compared to OpenMP Offloading. Additionally, as depicted in Figure~\ref{fig:model}, CUDA, HIP, and SYCL also exhibit slightly higher measured bandwidth between the CPU and GPU, as well as among GPUs. This enhanced bandwidth contributes to their overall better performance relative to OpenMP Offloading.

\item The use of HHT significantly enhances the overall performance of multi-GPU OpenMP Offloading implementations. The LLVM backend enables asynchronous operations of OpenMP Offloading to be executed concurrently with the assistance of HHT threads. In this research, setting HHT to 8 resulted in improved performance; however, the HHT thread size can be adjusted according to the specific requirements of the problem at hand.

\item Our performance model scales well with the measured time; nevertheless, it does not account for penalties associated with kernel launches, synchronization, and communication latencies, resulting in some discrepancies. Despite this, our model provides reasonable scaling predictions across all three GPUs and can be applied to other similar problems.

\item For larger or more complex codebases, utilizing CUDA may prove advantageous, although it typically comes at the cost of increased development time. Furthermore, even if a conversion to another programming model is attempted, it may not yield significant performance benefits. OpenMP Offloading, on the other hand, offers a more user-friendly approach, making it easier for individuals with backgrounds in physics, mathematics, or engineering to adopt. As such, the same codebase can be executed without further modifications.

\item To support the general scalability of OpenMP Offloading, the tested CG computation demonstrates speedups of approximately 2x and 4x when using 2 and 4 GPUs, respectively. This indicates that OpenMP Offloading can be effectively applied to various problem sets, as it maintains good scalability, provided that communication and computations are overlapped.

\item However, we believe that further compiler optimizations for kernel computations in OpenMP Offloading could enable performance levels comparable to those of CUDA on NVIDIA, HIP on AMD, and SYCL on Intel. Achieving such optimizations could significantly contribute to effective exascale computing that rivals other low-level GPU programming models.
\end{itemize}

\section{Conclusions}
In this research, we investigated four distinct OpenMP Offloading programming models targeting 2 and 4 GPUs, specifically those developed for NVIDIA, AMD, and Intel architectures. We compared these four OpenMP Offloading variants, operating on 2 and 4 GPUs, against native GPU programming models, including CUDA, HIP, and SYCL. Among the multi-GPU OpenMP Offloading configurations, the \texttt{OpenMP Offloading \{2,4\}-GPU Version 4} exhibited the best performance compared to the other three variants. This performance enhancement can largely be attributed to several factors: the concurrent execution of GPU kernels using multiple host threads, the utilization of P2P data copies for halo data exchanges, and efficient workload distribution across the available GPUs. Notably, the \texttt{OpenMP Offloading \{2,4\}-GPU Version 4} achieved approximately 2x speedup with 2 GPUs and around 4x speedup with 4 GPUs compared to the single GPU implementation of the \texttt{OpenMP Offloading 1-GPU Baseline}. The Intel configuration with 4 GPUs, however, achieved slightly less than the 4x speedup, at about 3.5x. 

The proposed parallelization strategy achieved better performance under configurations with 2 and 4 GPUs, demonstrating effective communication overlap. This overlap improves the efficiency of our parallel implementation. Utilizing 8 HHT appears to be optimal; however, we anticipate that this number can be increased based on the number of asynchronous operations executed on a single GPU. Additionally, our bandwidth analysis informs the development of a performance model to evaluate the efficiency of our parallel implementations. Furthermore, to substantiate the generic scalability claim of OpenMP Offloading, the tested CG computation provides nearly 2x and 4x speedup for 2 and 4 GPUs, respectively.

While OpenMP Offloading did not achieve performance levels competitive with established native programming models such as NVIDIA's CUDA, AMD's HIP, and Intel's SYCL, the significant advantage lies in having a single source code that is compatible across various GPU vendors. Adapting extensive scientific codes to different architectures can be a complex task. Even after conversion, the resulting code often requires substantial modifications to achieve satisfactory performance. Therefore, this work does not advocate for OpenMP Offloading as a replacement for CUDA, HIP, or SYCL. Instead, it outlines potential performance expectations for OpenMP Offloading in scientific domains, exemplified through a 3D heat transfer problem.

Our parallelization methodology, enhanced by the use of multiple host threads, a performance model, and bandwidth measurements, provides a robust framework for achieving optimal performance in OpenMP Offloading for multi-GPU computations on a single compute node. This research offers clear insights and a deeper understanding of the application of OpenMP Offloading for multi-GPU programming on single compute nodes, paving the way for broader adoption of OpenMP Offloading in various scientific computing applications, independent of specific GPU architectures.

\vspace{6pt} 

\section*{Author Contributions}
Conceptualization, methodology, software, validation, formal analysis,
investigation, resources, data curation, writing---original draft preparation,
writing---review and editing, visualization, supervision, project administration,
and funding acquisition, Ezhilmathi Krishnasamy.
The author has read and agreed to the published version of the manuscript.

\section*{Funding}
This research was carried out within the \textbf{QCLS} project---%
\textbf{Quantum Computing for Scientific Computing (ID OP27.01207)}---%
co-funded by the European Union through the European Regional Development
Fund (ERDF) and by the Republic of Slovenia through the Ministry of Higher
Education, Science and Youth, under the JR RZK call implemented by the
Slovenian Research and Innovation Agency (ARIS).

\section*{Institutional Review Board Statement}
Not applicable.

\section*{Informed Consent Statement}
Not applicable.

\section*{Data Availability Statement}
The source code and supporting materials underlying the results presented in this research are publicly available on Zenodo and GitHub. The archival version corresponding to release \texttt{v1.0.0} is available on Zenodo at \url{https://doi.org/10.5281/zenodo.21878593}, and the actively maintained version is available on GitHub at \url{https://github.com/ezhilmathik/OpenMP-Offloading-Intranode-Artifact/}.

\section*{Acknowledgments}
The author gratefully acknowledges the Leibniz Supercomputing Centre (LRZ) for providing computing resources on the SuperMUC-NG supercomputer. The author also gratefully acknowledges the EuroHPC Joint Undertaking (EuroHPC JU) for providing access to the MareNostrum 5 supercomputer at the Barcelona Supercomputing Center (BSC) and to the LUMI supercomputer through a Development Access call.

\section*{Conflicts of Interest}
The author declares no conflicts of interest. The funder had no role in the
design of the research; in the collection, analysis, or interpretation of data;
in the writing of the manuscript; or in the decision to publish the results.

\section*{Abbreviations}
The following abbreviations are used in this manuscript:\\

\noindent
\begin{tabular}{@{}ll}
API     & Application Programming Interface\\
BSC     & Barcelona Supercomputing Center\\
CFD     & Computational Fluid Dynamics\\
CPU     & Central Processing Unit\\
GPU     & Graphics Processing Unit \\
CU      & Compute Unit\\
CUDA    & Compute Unified Device Architecture\\
FLOP/s  & Floating-point operations per second\\
FP64    & Double-precision, 64-bit floating-point format\\
GB/s    & Gigabytes per second\\
GFLOP/s & Billions of floating-point operations per second\\
GPU     & Graphics Processing Unit\\
HIP     & Heterogeneous-compute Interface for Portability\\
HPC     & High-Performance Computing\\
IF      & Infinity Fabric\\
LUMI    & Large Unified Modern Infrastructure\\
LUMI-G  & GPU partition of the LUMI supercomputer\\
OpenACC & Open Accelerators\\
OpenCL  & Open Computing Language\\
OpenMP  & Open Multi-Processing\\
P2P     & Peer-to-Peer\\
PCIe    & Peripheral Component Interconnect Express\\
SM      & Streaming Multiprocessor\\
SYCL    & SYCL heterogeneous programming model \\
HtoD    & Host-to-Device \\
DtoH    & Device-to-Host \\
Meas.BW & Measured Bandwidth \\
GCDs    & Graphics Compute Dies \\
SIMT    & Single Instruction Multiple Threads \\
SIMD    & Single Instruction Multiple Data \\
HHT     & Hidden Helper Threads \\
$\bar{k}$ & Average Computational Constant Ratio \\
$\beta_{\mathrm{P2P}}$ & P2P Measured Bandwidth
\end{tabular}


\appendix

\section{Artifact Description}
\label{app:artifact}

\subsection{Availability}
\label{app:availability}

All source code, run scripts, raw measurement data, and figure-generation scripts are publicly available at \url{https://github.com/ezhilmathik/OpenMP-Offloading-Intranode-Artifact} and permanently archived on Zenodo (DOI: \href{https://doi.org/10.5281/zenodo.21878593}{10.5281/zenodo.21878593}, release tag: \texttt{v1.0.0}, commit: \texttt{e41a873563d1}). Source code is licensed under the European Union Public Licence v1.2 (EUPL-1.2); measurement data and generated figures are licensed under the Creative Commons Attribution 4.0 International licence (CC BY 4.0).

Every directory ships both the sources and the results they produced. All
figures and tables in this paper can be regenerated from the recorded data
\emph{without} access to the GPUs; re-running the measurements requires the
hardware listed in Tables~\ref{tab:compiler-support} and ~\ref{tab:artifact-environment}.

\subsection{Repository structure}
\label{app:structure}

\begin{lstlisting}[
    basicstyle=\ttfamily\scriptsize,
    frame=none,
    breaklines=true,
    breakatwhitespace=true,
    columns=fullflexible,
    keepspaces=true,
    showstringspaces=false
]
OpenMP_Single_Source/    nine OpenMP sources + serial CPU reference,
                         compiled unmodified on all three vendors
3D_Heat_Benchmark/       main results; one directory per machine
  <MACHINE>/Source-Results/
    OpenMP/              build/run/verify/aggregate pipeline + raw results
    HIP/ CUDA/ SYCL/     native-model counterpart, same pipeline
    model/               analytical performance model
    plots/               final figures
HHT/                     hidden-helper-thread sensitivity study
  <MACHINE>-{01,02,03}/  one full pipeline copy per HHT configuration
  gain.py                cross-machine comparison and figures
Bandwidth_Measurements/  intra-node bandwidth and latency microbenchmarks
\end{lstlisting}

Each directory carries its own \texttt{README.md} with the machine-specific
modules, compiler flags, device masking and submission commands.

\subsection{Experimental environment}
\label{app:environment}

\begin{table}[htbp]
\caption{Hardware and software environment. One OpenMP device is one GCD on
MI250X, one GPU on H100, and one whole card (two stacks, \texttt{COMPOSITE}
hierarchy) on Max~1550.}
\label{tab:artifact-environment}
\centering
\small
\begin{tabularx}{\columnwidth}{@{}lXX@{}}
\toprule
& \textbf{Compiler / runtime} & \textbf{Device selection} \\
\midrule
AMD MI250X (gfx90a) & Cray \texttt{cc} (PrgEnv-amd), ROCm 6.4.4, LUMI/25.09
  & \texttt{ROCR\_VISIBLE\_DEVICES} = 0 / 0,2 / 0,2,4,6 \\
NVIDIA H100 (sm\_90) & clang 18.1.8, CUDA 12.8, GCC 13.2.0
  & \texttt{CUDA\_VISIBLE\_DEVICES} = 0 / 0,1 / 0,1,2,3 \\
Intel Max 1550 (PVC) & \texttt{icx} (oneAPI)
  & \texttt{ZE\_AFFINITY\_MASK} = 0 / 0,1 / 0,1,2,3 \\
\bottomrule
\end{tabularx}
\end{table}

\subsection{Problem parameters}
\label{app:parameters}

Explicit finite-difference solve of $\partial u/\partial t = \nabla \cdot
(\kappa(x,y,z)\,\nabla u)$ on an $N^3$ grid, decomposed in the slowest-varying
dimension, with one halo-face exchange per internal boundary per timestep. The
timestep is derived identically in \texttt{submit.sh} and \texttt{verify.sh}:

\begin{equation}
h = \frac{1}{N-1}, \qquad
\Delta t = \frac{h^2}{12\,\kappa_{\max}}, \qquad
T = N_s \cdot \Delta t ,
\end{equation}

with $\kappa_{\max} = 0.95$. The production sweep uses
$N \in \{512, 640, 768, 896, 1024, 1152, 1280\}$, $N_s = 500$ timesteps, and
five independent repetitions per grid size, each submitted as a separate Slurm
job. Reported times are medians over the repetitions, taken from the solver's
own timer rather than the harness wall clock.

\subsection{Reproducing the measurements}
\label{app:reproducing}

The pipeline is identical on every machine and in every model directory:

\begin{lstlisting}[basicstyle=\ttfamily\scriptsize, frame=none]
cd 3D_Heat_Benchmark/<MACHINE>/Source-Results/OpenMP
./submit.sh 5        # one build job + 5 repetitions x 7 grid sizes
squeue -u $USER      # wait for completion
./aggregate.sh       # merge results_*/ into summary.csv (medians)
./plot.sh -a         # terminal summary of the aggregated data
\end{lstlisting}

The sweep can be narrowed with \texttt{NS}, \texttt{FAMILIES},
\texttt{BASELINE} and \texttt{STEPS}; \texttt{SKIP\_BUILD=1} reuses existing
binaries. The first argument of \texttt{submit.sh} is the repetition count, not
a grid size.

Correctness is checked separately, because timing runs never write solution
files:

\begin{lstlisting}[basicstyle=\ttfamily\scriptsize, frame=none]
sbatch --export=ALL,N=128,STEPS=50 verify.sh
\end{lstlisting}

This builds the serial CPU reference and compares every variant against it
element-wise. Timings produced by \texttt{verify.sh} are not meaningful, as
file I/O falls inside the measured region.

\subsection{Mapping results to the artifact}
\label{app:mapping}

\begin{table}[htbp]
\caption{Source files associated with each reported result.}
\label{tab:artifact-mapping}
\centering
\footnotesize

\begin{tabularx}{\columnwidth}{
  @{}
  >{\raggedright\arraybackslash}p{0.34\columnwidth}
  >{\raggedright\arraybackslash}X
  @{}
}
\toprule
\textbf{Result} & \textbf{Source} \\
\midrule

Figure~\ref{fig:diffusion_comparison}
(OpenMP Off. {2,4}-GPU Versions)
&
\path{3D_Heat_Benchmark/*/Source-Results/*/summary.csv}
via \path{speedup.py}
\\

Figure~\ref{fig:p2p}
(Bandwidth benchmark)
&
\path{Bandwidth_Measurements/}
\\

Figure~\ref{fig:gains_openmp}
(Machines, HHTs, and $N^3$)
&
\path{HHT/plots-approach/OpenMP/}
\path{allN_gain_OpenMP.pdf}
via \path{gain.py}
\\

Figure~\ref{fig:decision-openmp}
(HHT recommendation)
&
\path{HHT/plots-approach/OpenMP/}
\path{decision_OpenMP.pdf}
via \path{gain.py}
\\

Figure~\ref{fig:model}
(Model versus measurements)
&
\path{3D_Heat_Benchmark/*/}
\path{Source-Results/model/plot_p2p.py}
\\

\bottomrule
\end{tabularx}
\end{table}

The HHT figures are regenerated with:

\begin{lstlisting}[basicstyle=\ttfamily\scriptsize, frame=none]
cd HHT
python3 gain.py --list        # verify discovery of the nine directories
python3 gain.py --format pdf
\end{lstlisting}

\subsection{Resource requirements}
\label{app:resources}

A full sweep is 1 build job plus 35 run jobs per machine and model directory,
each requesting one exclusive node for up to 90~minutes. Aggregation and
plotting run on a laptop in seconds and require \texttt{numpy},
\texttt{pandas} and \texttt{matplotlib}; the terminal summaries produced by
\texttt{plot.sh}.

\section{Schematic Execution Workflows}
\input{pictures/ASCII-plot}
To improve understanding of the pseudocode and parallelization strategies used in this research with various versions of OpenMP Offloading, the following figures provide a schematic overview of each method.
\clearpage

\bibliographystyle{unsrtnat}
\bibliography{article-bibliography}

@misc{nvidia_h100,
  author       = {{NVIDIA Corporation}},
  title        = {{NVIDIA H100 Tensor Core GPU}},
  howpublished = {Available online: \url{https://www.nvidia.com/en-us/data-center/h100/}},
  year         = {2026}, 
  note         = {(accessed on 7 July 2026)}
}

@misc{bsc_mn5,
  author       = {{Barcelona Supercomputing Center}},
  title        = {{MareNostrum 5}},
  howpublished = {Available online: \url{https://www.bsc.es/marenostrum/marenostrum-5}},
  year         = {2026}, 
  note         = {(accessed on 7 July 2026)}
}

@misc{amd_mi250x,
  author       = {{Advanced Micro Devices, Inc.}},
  title        = {{AMD Instinct MI250X Accelerator}},
  howpublished = {Available online: \url{https://www.amd.com/en/products/accelerators/instinct/mi200/mi250x.html}},
  year         = {2026}, 
  note         = {(accessed on 7 July 2026)}
}

@misc{lumi_g,
  author       = {{LUMI Consortium}},
  title        = {{LUMI-G} Hardware},
  howpublished = {Available online: \url{https://docs.lumi-supercomputer.eu/hardware/lumig/}},
  year         = {2026}, 
  note         = {(accessed on 7 July 2026)}
}

@misc{intel_max1550,
  author       = {{Intel Corporation}},
  title        = {{Intel Data Center GPU Max 1550}---Product Specifications},
  howpublished = {Available online: \url{https://www.intel.com/content/www/us/en/products/sku/232873/intel-data-center-gpu-max-1550/specifications.html}},
  year         = {2026}, 
  note         = {(accessed on 7 July 2026)}
}

@misc{lrz_supermucng2,
  author       = {{Leibniz Supercomputing Centre}},
  title        = {Hardware of {SuperMUC-NG} Phase 2},
  howpublished = {Available online: \url{https://doku.lrz.de/hardware-of-supermuc-ng-phase-2-222891050.html}},
  year         = {2026}, 
  note         = {(accessed on 7 July 2026)}
}

@misc{hpe_openacc_2024,
  author       = {{Hewlett Packard Enterprise}},
  title        = {Introduction to {OpenACC}},
  howpublished = {HPE Cray Programming Environment, version 24.03. Available online: \url{https://cpe.ext.hpe.com/docs/24.03/cce/man7/intro_openacc.7.html}},
  year         = {2024},
  note         = {(accessed on 14 July 2026)}
}

@article{Krishnasamy2015MultiGPU3DSweeping,
  author      = {Krishnasamy, Ezhilmathi and Sourouri, Mohammed and Cai, Xing},
  title       = {Multi-{GPU} Implementations of Parallel {3D} Sweeping Algorithms with Application to Geological Folding},
  journal     = {Procedia Comput. Sci.},
  volume      = {51},
  pages       = {1494--1503},
  year        = {2015},
  doi         = {10.1016/j.procs.2015.05.339}
}

@unpublished{prior2026,
  author      = {Krishnasamy, Ezhilmathi and Trotter, James D. and Cai, Xing and Bouvry, Pascal},
  title       = {Performance Evaluation of the OpenACC API Targeting Multi-GPU Data Transfers: A
Case Study of Porting Parallel 3D Sweeping Algorithms for Geological Folding},
  year        = {2025},
  note        = {Manuscript submitted for publication (under review)}
}

@inproceedings{Krishnasamy2026MPIXIntegration,
  author      = {Krishnasamy, Ezhilmathi and Trotter, James and Cai, Xing and Pleiter, Dirk and Kos, Leon and Saavedra, Laura and Bouvry, Pascal},
  title       = {Performance and Programmability of {MPI+X} Integration with {CUDA}, {HIP}, {SYCL}, {OpenACC}, and {OpenMP} Offloading for Supercomputing: A Case Study on Dense Matrix--Vector Multiplication},
  booktitle   = {Proceedings of the Supercomputing Asia and International Conference on High Performance Computing in Asia Pacific Region Workshops},
  address     = {Osaka, Japan},
  month       = {26--29 January},
  pages       = {457--468},
  year        = {2026},
  doi         = {10.1145/3784828.3786264}
}

@inproceedings{Krishnasamy2025OpenMPAMDandNVIDIA,
  author      = {Krishnasamy, Ezhilmathi and Bouvry, Pascal},
  title       = {{OpenMP} Offloading on {AMD} and {NVIDIA} {GPUs}: Programmability and Performance Analysis},
  booktitle   = {Proceedings of the 2025 9th International Conference on High Performance Compilation, Computing and Communications},
  address     = {Jinan, China},
  month       = {27--29 August},
  pages       = {44--56},
  year        = {2025},
  doi         = {10.1145/3774949.3774956}
}

@inproceedings{Krishnasamy2026ComparativeCUDAOpenMP,
  author      = {Krishnasamy, Ezhilmathi and Bouvry, Pascal},
  title       = {Comparative Performance Analysis of {CUDA} and {OpenMP} Offloading for {BLAS} Operations on {GPU}},
  booktitle   = {ICT for Intelligent Systems: Proceedings of ICTIS 2025},
  address     = {New York, NY, USA},
  month       = {23--24 May 2025},
  volume      = {126},
  pages       = {9--35},
  year        = {2026},
  doi         = {10.1007/978-981-95-1361-1_2}
}

@inproceedings{Krishnasamy2025OpenACCCUDA,
  author      = {Krishnasamy, Ezhilmathi and Bouvry, Pascal},
  title       = {A Study of {GPU} Programming Paradigms: {OpenACC} vs. {CUDA} in Linear Algebra Computations},
  booktitle   = {Proceedings of the 2025 5th International Conference on Advances in Electrical, Electronics and Computing Technology (EECT)},
  address     = {Guangzhou, China},
  month       = {21--23 March},
  year        = {2025},
  doi         = {10.1109/EECT64505.2025.10966957}
}

@inproceedings{fridman2023portability,
  author      = {Fridman, Yehonatan and Tamir, Guy and Oren, Gal},
  title       = {Portability and Scalability of {OpenMP} Offloading on State-of-the-Art Accelerators},
  booktitle   = {High Performance Computing},
  address     = {Hamburg, Germany},
  month       = {21--25 May},
  volume      = {13999},
  pages       = {378--390},
  year        = {2023},
  doi         = {10.1007/978-3-031-40843-4_28}
}

@incollection{Krishnasamy2024TsunamiOpenMP,
  author      = {Krishnasamy, Ezhilmathi and Harig, Sven and Bouvry, Pascal},
  title       = {Accelerating Tsunami Computation on a {GPU} Using {OpenMP} Offloading},
  booktitle   = {Computer and Information Science and Engineering},
  publisher   = {Springer},
  address     = {Cham, Switzerland},
  volume      = {1156},
  pages       = {163--178},
  year        = {2024},
  doi         = {10.1007/978-3-031-57037-7_12}
}

@mastersthesis{Krishnasamy2014HybridCPUGPU,
  author      = {Krishnasamy, Ezhilmathi},
  title       = {Hybrid {CPU--GPU} Parallel Simulations of {3D} Front Propagation},
  school      = {Link{\"o}ping University},
  address     = {Link{\"o}ping, Sweden},
  year        = {2014},
  note        = {Available online: \url{https://urn.kb.se/resolve?urn=urn:nbn:se:liu:diva-114935} (accessed on 14 July 2026)}
}

@misc{nvidia_cuda_programming_guide,
  author       = {{NVIDIA Corporation}},
  title        = {{CUDA C++ Programming Guide}},
  year         = {2026},
  howpublished = {\url{https://docs.nvidia.com/cuda/cuda-c-programming-guide/}},
  note         = {Release 13.3. Accessed: 2026-07-19}
}

@misc{amd_hip_programming_guide,
  author       = {{Advanced Micro Devices, Inc.}},
  title        = {{HIP Programming Guide (ROCm Documentation)}},
  year         = {2026},
  howpublished = {\url{https://rocm.docs.amd.com/en/latest/how-to/programming_guide.html}},
  note         = {Heterogeneous-Compute Interface for Portability. Accessed: 2026-07-19}
}

@misc{khronos_opencl_api_spec,
  author       = {{Khronos OpenCL Working Group}},
  title        = {{The OpenCL Specification, Version 3.0}},
  year         = {2025},
  howpublished = {\url{https://registry.khronos.org/OpenCL/specs/3.0-unified/html/OpenCL_API.html}},
  note         = {Revision 3.0.19. Accessed: 2026-07-19}
}

@misc{khronos_sycl_2020_spec,
  author       = {{Khronos SYCL Working Group}},
  title        = {{SYCL 2020 Specification (Revision 11)}},
  year         = {2025},
  howpublished = {\url{https://registry.khronos.org/SYCL/specs/sycl-2020/html/sycl-2020.html}},
  note         = {Published 2025-11-07. Accessed: 2026-07-19}
}

@misc{openmp_6_offload_overview,
  author       = {{OpenMP Architecture Review Board}},
  title        = {{OpenMP 6.0: Offloading and Device Features}},
  year         = {2024},
  howpublished = {\url{https://www.openmp.org/articles/openmp-6/}},
  note         = {Accessed: 2026-07-19}
}

@misc{openacc_34_spec,
  author       = {{OpenACC Organization}},
  title        = {{The OpenACC Application Programming Interface, Version 3.4}},
  year         = {2025},
  howpublished = {\url{https://www.openacc.org/sites/default/files/inline-images/Specification/OpenACC-3.4.pdf}},
  note         = {Accessed: 2026-07-19}
}

@misc{openacc_specification,
  author       = {{OpenACC Organization}},
  title        = {{OpenACC Specification and Resources}},
  year         = {2025},
  howpublished = {\url{https://www.openacc.org/specification}},
  note         = {Accessed: 2026-07-19}
}

@inproceedings{Ferat2022EnhancingMPI,
  author    = {Ferat, Manuel and Pereira, Romain and Roussel, Adrien and
               Carribault, Patrick and Steffenel, Luiz-Angelo and
               Gautier, Thierry},
  title     = {Enhancing {MPI+OpenMP} Task Based Applications for
               Heterogeneous Architectures with {GPU} Support},
  booktitle = {{OpenMP} in a Modern World: From Multi-Device Support
               to Meta Programming ({IWOMP} 2022)},
  editor    = {Klemm, Michael and de Supinski, Bronis R. and
               Klinkenberg, Jannis and Neth, Brandon},
  series    = {Lecture Notes in Computer Science},
  volume    = {13527},
  pages     = {3--16},
  publisher = {Springer},
  address   = {Cham, Switzerland},
  year      = {2022},
  doi       = {10.1007/978-3-031-15922-0_1}
}

@inproceedings{Tian2022HiddenHelperThreads,
  author    = {Tian, Shilei and Doerfert, Johannes and Chapman, Barbara},
  title     = {Concurrent Execution of Deferred {OpenMP} Target Tasks
               with Hidden Helper Threads},
  booktitle = {Languages and Compilers for Parallel Computing
               ({LCPC} 2020)},
  editor    = {Chapman, Barbara and Moreira, Jos{\'e}},
  series    = {Lecture Notes in Computer Science},
  volume    = {13149},
  pages     = {41--56},
  publisher = {Springer},
  address   = {Cham, Switzerland},
  year      = {2022},
  doi       = {10.1007/978-3-030-95953-1_4}
}

@article{lindholm2008nvidia,
  title={NVIDIA Tesla: A Unified Graphics and Computing Architecture},
  author={Lindholm, Erik and Nickolls, John and Oberman, Stuart and Montrym, John},
  journal={IEEE Micro},
  volume={28},
  number={2},
  pages={39--55},
  year={2008},
  doi={10.1109/MM.2008.31}
}

@article{flynn1972some,
  author  = {Flynn, Michael J.},
  title   = {Some Computer Organizations and Their Effectiveness},
  journal = {IEEE Transactions on Computers},
  volume  = {C-21},
  number  = {9},
  pages   = {948--960},
  month   = sep,
  year    = {1972},
  doi     = {10.1109/TC.1972.5009071}
}

@inproceedings{williams2007,
  author    = {Williams, Samuel and Oliker, Leonid and Vuduc, Richard and
               Shalf, John and Yelick, Katherine and Demmel, James},
  title     = {Optimization of Sparse Matrix-Vector Multiplication on
               Emerging Multicore Platforms},
  booktitle = {Proceedings of SC'07}, year = {2007}
}

@book{saad2003,
  author    = {Saad, Yousef},
  title     = {Iterative Methods for Sparse Linear Systems},
  edition   = {2}, publisher = {SIAM}, year = {2003}
}

@article{tinney1967,
  author  = {Tinney, William F. and Walker, John W.},
  title   = {Direct Solutions of Sparse Network Equations by Optimally
             Ordered Triangular Factorization},
  journal = {Proceedings of the IEEE},
  volume  = {55}, number = {11}, pages = {1801--1809}, year = {1967}
}

@book{heath2018,
  author    = {Heath, Michael T.},
  title     = {Scientific Computing: An Introductory Survey},
  edition   = {2, revised},
  series    = {Classics in Applied Mathematics},
  publisher = {SIAM},
  address   = {Philadelphia, PA},
  year      = {2018},
}

\end{document}